\documentclass[12pt,thmsa]{article}
\usepackage{amssymb}

\usepackage{sw20lart}

\input{tcilatex}
\begin{document}

\author{Tor Fl\aa \and Institute of Mathematical and Physical Sciences \and
University of Troms\o\ \and N-9037 Troms\o , Norway\thanks{%
E-mail: tor@math.uit.no}}
\title{A Hybrid Fluid-Kinetic Theory for Plasma Physics }
\maketitle

\begin{abstract}
We parameterize the phase space density by time dependent diffeomorphic,
Poisson preserving transformations on phase space acting on a reference
density solution. We can look at these as transformations which fix time on
the extended space of phase space and time. In this formulation the Vlasov
equation is replaced by a constraint equation for the above maps. The new
equations are formulated in terms of hamiltonian generators of one parameter
families of diffeomorphic, Poisson preserving maps e.g. generators with
respect to time or a perturbation parameter. We also show that it is
possible to parameterize the space of solutions of the Vlasov equation by
composition of maps subject to certain compatibility conditions on the
generators. By using this composition principle we show how to formulate new
equations for a hybrid fluid kinetic theory. This is done by observing that
a certain subgroup of the group of phase space maps with generators which
are linear in momentum correspond to the group of diffeomorphic maps
parameterizing the continuity equation in fluid theory.
\end{abstract}

\section{Introduction}

The development of collisionless plasma physics in recent years have had
some interesting breakthroughs. Among these should especially be mentioned
the recently found Eulerian action principles for plasma physics(Larsson$^{%
\text{1,2}}$, Ye and Morrison$^{\text{3}}$, and Fl\aa{}$^{\text{4}}$). The
intention of this work is to demonstrate that some of the substantial
problems in developping good models in plasma physics comes partially from
purely formal problems with how to formulate model equations in an efficient
, invariant language. A key problem is how to discriminate between the
incoherent(containing heat fluctuations and resonant particle interactions)
and coherent, fluid part of the plasma fluctuations. For this purpose we
will use the flexibility of the generator approach to separate between
generators of the fluid motion to first order in momentum and an incoherent
part of the generators to higher order in momentum.

We will be interested in reformulating the continuity equation and the
Vlasov equation in terms of the action of infinite dimensional
transformations on space and phase space on the density and phase space
density. By parameterizing the densities by these transformations with
respect to reference densities, we find that the continuity equation for
densities in fluid and kinetic theory are consistent with that the action of
the transformations on reference fluid velocity and hamiltonian vectorfield
plus a timelike generator of the transformations are constrained to be equal
to the actual fluid velocity and hamiltonian vectorfield. We thus replace
the continuity equations by constraint equations for the time dependent
transformations. Moreover, we demonstrate in App. A that the continuity
fluid equation and the Vlasov equation can be looked upon as the defining
equations for an infinite dimensional pseudogroup on space and phase space
extended with time, but constrained to transformations which fixes time. The
moment we have realized this an interesting composition principle pops up.
Namely, we have not only discovered one new equation, but an infinitely many
corresponding to different compositions of tranformations which are
compatible with the above density structures. Our philosophy is that we
encode a priori information into the choice of composition transformations.
The a priori information restrict the class of experiments or processes
which the constructed theory is intended to describe, and every composition
will indeed give new equations. In this respect our philosophy is analogous
to the theory of measurements in quantum mechanics where every measurement
is related to a new class of operators. In our case, for every measurement
or class of experiments there correspond a composition principle and
corresponding equations. In our opinion this principle put the construction
of invariant wave equations and field theory into a new light (and we
believe not only in plasma physics) since to every a priori information
encoded there correspond new wave operators with corresponding new spectrum
and dispersion relations.

In this work we will try this new principle by making a hybrid fluid-kinetic
theory for collisionless plasma physics. Thus we restrict our attention to a
class of experiments where it makes sense to measure fluid density and
velocity. If these quantities are not reasonable to measure, the theory of
course does not apply. The moment we have decided on this extra bit of a
priori information, we have a new theory since we can explicitly construct
and separate out the fluid generator from the incoherent generator
containing heat fluctuations and resonant particle interactions.\footnote{%
This notation is however not quite precise since for an experimental
situation where it is reasonable to define and measure also the stress
tensor as a fluid variable, one would try to separate out a fluid equation
also for this variable by an additional transformation with a generator
which is second order in momentum. Similar developments and introduction of
a priori information can in principle be done to any order in momentum and
thus put tighter and tighter constraints on the class of experiments and
measurements which can be described by the corresponding theories. The
resonant particle effects will for such theories be described by higher
order generators (in momentum) in interaction with the the corresponding
fluid variables..} \emph{It must be noted that in the ordinary Vlasov
equation the heat fluctuations are not explicitly described as mode of
fluctuation like e.g. in fluid theory. Even in the constant background case
the heat fluctuations will be hidden in the continous spectrum. With our
invention of hybrid fluid kinetic theory the heat fluctuations will appear
as a mode in any background or size of fluctuations due to the new operators
appearing.}

The plan of the paper is that we will use canonical coordinates to separate
the fluid generators from the the incoherent kinetic generators. This is
done to obtain as simple presentation as possible. Then we introduce what we
call interaction physical coordinates in which the Poisson bracket is
noncanonical, but still fixed by reference electromagnetic fields . These
coordinates are essential since otherwise we would have to perturb the
brackets also. Then we introduce kinetic fluid generators which is related
to the above, but which can be interpreted as near identity transformations
and moreover coincide with the fluid generators when we integrate the phase
space density over momentum.

\section{Parameterization of the Vlasov phase space density}

For a given reference distribution $f^{0}$, it turns out that the accessible
part of the space of distributions can be traced by canonical displacements,
i.e. $\frac{\partial f}{\partial \epsilon }=\left\{ S,f\right\} $, where $S$
is any hamiltonian (e.g. Larsson$^{\text{1,2}}$). These are the only
allowable displacements in a collisionless plasma. Another way to
parameterize such displacements are by near identity transformations (see
the references 3,4). 
\begin{eqnarray*}
f &=&\exp (\mathcal{L}_{w})f^{0}, \\
\mathcal{L}_{w} &\equiv &\{w,\cdot \}\;.
\end{eqnarray*}
As we will see a more general way of expressing the above result is in terms
of the action of Poisson preserving maps $f=\psi \bullet f^{0}$. (The
notation is explained in Appendix A and in the text below.)

In the litterature one has often considered this parameterization of
solutions of the Vlasov equation to be a result of the canonical
transformations resulting from the underlying particle orbits. A different ,
may be more natural point of view, is to consider the above parameterization
as a result of that the Vlasov equation is a Lie equation$^{\text{5,6}}$
having an infinite dimensional symmetry group preserving density on extended
phase space. In App. A we have elaborated on this point of view.

The Vlasov equation in canonical coordinates has the form 
\begin{equation}
\frac{\partial f}{\partial t}+\{f,H\}=0.\;  \label{eq1}
\end{equation}

This equation simply expresses conservation of phase space density, $\omega
(t)=f(z,t)d^{6}z$, in a Hamiltonian flow. In this report we will basically
use canonical phase space coordinates to derive our hybrid fluid kinetic
theory since this leads to that the brackets are not perturbed and the
Jacobian is unity for canonical transformations.This is approach give simple
derivations, but leads to no loss of generality. We will show how to apply
the method in physical euclidean coordinates\footnote{%
Since the fluid and electromagnetic fluctuations naturally are divided into
divergent and divergencefree parts it is actually necessary to introduce
these concepts on Riemannian manifold with respect to a metric even if the
laboratory frame is euclidean. The reason why is that the above physical
division of fluctuations are related to Hodge decomposition which transforms
in a nontrivial way with respect to diffeomorphisms. We will return to this
invariant presentation of the fluid and electromagnetic theory in coming
papers.} also by introducing a fixed Poisson bracket in physical coordinates
which is not perturbed (see App. A and below). In effect, if one restricts
to canonical coordinates it is possible to use the canonical distribution
function instead of the density volumeform as the basic entity. We
demonstrate in App. A that eq. (\ref{eq1}) is equivalent to that
distribution function is parameterized by canonical transformations on the
phase space with respect to a reference solution of the Vlasov equation 
\begin{equation}
f=\psi ^{-1*}f^{0}\equiv f^{0}\circ \psi ^{-1}.  \label{eq2}
\end{equation}
Here $\psi $ is a canonical (i.e. Poisson bracket preserving) transformation
of the phase space P, i.e. $\psi ^{-1*}\{g,h\}=\{\psi ^{-1*}g,\psi ^{-1*}h\}$%
. The infinitesimal version eq. (\ref{eq2}) is expressed by hamiltonian
generators. The hamiltonian generator with respect to the time parameter is
given by

\begin{equation}
f,_{t}=\{\psi _{t},f\}+\psi ^{-1*}(f^{0},_{t})=\{\psi _{t}+\psi
^{-1*}H_{0},f\}  \label{eq3}
\end{equation}
Here we have assumed that the distribution function and that the canonical
transformation depend parameterically on $t$, i.e. $f(t)$ and $\psi (t).$ If
we in addition assume that they depend on one (or several) additional
parameter $\epsilon $, i.e. $f(t,\epsilon )$ and $\psi (t,\epsilon )$, we
can also define a hamiltonian generator, $\psi _{\epsilon }$ , with respect
to $\epsilon $ as

\begin{equation}
f,_\epsilon =\{\psi _\epsilon ,f\}  \label{eq4}
\end{equation}

We could think of this additional parameter as a formal perturbation
parameter which vary say between 0 to 1 corresponding to $f^{0},H_{0}$ and $%
f(t),\,H$ respectively. It could, however, have other interpretations
(e.g.describing a one parameter symmetry). In the case that the
transformation is composed of two canonical transformations we have

\begin{eqnarray}
\psi &=&\bar{\psi}\circ \tilde{\psi},  \label{eq5} \\
\psi ^{*} &=&(\bar{\psi}\circ \tilde{\psi})^{-1*}=\bar{\psi}^{-1*}\circ 
\tilde{\psi}^{-1*},  \nonumber \\
f &=&\bar{\psi}^{-1*}\circ \tilde{\psi}^{-1*}f^0.  \nonumber
\end{eqnarray}

For composite transformations it is realized that the generators has to obey
the following rule since they are derived from derivatives with respect to
parameters(see App. A) 
\begin{eqnarray}
\psi _{t} &=&\overline{\psi }_{t}+\overline{\psi }^{-1*}\widetilde{\psi }%
_{t},  \label{eq6} \\
\psi _{\epsilon } &=&\overline{\psi }_{\epsilon }+\overline{\psi }^{-1*}%
\widetilde{\psi }_{\epsilon }.  \nonumber
\end{eqnarray}

We also know from the assumption that we can interchange the $\epsilon ,t$
derivatives for the distribution function that the following Maurer-Cartan
relation must hold [App. A]\footnote{%
Notice that the $t$ and $\epsilon $ index in e.g.$\psi _{t}$ and $\psi
_{\epsilon }$ are not derivatives. Rather these are the hamiltonian
generators corresponding to the hamiltonian vectorfields $\mathbf{\psi }_{t}$
and $\mathbf{\psi }_{\epsilon }$ defined in App. A. Derivatives will be
distinguished from an index by a comma or by explicit derivation symbols.}

\begin{eqnarray}
f,_{t\epsilon } &=&f,_{\epsilon t}\;,  \label{eq7} \\
\psi _t,_\epsilon -\psi _\epsilon ,_t+\{\psi _t,\psi _\epsilon
\}+k_{t\epsilon } &=&0,  \nonumber \\
k_{t\epsilon } &=&\psi ^{-1*}k_{t\epsilon }^0,  \nonumber \\
\{k_{t\epsilon },f\} &=&0,\,\{k_{t\epsilon }^0,f^0\}=0.  \nonumber
\end{eqnarray}

Here we obviously can extend our notation by treating the phase space
coordinates $z_{i},\,i=1,..,6$ as parameters and define the hamiltonian
generators $\psi _{i}$ \footnote{%
Our definition of hamiltonian generators with respect to different parameter
variations reflects the fact that our parameterization of the Vlasov
equation are Poisson preserving maps. Only in the case that the evolution of
the background distribution is hamiltonian or independent with respect to
the parameter in question. Hamiltonian evolution will therefore be realized
for time and perturbation parameters which we already have observed. We also
realize that in case the background distribution has an ignorable
coordinate- i.e. a symmetry, the evolution with respect to this coordinate
of the phase space density will be hamiltonian.} (Similarly one can
generalize to generators for other parameters like the noncanonical guiding
center coordinates or oscillation center coordinates. See a brief discussion
of presentation in other coordinates in App. B).

\[
f,_i=\{\psi _i,f\}+\psi ^{-1*}f^0,_i\,\;,\;i=1,..,6. 
\]

The compatibility condition or the Maurer-Cartan relation then takes the
form for the seven coordinates $(z,t)$ (here we don't use the $\epsilon $
coordinate)

\begin{eqnarray}
\psi _i,_j-\psi _j,_i+\{\psi _i,\psi _{j\}}+k_{ij} &=&0,\;i,j=1,..,7,
\label{eq8} \\
k_{ij} &=&\psi ^{-1*}k_{ij}^0,\;k_{ij}=-k_{ji},  \nonumber \\
\{k_{ij},f\} &=&0,\,\{k_{ij,}^0f^0\}=0.  \nonumber
\end{eqnarray}

The Hamiltonian generators and the Maurer-Cartan relation can only be
understood invariantly in the language of forms. We will not describe this
more general formalism here.

\section{Hybrid fluid-kinetic theory}

In an earlier paper$^{\text{4}}$ we have elaborated on the formal expansions
of the distribution function with respect to near identity
symplectomorphisms parameterized by exponential maps. We purposely did not
use the term pull back map in that paper, but all the results can be
verified by interchanging the exponential map with the pull back map%
\footnote{%
In ref.4 we used partial integration in the variational functionals to
invert the action of the exponential symplectic transformations from one
object to another. This trick cannot be done with generic pull back maps
instead of exponential symplectic maps, but the end results are still valid.
The reason is that one obtain variational equivalent functionals after doing
partial integration with respect to the exponential symplectic
transformations. We define variational equivalent functionals to mean all
functionals which gives the same result after variation. The trick of
partial integration can still be performed after performing variations (see
App. B).}. In this work we found the following equation for the the
Hamiltonian generator in the time direction and consequently the
compatibility condition if we have a additional parameter $\epsilon $

\begin{eqnarray}
\psi _{t} &=&H-\psi ^{-1*}H_{0}\;,  \label{eq9} \\
\psi _{\epsilon ,t}+\{\psi _{\epsilon },H\}-H,_{\epsilon } &=&0  \nonumber
\end{eqnarray}
Let us now use our formalism for composite transformations to try to develop
interacting equations for fluid and kinetic degrees of freedom. We now
assume that our distribution are described by 
\begin{eqnarray}
f &=&\overline{\psi }^{-1*}\widetilde{f}=\overline{\psi }^{-1*}\circ 
\widetilde{\psi }^{-1*}\tilde{f}^{0}\,,  \label{eq10} \\
f^{0} &=&(\bar{\psi}^{0})^{-1*}\tilde{f}^{0}\;.
\end{eqnarray}
Here $\overline{\psi }$ and $\widetilde{\psi }$ is supposed to contain the
fluid degrees of freedom (i.e. mass density and momentum) and the
'incoherent' kinetic degrees of freedom respectively. $\bar{\psi}_{0}$ is
the map $\bar{\psi}\,$in the reference state. Notice that the order of the
composition of the fluctuating and the averaged transformations are in the
opposite order than usually used in passive coordinate transformations where
one want to define hypothetical averaged coordinates.(c.f. Fl\aa $^{\text{4}%
} $ where we briefly discussed the opposite ordering.) The reason for our
choice is our goal of separating out the fluid generators from the resonant
distribution $\tilde{f}$ and obtain a new Liouville equation for this
distribution. With another goal in mind other choices could very well be
preferrable. To avoid additional tranformations due to the perturbations of
the generators this is more suitable.

In App. A we have discussed a similar formalism as the above for ideal fluid
theory and time dependent volume preserving transformations on space-time
which fixes time (see discussion in App. A). In this case one has to use
vectorfield generators $\mathbf{\psi }_t$ and $\mathbf{\psi }_\epsilon $ .
The parameterisation of the mass density and the constraint equation for the
fluid generator given reference density and velocity are 
\begin{eqnarray}
\widehat{\mathbf{u}} &=&\mathbf{\psi }_t+\mathbf{\psi }_{*}\mathbf{u}_0\;,
\label{eq11} \\
\rho ,_t &=&-\nabla \cdot (\widehat{\mathbf{u}}\rho )\;,  \nonumber \\
\rho ,_\epsilon &=&-\nabla \cdot (\mathbf{\psi }_\epsilon \rho )  \nonumber
\\
\delta \rho &=&-\nabla \cdot (\delta \mathbf{\bar{\psi}}\rho ), \\
\mathbf{\psi }_t &=&\frac{\partial \mathbf{\psi }}{\partial t}\circ \mathbf{%
\psi }^{-1}\,,\;\mathbf{\psi }_\epsilon =\frac{\partial \mathbf{\psi }}{%
\partial \epsilon }\circ \mathbf{\psi }^{-1}, \\
\delta \mathbf{\bar{\psi}} &=&\delta \mathbf{\psi \circ \psi }^{-1}\,.
\end{eqnarray}
The parameterized velocity field $\mathbf{\hat{u}}$ give a constraint
equation for the above diffeomorphisms when we give the velocity field $%
\mathbf{u=\hat{u}}$ , e.g. as in our case from the momentum equation. Here
the invariant object is not mass density, but the density form $\omega =\rho
d^3xdt$ which the pull back map acts properly on. The volume density
preserving pull back maps can also be presented as an action directly on the
density by taking into account the Jacobian of the mapping. We introduce the
symbol $\mathbf{\psi }\bullet $ for this action which also can be
parameterized by a volume density preserving near identity transformation in
the following way (see App.A)

\begin{equation}
\rho (\mathbf{x},t)=\mathbf{\psi }\bullet \rho ^{0}(\mathbf{x},t)\equiv \rho
^{0}(\mathbf{\psi }^{-1}(\mathbf{x,}t\mathbf{),}t)J\;.  \label{eq12}
\end{equation}
In the case of near identity transformations one can further express the
action in terms of the near identity generator $\mathbf{w}$ as 
\begin{eqnarray}
\mathbf{\psi \bullet (\cdot )} &\equiv &\exp (-\nabla \cdot (\mathbf{w\cdot
))(\cdot )\;,}  \label{eq12b} \\
\mathbf{\psi }_{*} &=&\exp (ad(\mathbf{w)),} \\
\mathbf{\psi } &=&\exp (\mathbf{w).}  \nonumber
\end{eqnarray}

Here $ad(\mathbf{w)\equiv [\cdot ,w]}$ in terms of the standard bracket for
vectorfields(see definition in App. A). The proof of these expressions
follows from using $\mathbf{\psi }=\exp (\mathbf{w)}$ in the above
definitions of the actions of diffeomorphisms on densities and vectorfields
and compare it with the action of the above operators. Notice, that the
fundamental operators in the exponential is Lie derivatives of the
corresponding object with respect to the near identity generator
vectorfield. Further, we can express the vectorfield generators in terms of
near identity generators as (see the definition in App.A)

\begin{eqnarray}
\mathbf{\psi }_{t} &=&(\mathbf{\psi }_{*}-Id)\frac{\partial }{\partial t}%
=i\exp (ad(\mathbf{w))}\frac{\partial \mathbf{w}}{\partial t},  \label{eq12c}
\\
\mathbf{\psi }_{\epsilon } &=&(\mathbf{\psi }_{*}-Id)\frac{\partial }{%
\partial \epsilon }=i\exp (ad(\mathbf{w))}\frac{\partial \mathbf{w}}{%
\partial \epsilon }  \nonumber \\
\delta \mathbf{\bar{\psi}} &=&i\exp (ad(\mathbf{w)})\delta \mathbf{w,} \\
i\exp (x) &\equiv &\frac{\exp (x)-1}{x}\,.
\end{eqnarray}

If we use the identities established in App.A to transform from the above
type of expressions on phase space and Hamiltonian vectorfields to
Hamiltonian near identity generators, $w$, we obtain that 
\begin{eqnarray}
\psi &=&\exp (X_{w})\,,  \label{eq12d} \\
\psi ^{*} &=&\exp (-\mathcal{L}(X_{w}))\,,  \nonumber \\
\psi _{t} &=&i\exp (\mathcal{L}_{w})\frac{\partial w}{\partial t\,}\,, 
\nonumber \\
\psi _{\epsilon } &=&i\exp (\mathcal{L}_{w})\frac{\partial w}{\partial
\epsilon }\,,  \nonumber \\
\delta \psi &=&i\exp (\mathcal{L}_{w})\delta w \\
\{w,\cdot \} &\equiv &\mathcal{L}_{w}\,\,.  \nonumber
\end{eqnarray}
Here, the Lie derivative $\mathcal{L}(X_{w})=-\mathcal{L}_{w};$ act as the
operator $X_{w}=-\mathcal{L}_{w}$ on functions, but act of course
differently on other objects. The above interpretation corresponds exactly
to the point of view we proposed at an earlier stage in Fl\aa $^{\text{4}}.$

The above formal expansions in terms of near identity generators can be
thought of to correspond to Larssons perturbation expansion when an ordering
is given to operators (see App.C and Larsson$^{\text{1,2}}$. However,
because of our group composition concept, we have considerable freedom when
it comes to modelling of specific physical processes. At all steps in our
theory it will be possible to specialize to near identity generators, but we
will not stress this below.

We can also think of the above maps as a family of diffeomorphisms on space
parameterized by time. Here $J$ is the Jacobian of the transformation. The
negative sign in the near identity transformation is used to obtain an
adequate sign in the continuity equation. We used the same reason for
positive sign in the parametrization of the phase space density which
corresponds to negative sign in the corresponding Hamiltonian vectorfield.
The composition of the above maps is similar as for symplectic pull back
maps. The compatibility condition with respect to an additional parameter $%
\epsilon $, is given by (see App. A)

\begin{eqnarray}
\mathbf{\psi }_{\epsilon ,t}-\mathbf{\psi }_{t,\epsilon }-[\mathbf{\psi }%
_{\epsilon },\mathbf{\psi }_{t}]+\mathbf{k}_{\epsilon t} &=&\mathbf{0\;,}
\label{eq13} \\
\nabla \cdot (\mathbf{k}_{\epsilon t}\rho ) &=&0,\,\nabla \cdot (\mathbf{k}%
_{\epsilon t}^{0}\rho ^{0})=0\;,  \nonumber \\
\mathbf{k}_{\epsilon t} &=&\mathbf{\psi }_{*}\mathbf{k}_{\epsilon t}^{0}\;. 
\nonumber
\end{eqnarray}

Since the velocity field can be parameterized as in eq. (\ref{eq11}), we
obtain that the compatibility relation can also be written as 
\begin{equation}
\mathbf{\psi }_{\epsilon ,t}-\widehat{\mathbf{u}},_{\epsilon }-[\mathbf{\psi 
}_{\epsilon },\widehat{\mathbf{u}}]=\mathbf{0\func{mod}(k}_{\epsilon t})\;.
\label{eq14}
\end{equation}

It turns out that it is possible to lift the fluid generating maps to
kinetic theory and consider them as a subgroup of kinetic,canonical
transformations through the definition of the following canonical generators 
\footnote{%
These symplectic transformations cannot be defined through near identity
transformations since the resonant particle distribution will be following
fluid orbits. This is done for mathematical convenience of the separation
procedure. Later on when we linearize the hybrid fluid kinetic equations, we
will relate these symplectic transformations to the fluid generators with
respect to a reference state.} 
\begin{eqnarray}
\overline{\psi }_{t}(z,t) &=&\overline{\psi }_{1t}(z,t)+\overline{\psi }%
_{2t}(\mathbf{x},t)=\mathbf{p}\cdot \widehat{\mathbf{u}}+\overline{\psi }%
_{2t}\;,  \label{eq15} \\
\overline{\psi }_{2t} &=&-m\frac{\widehat{\mathbf{u}}^{2}}{2}-\frac{e}{c}%
\mathbf{A}\cdot \widehat{\mathbf{u}}\;, \\
\overline{\psi }_{\epsilon }(z,t) &=&\overline{\psi }_{1\epsilon }(z,t)+\bar{%
\psi}_{2\epsilon }(\mathbf{x},t)=\mathbf{p}\cdot \mathbf{\psi }_{\epsilon }+%
\overline{\psi }_{2\epsilon }\;,  \nonumber \\
\overline{\psi }_{1t} &\equiv &\mathbf{p}\cdot \mathbf{\widehat{u}\;,\;}%
\overline{\psi }_{1\epsilon }\equiv \mathbf{p}\cdot \mathbf{\psi }_{\epsilon
}\;.
\end{eqnarray}
Note that we have assumed that the generator $\overline{\psi }%
_{2a}\,,a=t,\epsilon $ depend only on $\mathbf{x},t.$ With the above
definitions one easily convince oneself that the compatibility conditions
for the barred symplectic transformation, $\bar{\psi}_{1a},a=t,\epsilon $ is
consistent with the fluid compatibility condition in inner product with $%
\mathbf{p.}$

The explicit form of the above kinetic fluid generating maps can be
described in terms of cotangent lift $\bar{\Psi}(t)(t)\equiv T^{*}\mathbf{%
\phi }(t)$ in composition with the fibertranslation by an exact form
described below.$.$ In Abraham and Marsden it is proven that the cotangent
lift is a symplectic map which preserves the canonical oneform $\theta $ on
phase space $T^{*}M$ $(=\mathbf{p\cdot }d\mathbf{x}$ in euclidean canonical
coordinates). For a point $\alpha _{q}\in T^{*}M=P$ and $\mathbf{v\in }T_{%
\mathbf{\phi (}t)^{-1}(q)}M$ one find that $T^{*}\mathbf{\phi }(t)\alpha
_{q}(\mathbf{v)=}\alpha _{q}(T\mathbf{\phi }(t)\mathbf{v).}$ Define the
coordinate functions on $T^{*}M$ by $\pi _{P}:\alpha _{q}\rightarrowtail \pi
_{P}(\alpha _{q})=z=(\mathbf{x,p).}$ The action on coordinate functions is
therefore $(T^{*}\mathbf{\phi (}t))^{*}\pi _{P}$ which for a euclidean
metric means that the action of the cotangent lift is $\bar{\Psi}%
(t)(z)\equiv \bar{Z}=T^{*}\mathbf{\phi (}t)(z)=(\mathbf{\phi }(t)^{-1}(%
\mathbf{x),(}\nabla \mathbf{\phi (}t)(\mathbf{x))}\cdot \mathbf{p).}$ A
simple way to find invariant properties of the above map is to use that the
canonical oneform is preserved under cotangent lift, i.e. (here we treat $%
p_{i},x^{i}$ as coordinate functions) 
\begin{eqnarray*}
\theta &=&\bar{\Psi}(t)^{-1*}\theta , \\
&\Rightarrow &p_{j}dx^{j}=\bar{P}_{j}d\bar{\Psi}(t)^{-1*}x^{j}, \\
\bar{P}_{j} &=&\bar{\Psi}(t)^{-1*}p_{j}=((\nabla \mathbf{\phi (}t\mathbf{%
)(x)\cdot p)}_{j}\,.
\end{eqnarray*}
Another map which will interest us is fibertranslations by oneforms $\mathbf{%
A}^{(1)}(t)$ on $M$ , defined by $\psi _{\mathbf{A}}^{-1*}(t)\theta =\theta
-\pi _{M}^{*}\mathbf{A}^{(1)}(t)$ where $\pi _{M}:T^{*}M\rightarrow M$. In
this case the map is only preserving the symplectic tensor $\omega =-\hat{d}%
\theta $ if the oneform $\mathbf{A}^{(1)}(t)$ is exact since $\psi _{\mathbf{%
A}}^{-1*}(t)\omega =\omega +\pi _{M}^{*}\hat{d}\mathbf{A}^{(1)}(t).$ In
euclidean canonical coordinates the fibertranslations correspond to the map $%
\psi _{\mathbf{A}}:\mathbf{(x,p)\rightarrow (x,p-A(x,}t)\mathbf{)}$ which up
to a numerical factor is the transformation to euclidean physical
coordinates. The fibertranslation give rise to a generatorvectorfield on
phase space which can be described as $\mathbf{X}_{t}^{\mathbf{A}}=-\frac{%
\partial \mathbf{A}}{\partial t}$ in euclidean coordinates, but has an
invariant description given below (put the fourth component of the four
oneform $A^{(1)}$ to zero and specialize to Euclidean coordinates in the
invariant description).

If we extend the canonical oneform to a oneform $\Theta $ on extended phase
space $T^{*}X$ , $X=M\times \Bbb{R}$ , $\Theta \mid _{T^{*}M}\equiv \theta $
, the extended map$\,\,\psi $ which fixes time will generate a component in
the time direction.

\begin{theorem}
$\psi ^{-1*}\Theta =\Theta +\hat{d}S+J_{t}dt=\Theta +dS+\psi _{t}dt,$
\end{theorem}

$i_{\frac{\partial }{\partial t}}(\psi ^{-1*}\Omega )=-i_{X_{\psi
_{t}}}(\psi ^{-1*}\Omega )\equiv \hat{d}\psi _{t}$

$\psi ^{-1*}\Omega =\Omega -\hat{d}\psi _{t}\wedge dt$ $\,,\,\Omega \equiv
-d\Theta ,$

$i_{\frac{\partial }{\partial t}}(\psi ^{-1*}\Theta )=-i_{X_{\psi
_{t}}}(\psi ^{-1*}\Theta )\equiv J_{t\,\ },$

$\frac{\partial S}{\partial t}=J_{t}-\psi _{t}.$

\proof%
The proof of the above lemma is simply by observing that $i_{(\frac{\partial 
}{\partial t}+X_{\psi _{t}})}(\psi ^{-1*}\Theta )=\psi ^{-1*}(i_{\frac{%
\partial }{\partial t}}\Theta )=0$ since the extended oneform by definition
has no time components. Similarly, the above relation between the generators
and the gaugefields is a consistency requirement coming from $\mathcal{L}(%
\frac{\partial }{\partial t})(\psi ^{-1*}\Omega )=-\mathcal{L}(X_{\psi
_{t}})(\psi ^{-1*}\Omega )=-\mathcal{L}(X_{\psi _{t}})\Omega =-d(\hat{d}\psi
_{t})$

$=-\mathcal{L}(\frac{\partial }{\partial t})\hat{d}\psi _{t}\wedge dt=-%
\mathcal{L}(\frac{\partial }{\partial t})\hat{d}(J_{t}-\frac{\partial S}{%
\partial t})\wedge dt\,.$%
\endproof%

In coordinates $J_{t}=\mathbf{p\cdot }\frac{\partial \psi _{t}}{\partial 
\mathbf{p}}.$ In the case of cotangent lift we find that $\bar{J}_{t}=%
\mathbf{p\cdot }\frac{\partial \bar{\Psi}_{t}}{\partial \mathbf{p}}=\bar{\Psi%
}_{t}$ $=\mathbf{p\cdot \phi }_{t}\,,\,\mathbf{\phi }_{t}\equiv (\mathbf{%
\phi }_{*}-Id)\frac{\partial }{\partial t}$.\footnote{%
We can consider our results for $J_{t}$ to be an extension to infinite
dimensional pseudogroups of the momentum map for finitely generated groups
and Banach Lie groups (see e.g. Abraham and Marsden). Earlier results has
been directed towards reduction while we concentrate our efforts towards the
composition principle for in principle infinitely generated groups. Similar
results with respect to reduced actions also holds for pseudogroups, but we
will not study it here.}

The above extension of the canonical one and twoforms to extended phase
space, give an opportunity to consider the action of noncanonical
transformations on these forms. Therefore we can e.g. consider
transformations generated by oneforms on extended space $X.$%
\begin{eqnarray*}
\Theta _{A} &\equiv &\psi _{A}^{-1*}\Theta =\Theta -\pi _{X}^{*}A^{(1)}, \\
\Omega _{A} &\equiv &\psi _{A}^{-1*}\Omega =\Omega -\pi _{X}^{*}dA^{(1)}.
\end{eqnarray*}
The generating vectorfield corresponding to the above generalized
fibertranslation is given by $X_{t}^{A}=-\mathbf{J}_{A}\cdot \pi
_{X}^{*}(Y^{(1)})\,,\;Y^{(1)}\equiv i_{\frac{\partial }{\partial t}%
}(dA^{(1)})$ where $\mathbf{J}_{A}=\psi _{A*}\mathbf{J}$ and $\mathbf{J}$ is
the Poisson tensors corresponding to $\omega _{A}=\Omega _{A}\mid _{T^{*}M}$
and $\omega =\Omega \mid _{T^{*}M}.$ In the special case of translations by
an exact form $A^{(1)}=df\,($here $f$ is a function on space/time not phase
space$,$we find that the generating vectorfield is hamiltonian and the
transformation trivially symplectic, i.e. $X_{t}^{df}=-$ $\mathbf{J}\Bbb{%
\cdot (}df_{,t}).$ In fact, a large class of transformations which is not
necessarily preserving the the symplectic tensor, but is still
divergencefree, is generated as $X_{t}^{a}=\mathbf{J}_{a}\cdot (A^{a}),\,%
\mathbf{J}_{a}=\psi _{a*}\mathbf{J\,},\Omega _{a}=\psi _{a}^{-1*}\Omega \,$
and $A^{a}$ is some oneform given on $T^{*}M\,.$

It is possible to extend the above theorem to a more general setting based
on noncanonical twoforms $\Omega _{n}\equiv \psi _{n}^{-1*}\Omega $ and
oneforms $\Theta _{n}=\psi _{n}^{-1*}\Theta $ by introducing the convective
derivative\ $\frac{d}{d\tau }\equiv \frac{\partial }{\partial t}%
+X_{t}^{n}\,,X_{t}^{n}\equiv (\psi _{n*}-Id)\frac{\partial }{\partial t}$ .

\begin{theorem}
($\psi ^{n})^{-1*}\Theta ^{n}=\Theta ^{n}+\hat{d}S^{n}+J_{\tau }^{n}dt,$
\end{theorem}

($\psi ^{n})^{-1*}\Omega ^{n}=\Omega ^{n}-\hat{d}(\psi _{\tau }^{n}-\frac{%
dS^{n\prime }}{d\tau })\wedge dt$ $\,,\,\Omega ^{n}\equiv -d\Theta ^{n},$

$i_{\frac{d}{d\tau }}((\psi ^{n})^{-1*}\Theta ^{n})=-i_{X_{\psi _{\tau
}^{n}}}((\psi ^{n})^{-1*}\Theta ^{n})\equiv J_{\tau \,\ }^{n},$

$\frac{dS^{n}}{d\tau }=J_{\tau }^{n}-\psi _{\tau }^{n}\ ,\,$

$\psi ^{n}\equiv \psi _{n}\circ \psi \circ \psi _{n}^{-1}\,,\,X_{\psi _{\tau
}^{n}}\equiv (\psi _{n*}-Id)\frac{d}{d\tau }\,,\,S^{n}\equiv \psi
_{n}^{-1*}S.$

We are now in a position to explicitly describe the above fluid kinetic
generators in canonical, euclidean coordinates as the composition of the
following cotangent lift and translation by an exact form $df$ , i.e. $\bar{%
\psi}=\bar{\Psi}\circ \psi _{df}$ with generator $\bar{\psi}_{t}=\bar{\Psi}%
_{t}+\bar{\Psi}^{-1*}f,_{t}=\mathbf{\phi }_{t}\cdot \mathbf{p+}%
f_{t}\,,\,\,f_{t}\equiv \Psi ^{-1*}f,_{t}.$ Here we must choose $\mathbf{%
\phi }_{t}\equiv \mathbf{\hat{u}=\psi }_{t}+\mathbf{\psi }_{*}\mathbf{u}_{0}$
which for $\mathbf{u}_{0}\equiv \mathbf{\psi }_{0t}$ defines the unperturbed 
$\mathbf{\phi }_{0\text{ }}$such that $\mathbf{\phi =\psi \circ \psi }_{0}.$
The gaugefunction has to be chosen as $f_{t}=\bar{\psi}_{2t}$ to agree with
the above. Since we have assumed $\mathbf{u}_{0}$ to be unperturbed, we find
that $\bar{\psi}_{t}=\mathbf{\psi }_{\epsilon }\cdot \mathbf{p+}f_{\epsilon
} $ ,\thinspace \thinspace $f_{\epsilon }\equiv \Psi ^{-1*}f,_{t}$.

We now have the following theorem for the action of fluid kinetic maps in
canonical coordinates defined as a composition between a cotangent lift and
translation by an exact form

\begin{theorem}
Let $f=\bar{\psi}^{-1*}\tilde{f}$ such that $\tilde{\rho}_{0}=\int \tilde{f}%
d^{3}\mathbf{p\,}$\thinspace and $\rho =\int fd^{3}\mathbf{p\,.}$ Then one
has that

$\rho =\mathbf{\phi }\bullet \tilde{\rho}_{0}=\mathbf{\psi }\bullet \rho
_{0} $ $,\,\rho _{0}\equiv \mathbf{\psi }_{0}\bullet \tilde{\rho}_{0}\,.$
\end{theorem}

\proof%
The proof follows from that $\rho =\int \tilde{f}\,\,\bar{\psi}^{*}d^{3}%
\mathbf{\bar{p}=}$

$J(\mathbf{\phi }^{-1})\int \tilde{f}\,\,d^{3}\mathbf{\bar{p}=}J(\mathbf{%
\phi }^{-1})\mathbf{\phi }^{-1*}\tilde{\rho}_{0}=\mathbf{\phi \bullet }%
\tilde{\rho}_{0}=\mathbf{\psi \bullet \psi }_{0}\bullet \tilde{\rho}_{0}=%
\mathbf{\psi \bullet }\rho _{0}. $%
\endproof%

With the above relation between the Hamiltonian generator and $\overline{%
\psi }_{t}$ and macroscopic fields we obtain the reduced Hamiltonian which
can be identified with the Hamiltonian in the fluid reference frame. We also
find a reduced Vlasov equation for the fictive phase space density $%
\widetilde{f}$ 
\begin{eqnarray}
\widehat{H} &=&H-\overline{\psi }_{t}=(\widehat{\mathbf{p}}-m\mathbf{\hat{u})%
}^{2}/2m\;  \label{16b} \\
\widehat{\mathbf{p}} &=&\mathbf{p}-\frac{e}{c}\mathbf{A\;,}  \nonumber \\
\frac{\partial \widetilde{f}}{\partial t}+\{\widetilde{f},\widetilde{H}\}
&=&0\;,  \nonumber \\
\widetilde{H}\; &\equiv &\overline{\psi }^{*}\widehat{H}\;.
\end{eqnarray}
The motivation for the special choice of $\psi _{2t}(\mathbf{x,}t)$ is to
obtain a reduced Hamiltonian of the above form corresponding to the fluid
reference frame.

We will now discuss the physical consequences of the compatibility
conditions for the distribution function and mass density. The fluid density
is given by $\rho =\int fd^{3}\mathbf{p}$ such that the relation between the
fluid and kinetic generators becomes 
\begin{eqnarray}
\rho ,_{t} &=&\int (\{\overline{\psi }_{t},f\}+\overline{\psi }^{-1*}\frac{%
\partial \widetilde{f}}{\partial t})d^{3}\mathbf{p=-}\nabla \cdot (\widehat{%
\mathbf{u}}\rho )+\int \overline{\psi }^{-1*}\frac{\partial \widetilde{f}}{%
\partial t}d^{3}\mathbf{p\;,}  \label{eq17} \\
\rho ,_{\epsilon } &=&\int (\{\overline{\psi }_{t},f\}+\overline{\psi }^{-1*}%
\frac{\partial \widetilde{f}}{\partial \epsilon })d^{3}\mathbf{p=}-\nabla
\cdot (\mathbf{\psi }_{\epsilon }\rho )+\int \overline{\psi }^{-1*}\frac{%
\partial \widetilde{f}}{\partial \epsilon }d^{3}\mathbf{p\;.}  \nonumber
\end{eqnarray}
Here we have used the fact that $\int \{\phi (\mathbf{x},t),f(z,t)\}d^{3}%
\mathbf{p=}\frac{\partial \phi }{\partial \mathbf{x}}\cdot \int \frac{%
\partial f(z,t)}{\partial \mathbf{p}}d^{3}\mathbf{p=}0$ with suitable decay
properties on $f$ for large momentas. We immediately conclude by comparing
with the above fluid theory that the following additional restrictions on
the hypothetical density $\widetilde{f}$ has to be imposed if the barred
symplectic generators should correspond to a fluid subgroup with respect to
its moments 
\begin{eqnarray}
\int \overline{\psi }^{-1*}\frac{\partial \widetilde{f}}{\partial t}d^{3}%
\mathbf{p} &=&0\;,  \label{eq18} \\
\int \overline{\psi }^{-1*}\frac{\partial \widetilde{f}}{\partial \epsilon }%
d^{3}\mathbf{p} &=&0.  \nonumber
\end{eqnarray}
It is now obvious that we could add any additional term $\phi _{t}(\mathbf{x}%
,t) $ to the generator $\overline{\psi }_{t}$ without changing the form of
eq.(\ref{eq17}). The trick of using the above form of barred generators is
however that it is possible to obey eq. (\ref{eq18}) easily.

We find the following theorem

\begin{theorem}
With the earlier definitions one obtains $\int \overline{\psi }^{-1*}\frac{%
\partial \widetilde{f}}{\partial t}d^{3}\mathbf{p=}0.$ Moreover, the
definition of the fluid density from the composed fluid density is
compatible with fluid theory if we select $\widetilde{f}$ such that $\int 
\overline{\psi }^{-1*}\frac{\partial \widetilde{f}}{\partial \epsilon }d^{3}%
\mathbf{p=}0.$ If we select the reduced distribution function such that $%
\int \overline{\psi }^{-1*}\frac{\partial \widetilde{f}}{\partial \epsilon }%
d^{3}\mathbf{p}\mid _{t=0}\mathbf{=}0$ and $\widetilde{\rho }(\mathbf{x}%
,t)\mid _{t=0}=\tilde{\rho}^{0}(\mathbf{x})\,\,$, it will also be an
identity at any other time.
\end{theorem}

\proof%
We need the following lemma

\begin{lemma}
For symplectomorphisms, $\overline{\psi },$ with generators which are not
more than linear in momentum the following applies for a phase space density 
$g$ with suitable decay properties for large momentas and $\int \bar{\psi}%
^{-1*}\frac{\partial g}{\partial t}=0$%
\begin{eqnarray}
\widehat{g} &=&\int \overline{\psi }^{-1*}gd^{3}\mathbf{p}=\mathbf{\bar{\psi}%
}\bullet \overline{g}\;,  \label{eq19} \\
\overline{g} &=&\int gd^{3}\mathbf{p\;.}  \nonumber
\end{eqnarray}
\end{lemma}

\proof%
In the case of near identity transformations the lemma can be proved by
using near identity generators which are linear in momentum. Since it
follows that the deviation from identity is also linear in momentum, one
find the above relation after integration over momentum. If one does not
assume near identity mappings the proof is a little more involved. The
infinitesimal version of the above equation is since the generators are
supposed to linear in momentum and with suitable decay properties of $g$ in
momentum such that the following holds 
\begin{eqnarray*}
\widehat{g}(t),_{t} &=&\int \{\overline{\psi }_{t},\overline{\psi }%
^{-1*}(t)g\}d^{3}\mathbf{p+}\int \bar{\psi}^{-1*}\frac{\partial g}{\partial t%
}d^{3}\mathbf{p} \\
&=&-\nabla \cdot (\int \mathbf{\hat{u}}\overline{\psi }^{-1*}(t)gd^{3}%
\mathbf{p)+}\int \bar{\psi}^{-1*}\frac{\partial g}{\partial t}d^{3}\mathbf{p}%
, \\
\hat{g}(t),_{t} &=&-\nabla \cdot (\mathbf{\hat{u}}\hat{g}(t)).
\end{eqnarray*}

Therefore if $\int \bar{\psi}^{-1*}\frac{\partial g}{\partial t}=0,$ we can
parameterize $\hat{g}(t)=\mathbf{\bar{\psi}(}t)\mathbf{\bullet }\bar{g}.$
With this parameterization it also follows that $\mathbf{\bar{\psi}\bullet }%
\frac{\partial \bar{g}}{\partial t}=0$ from which we find that $\frac{%
\partial \bar{g}}{\partial t}=0$ since $\mathbf{\bar{\psi}}$ is invertible
and takes zero to zero. 
\endproof%

\begin{remark}
The above parameterization is performed with respect to a reference $\bar{g}$
such that $\frac{\partial \bar{g}}{\partial t}=0.$ We can change this
reference by doing an active transformation (or alternatively a passive
coordinate transformation) $\bar{g}=(\mathbf{\bar{\psi}}_{0}{}^{-1})\bullet 
\bar{g}_{0}$ such that $\frac{\partial \bar{g}_{0}}{\partial t}+\nabla \cdot
(\mathbf{u}_{0}\bar{g}_{0})=0.$ Now it is possible to define transformations 
$\mathbf{\psi }=\mathbf{\bar{\psi}\circ \bar{\psi}}_{0}^{-1}$ which is more
suitable for near identity transformations with respect to reference state.
This is what we will eventually do in section 3.2.
\end{remark}

Using the above lemma we immediately find that 
\begin{eqnarray}
\int \overline{\psi }^{-1*}\widetilde{f},_{a}d^{3}\mathbf{p} &=&\mathbf{\bar{%
\psi}\bullet }\widetilde{\rho },_{a}\;,\;a=t,\epsilon \;,  \label{eq20} \\
\int \overline{\psi }^{-1*}\widetilde{f}d^{3}\mathbf{p} &=&\mathbf{\bar{\psi}%
\bullet }\widetilde{\rho \;.}  \nonumber
\end{eqnarray}
Now we use the reduced Vlasov equation and find 
\[
\int \overline{\psi }^{-1*}\frac{\partial \widetilde{f}}{\partial t}d^{3}%
\mathbf{p=}\int \{\widehat{H},f\}d^{3}\mathbf{p.} 
\]
By partial integration in momentum we find that 
\begin{equation}
\int \overline{\psi }^{-1*}\frac{\partial \widetilde{f}}{\partial t}d^{3}%
\mathbf{p}=-\nabla \cdot (\int \frac{\partial \widehat{H}}{\partial \mathbf{p%
}}fd^{3}\mathbf{p)\equiv }0  \label{eq21}
\end{equation}
since by definition $\int (\widehat{\mathbf{p}}-m\widehat{\mathbf{u}})fd^{3}%
\mathbf{p=}\rho (\mathbf{u-\widehat{u})=0.}$ (I.e., we naturally will
restrict the parametrization of the velocity to be equal to the velocity
moment of the distribution. In fact this restriction is the defining
equation for the volume density preserving transformation.)

The compatibility condition for the mass density parameterized by the above
composition of symplectic transformations acting on phase space density
leads by eq.'s (\ref{eq17}, \ref{eq18}) to the constraint 
\begin{eqnarray}
\frac{\partial ^{2}\rho }{\partial t\partial \epsilon } &=&\frac{\partial
^{2}\rho }{\partial \epsilon \partial t}  \label{eq22} \\
&\Rightarrow &\frac{\partial }{\partial t}(\int \overline{\psi }^{-1*}\frac{%
\partial \widetilde{f}}{\partial \epsilon }d^{3}\mathbf{p})=0.  \nonumber
\end{eqnarray}
Therefore we conclude that $\int \overline{\psi }^{-1*}\frac{\partial 
\widetilde{f}}{\partial \epsilon }d^{3}\mathbf{p=\psi }\bullet \int \frac{%
\partial \widetilde{f}}{\partial \epsilon }d^{3}\mathbf{p}$ is constant with
respect to time. Consequently if the $\epsilon $ parameterization is such
that the $\int \overline{\psi }^{-1*}\frac{\partial \widetilde{f}}{\partial
\epsilon }d^{3}\mathbf{p\mid }_{t=0}=0,$ it will continue to be so at all
times. We assume that both the transformations $\psi ^{-1}$ and $\psi $
exists. Since the action of the symplectomorphisms is such that it preserves
volume density forms, it is realized that the zero density must be
transported to zero density by the action of all symplectomorphisms. It
follows that $\int \frac{\partial \tilde{f}}{\partial \epsilon }d^{3}\mathbf{%
p=}\frac{\partial \widetilde{\rho }}{\partial \epsilon }=0$ if it is
initially chosen in such a way. Moreover, it is implied that $\widetilde{%
\rho }=\tilde{\rho}^{0}$ is given by the reference density independent of $%
\epsilon $ even if $\widetilde{f}$ is depending on $\epsilon .$ By a similar
argument we deduce that $\frac{\partial \tilde{\rho}^{0}}{\partial t}=0,$
and consequently corresponds to a spatial reference density given in the
fluid frame of reference.

\endproof%

The interpretation of the above result is that to obtain a composition
symplectomorphism where the volume density preserving transformation is
described as a subgroup of the group of all symplectomorphisms lead to that
density perturbations are parameterized in phase space by the fluid subgroup
consisting of the barred symplectomorphisms.

Let us now study the continuity and momentum equation more explicitly. The
Vlasov equation can be written 
\begin{eqnarray*}
\frac{\partial f}{\partial t}+\{f,-\overline{\psi }_{t}+\widehat{H}\} &=&0 \\
H &=&-\bar{\psi}_{t}+\hat{H}.
\end{eqnarray*}
The zeroth order moment integrated over momentum space now gives the
momentum equation with no contribution from the $\{f,\widehat{H}\}$ term.
The momentum equation can now be found as

$\frac{\partial }{\partial t}\int \mathbf{p}fd^{3}\mathbf{p+}\int \mathbf{p}%
(-\{\overline{\psi }_{t},f\}+\{\widehat{H},f\})d^{3}\mathbf{p}=0\;,$

\begin{eqnarray}
&&  \label{eq24} \\
\frac{\partial }{\partial t}(\rho \mathbf{u)+}\nabla \cdot (\rho \mathbf{%
\widehat{u}u+}\Bbb{P}\mathbf{)-}\rho \mathbf{f}_{L} &=&0\;,  \nonumber \\
\mathbf{f}_{L} &=&\frac{e}{m}(\mathbf{E+\widehat{u}\times B)}\;,  \nonumber
\\
\Bbb{P} &\equiv &\int \frac{1}{m}(\mathbf{p}_{p}-m\widehat{\mathbf{u}})(%
\mathbf{p}_{p}-m\widehat{\mathbf{u}})fd^{3}\mathbf{p\;.}
\end{eqnarray}
Here $\mathbf{f}_{L}$ and $\Bbb{P}$ are the Lorentz force and the
stresstensor of the fluid and the physical momentum is related to the
canonical momentum by $\mathbf{p}_{p}=\mathbf{p-}\frac{e}{c}\mathbf{A.}$ The
reduced Vlasov equation is given in eq.(\ref{16b}) and form together with
the continuity equation and the above momentum equation a new set of
equations for collisionless plasma physics, the\textit{\ hybrid
fluid-kinetic theory. }

\subsection{Parameterization of the hybrid fluid-kinetic theory}

Let us briefly discuss the parameterization of the continuity equation and
the reduced Vlasov equation. We recall that fluid generating vector is
related to the parameterized velocity by $\widehat{\mathbf{u}}=\mathbf{\psi }%
_{t}+\mathbf{\psi }_{*}\mathbf{u}_{0}$. We purposely have been operating
with the parameterized velocity $\widehat{\mathbf{u}}$ generated by the
diffeomorphisms different from the velocity vector derived from the moment
of the distribution. The reason why is that they are not a priori equal.
Indeed, the equality of these two quantities is the constraint equation
which together with the above fluid momentum and reduced Vlasov equations
determines diffeomorphism

\begin{equation}
\mathbf{u=\widehat{u}\;.}  \label{eq25}
\end{equation}
This equation replaces the continuity equation since the density can
immediately be mapped the moment we have e.g. the near identity
representation of the diffeomorphism. The reduced Vlasov equation can be
parameterized in a similar way as the Vlasov equation by

\begin{eqnarray}
\widetilde{\psi }_{t} &=&\widetilde{H}-\widetilde{\psi }^{-1*}H_{0},
\label{eq26} \\
H_{0} &=&\widehat{H}_{0}+\overline{\psi }_{t}^{0}.  \nonumber
\end{eqnarray}
If we parameterize our diffeomorphism by the perturbation parameter $%
\epsilon $, we use the compatiblity equation to obtain the the determining
equation for the generating function $\widetilde{\psi }_{\epsilon }$%
\begin{equation}
\widetilde{\psi }_{\epsilon ,t}-\{\widetilde{H},\widetilde{\psi }_{\epsilon
}\}-\widetilde{H}_{,\epsilon }=0\;.  \label{eq27}
\end{equation}
Eq.'s (\ref{eq24}-\ref{eq27}) now form a complete set of equations as an
alternative to the continuity and reduced Vlasov equation together with the
momentum equation.

We now have to discuss the momentum equation more carefully to obtain an
invariant description of the parameterization. To obtain such an invariant
description it is useful to describe the velocity field as a oneform $%
\mathbf{u}^{(1)}$ and the stresstensor as a symmetric, covariant twotensor $%
\mathbf{P}^{(2)}.$ It is now possible to parametrize the velocity field by
using Hodge decomposition with respect to the threedimensional metric $%
\mathbf{g}$ to split it in a rotational and divergent part, $\mathbf{u}%
^{(1)}=-\hat{d}\eta +*_{\mathbf{g}}\hat{d}\mathbf{A}_{r}^{(1)}=\mathbf{u}%
_{d}^{(1)}+\mathbf{u}_{r}^{(1)}.$ Here we have used the Hodge decomposition
with respect to the transformed metric (see below) in agreement with the
philosophy that the decomposition should be given with respect to the
standard metric when pulled back to the reference level, i.e. $\mathbf{u}%
^{(1)}=\mathbf{\psi (}t)^{-1*}\mathbf{\bar{u}}^{(1)}\,,\,\mathbf{\bar{u}}%
^{(1)}=-\hat{d}\bar{\eta}+*_{\mathbf{g}_{0}}\hat{d}\mathbf{\bar{A}}%
_{r}^{(1)}.$ It can be useful to further describe the rotational gaugefield
by Pfaff decomposition as $\mathbf{A}_{r}^{(1)}=\alpha \hat{d}\beta +\hat{d}%
\gamma $ , where the gaugefield does not depend on the gaugepotential $%
\gamma .$ We find this parameterization more convenient than the standard
Clebsch decomposition which does not separate into rotational and divergent
parts and does not use the metric structure. To split the equation with
respect to the above structure we write the momentum equation as 
\begin{eqnarray}
(\frac{\partial }{\partial t}+\mathcal{L}(\mathbf{\hat{u}))u}^{(1)} &=&-%
\frac{1}{\bar{\rho}}D_{\mathbf{g}}\mathbf{\bar{P}}^{(2)}+\hat{d}(\frac{1}{2}%
i_{\mathbf{\hat{u}}}\mathbf{u}^{(1)})+\mathbf{f}_{L}^{(1)},  \label{eq29} \\
D_{\mathbf{g}}\mathbf{\bar{P}}^{(2)} &\equiv &\sum_{j}*_{\mathbf{g}}\hat{d}%
*_{\mathbf{g}}(\mathbf{\bar{P}}_{j}^{(1)})\hat{d}(\mathbf{\psi (}t\mathbf{)}%
^{-1*}x_{j})=\sum_{j}div_{\mathbf{g}}(\mathbf{P}_{j})\hat{d}(\mathbf{\psi (}t%
\mathbf{)}^{-1*}x_{j}),  \nonumber \\
\,\mathbf{P}^{(2)} &=&J(t)\mathbf{\bar{P}}^{(2)},\,\,\mathbf{\bar{P}}%
_{j}^{(1)}\equiv \mathbf{\psi }(t)^{-1*}\mathbf{\hat{P}}_{j}^{(1)} \\
\mathbf{\bar{P}}^{(2)} &=&\mathbf{\psi (}t)^{-1*}\mathbf{\hat{P}}%
^{(2)}=\sum_{j}\mathbf{\bar{P}}_{j}^{(1)}\otimes \hat{d}(\mathbf{\psi (}t%
\mathbf{)}^{-1*}x_{j}) \\
\mathbf{g} &=&\mathbf{\psi }(t)^{-1*}\mathbf{g}_{0} \\
\bar{\rho}(t) &\equiv &\mathbf{\psi (}t)^{-1*}\rho _{0}(t),  \nonumber \\
\mathbf{f}_{L}^{(1)} &=&-\frac{e}{c}i_{\hat{u}}F^{(2)}.  \nonumber
\end{eqnarray}
This equation can also be written in an invariant way as an equation for the
momentum $\mathbf{M\equiv }\omega \otimes \mathbf{u}^{(1)}$ and/or written
in terms of covariant derivatives. In addition we can embed this it in
extended space by taking the wedge product with $dt$ as we did for the
density form, but we prefer to defer this formulation to another paper. Here
we have used that $D_{\mathbf{g}}(\cdot )=\frac{1}{J(t)}D_{\mathbf{g}%
_{0}}(J(t)\cdot \,)$ to obtain an invariant form of the equations suitable
for transformations by timedependent diffeomorphisms. We need the following
natural properties of the diffeomorphism action 
\begin{eqnarray}
(\frac{\partial }{\partial t}+\mathcal{L}\mathbf{(\hat{u}))\circ \psi }%
(t)^{-1*} &=&\mathbf{\psi (}t)^{-1*}\circ (\frac{\partial }{\partial t}+%
\mathcal{L}(\mathbf{u}_{0}))\,,  \label{eq30} \\
\ast _{\mathbf{g}}\circ \mathbf{\psi }(t)^{-1*} &=&\mathbf{\psi }%
(t)^{-1*}\circ *_{\mathbf{g}_{0}},  \nonumber \\
\hat{d}\circ \mathbf{\psi }(t)^{-1*} &=&\mathbf{\psi }(t)^{-1*}\circ \hat{d}%
,\,d\circ \psi ^{-1*}=\psi ^{-1*}\circ d,  \nonumber \\
i_{\mathbf{\hat{u}}}\circ \mathbf{\psi }(t)^{-1*} &=&\mathbf{\psi (}%
t)^{-1*}\circ i_{\mathbf{\check{u}}}\,\,\,,\,i_{\hat{u}}\circ \psi
^{-1*}=\psi ^{-1*}\circ i_{u_{0}},  \nonumber \\
\mathbf{\hat{u}} &=&\mathbf{\psi }(t)_{*}\mathbf{\check{u}\,,\,\check{u}=%
\hat{\psi}}_{t}+\mathbf{u}_{0},  \nonumber \\
\hat{u} &=&\psi _{*}u_{0}\,,\,\hat{u}\equiv \mathbf{\hat{u}+}\frac{\partial 
}{\partial t}\,,u_{0}\equiv \mathbf{u}_{0}+\frac{\partial }{\partial t}\,, 
\nonumber \\
D_{\mathbf{g}}(\mathbf{P}^{(2)}) &=&\mathbf{\psi (}t)^{-1*}D_{\mathbf{g}%
_{0}}(\mathbf{\hat{P}}^{(2)})\,.
\end{eqnarray}
Here $\psi $ is as before the extension of the timedependent map $\mathbf{%
\psi }(t)$ to maps in four space which fixes time. We now observe that our
momentum equation can be pulled back to the reference level where in fact
the Lie derivative and the fourdimensional interior multiplication will be
linear operators with respect to the background velocity field and all
quantities tranform in a natural way.

\begin{eqnarray}
(\frac{\partial }{\partial t}+\mathcal{L}(\mathbf{u}_{0}\mathbf{))\bar{u}}%
^{(1)} &=&-\frac{1}{\rho _{0}}D_{\mathbf{g}_{0}}\mathbf{\check{P}}^{(2)}+%
\hat{d}(\frac{1}{2}i_{\mathbf{\check{u}}}\mathbf{\bar{u}}^{(1)})+\mathbf{%
\check{f}}_{L}^{(1)},  \label{eq31} \\
\mathbf{\check{f}}_{L}^{(1)} &=&-\frac{e}{c}i_{u_{0}}\check{F}^{(2)}=-\frac{e%
}{c}(\mathbf{\check{E}}^{(1)}+*_{\mathbf{g}_{0}}(\mathbf{u}_{0}^{(1)}\wedge 
\mathbf{\check{B}}^{(1)}))\,\,_{,}\,  \nonumber \\
F^{(2)} &=&\psi ^{-1*}\check{F}^{(2)}\,\,,\,\,\mathbf{E}^{(1)}=\mathbf{\psi (%
}t)^{-1*}\mathbf{\check{E}}^{(1)},\,\mathbf{B}^{(1)}=\mathbf{\psi }(t)^{-1*}%
\mathbf{\check{B}}^{(1)},  \nonumber \\
\mathbf{B}^{(1)} &=&*_{\mathbf{g}}\mathbf{B}^{(2)}=*_{\mathbf{g}}\hat{d}%
\mathbf{A}^{(1)}\,,\,\mathbf{\check{B}}^{(1)}=*_{\mathbf{g}_{0}}\mathbf{%
\check{B}}^{(2)}\,,  \nonumber
\end{eqnarray}
In addition after the pullback all operators and physical quantities
according to the above will be specified with respect to the background
metric $\mathbf{g}_{0}.$ Our philosophy is that all equations and physical
fields should be represented such that they can be pulled back to the
reference level. The pulled back equation can now be compared with the
equation for the reference solution and an equation for deformations from
the reference fields can be formulated in a strikingly simple way in which
many of the terms are linear. This make our theory especially attractive
from a perturbation theory and complexity point of view, but we believe this
also has implications for the interpretation, predictions and formulation of
measurements for physical fields. E.g., according to us a linearized
equation and fields pulled back to the reference level in no way is linear
at the original level which even obtain an induced, nonlinear metric
structure.

We now introduce the deviations at the pullback level of the physical fields
from the background fields. Since the equation for the reference fields also
obey eq.\ref{eq30} taken at the reference metric $\mathbf{g}_{0},$ we can
subtract the background equation from eq.\ref{eq31}. We then obtain an
equation for the fluctuating quantities suitable for perturbation theory 
\begin{eqnarray}
(\frac{\partial }{\partial t}+\mathcal{L}(\mathbf{u}_{0}\mathbf{))\tilde{u}}%
^{(1)} &=&-\frac{1}{\rho _{0}}D_{\mathbf{g}_{0}}\mathbf{\tilde{P}}^{(2)}+%
\frac{1}{2}\hat{d}(i_{\mathbf{\check{u}}}\mathbf{\bar{u}}^{(1)}-i_{_{\mathbf{%
u}_{0}}}\mathbf{u}_{0}^{(1)})+\mathbf{\tilde{f}}_{L}^{(1)},  \label{eq32} \\
\mathbf{\bar{u}}^{(1)} &=&\mathbf{u}_{0}^{(1)}+\mathbf{\tilde{u}}^{(1)},\,%
\mathbf{\check{P}}^{(2)}=\mathbf{P}_{0}^{(2)}+\mathbf{\tilde{P}}^{(2)},\,%
\check{F}^{(2)}=F_{0}^{(2)}+\tilde{F}^{(2)},  \nonumber \\
\mathbf{\tilde{f}}_{L}^{(1)} &\equiv &i_{u_{0}}\tilde{F}^{(2)}\text{, e.t.c.}
\nonumber
\end{eqnarray}
Notice that even if we were only interested in linearization and linear
quantities at the original level, the distinction between the pullbacked
equations and the equations at the original level will still be essential
since the fluctuations in the pullbacked metric and the pullback map itself
will not affect first order quantities, but they will affect background
quantities to linear order presented at the original level.

Let us complete our discussion of the momentum equation by showing how
elegant it separates into equations for the acoustic and rotational
potentials. We split the right hand side of eq.(\ref{eq30},\ref{eq31}) into
rotational and divergent parts by using Hodge theorem to define the
potentials 
\begin{eqnarray}
-\hat{d}\kappa +*_{\mathbf{g}}\hat{d}\mathbf{R}^{(1)} &\equiv &\mathbf{f}%
_{L}^{(1)}-\frac{1}{\rho }D_{\mathbf{g}}\mathbf{\bar{P}}^{(2)},
\label{eq32b} \\
-\hat{d}\hat{\kappa}+*_{\mathbf{g}_{0}}\hat{d}\mathbf{\hat{R}}^{(1)} &\equiv
&\mathbf{\check{f}}_{L}^{(1)}-\frac{1}{\rho _{0}}D_{\mathbf{g}_{0}}\mathbf{%
\check{P}}^{(2)}.  \nonumber
\end{eqnarray}
If we now take the exterior derivative of eq.(\ref{eq30}), we obtain the
vorticity equation 
\begin{eqnarray}
(\frac{\partial }{\partial t}+\mathcal{L}(\mathbf{\hat{u}))}\hat{d}\mathbf{u}%
_{c}^{(1)} &=&\hat{d}*_{\mathbf{g}}\hat{d}\mathbf{R}^{(1)},  \label{eq32c} \\
\mathbf{u}_{c}^{(1)} &=&*_{\mathbf{g}}\hat{d}\mathbf{A}_{r}^{(1)}=*_{\mathbf{%
g}}(\hat{d}\alpha \wedge \hat{d}\beta ).  \nonumber
\end{eqnarray}
In fact, up to a potential we can even give the equation for the rotational
part of the velocity oneform(relative to the fluctuation metric) or
alternatively the vorticity twoform, $\mathbf{\pi }^{(2)}$ as 
\begin{eqnarray}
(\frac{\partial }{\partial t}+\mathcal{L(}\mathbf{\hat{u}))u}_{c}^{(1)}
&=&*_{\mathbf{g}}\hat{d}\mathbf{R}^{(1)},  \label{eq32d} \\
(\frac{\partial }{\partial t}+\mathcal{L(}\mathbf{\hat{u}))\pi }^{(2)} &=&%
\hat{d}*_{\mathbf{g}}\hat{d}\mathbf{R}^{(1)}, \\
\mathbf{\pi }^{(2)} &\equiv &\hat{d}\mathbf{u}_{c}^{(1)}.
\end{eqnarray}
Moreover, if we subtract this equation from the original momentum equation,
we otain an equation for the acoustic potential 
\begin{equation}
(\frac{\partial }{\partial t}+\mathcal{L(}\mathbf{\hat{u}))}\eta =\kappa -%
\frac{1}{2}i_{\mathbf{\hat{u}}}\mathbf{u}^{(1)}.  \label{q32e}
\end{equation}
All this equations transform in a natural way with respect to
diffeomorphisms, i.e. we can obviously formulate pullbacked equations and
equations for fluctuations with respect to corresponding background.

Let us now describe our diffeomorphisms more explicitly and relate them to
the above potentials. We do this by studying the constraint equation for
velocity vectorfield reformulated as a constraint on the momentum lifted to
a oneform; $\rho \mathbf{u}^{(1)}=\rho \mathbf{\hat{u}}^{(1)}.$ The
compatibility equation for the perturbed vectorfields now become a relation
between oneforms(lifted by the fixed metric) 
\[
(\rho \mathbf{\psi }_{\in }^{(1)}),_{t}-(\rho \mathbf{u}^{(1)}),_{\in }+*_{%
\mathbf{g}_{0}}\hat{d}(*_{\mathbf{g}_{0}}(\rho \mathbf{\psi }_{\in
}^{(1)}\wedge \mathbf{u}^{(1)}))=0, 
\]
which can be pullbacked to 
\[
(\rho _{0}\mathbf{\hat{\psi}}_{\in }^{(1)}),_{t}-(\rho _{0}\mathbf{\check{u}}%
^{(1)}),_{\in }+*_{\mathbf{g}_{0}}\hat{d}(*_{\mathbf{g}_{0}}(\rho _{0}%
\mathbf{\hat{\psi}}_{\in }^{(1)}\wedge \mathbf{\check{u}}^{(1)}))=0. 
\]
If we now take the divergence of the first equation, we obtain that 
\[
(*_{\mathbf{g}_{0}}\hat{d}*_{\mathbf{g}_{0}}(\rho \mathbf{\psi }_{\in
}^{(1)})),_{t}=-\rho _{,\in t}=(*_{\mathbf{g}_{0}}\hat{d}*_{\mathbf{g}%
_{0}}(\rho \mathbf{u}^{(1)}))_{,\in }\,. 
\]
This is in fact the perturbed, parameterized continuity equation.

We can also study the compatibility equation with respect to the oneform $J%
\mathbf{u}^{(1)}=J\mathbf{\hat{u}}^{(1)}.$ In this case we obtain 
\begin{eqnarray}
(J\mathbf{\psi }_{\in }^{(1)}),_{t}-(J\mathbf{u}^{(1)}),_{\in }+*_{\mathbf{g}%
_{0}}\hat{d}(*_{\mathbf{g}_{0}}(J\mathbf{\psi }_{\in }^{(1)}\wedge \mathbf{u}%
^{(1)})) &=&0,  \label{eq32f} \\
(*_{\mathbf{g}_{0}}\hat{d}*_{\mathbf{g}_{0}}(J\mathbf{\psi }_{\in
}^{(1)})),_{t} &=&-J,_{\in t}=(*_{\mathbf{g}_{0}}\hat{d}*_{\mathbf{g}_{0}}(J%
\mathbf{u}^{(1)}))_{,\in }.  \nonumber
\end{eqnarray}
Here we have used the continuity equations for the Jacobian $J,_{t}+\nabla
\cdot (J\mathbf{\psi }_{t})=0,\,\,J,_{\in }+\nabla \cdot (J\mathbf{\psi }%
_{\in })=0 $ lifted up to forms. In fact it is possible to find material
coordinates such that the Jacobian and the density is equal up to a constant
density, i.e. $\rho =J^{f}\rho _{0}^{f}.$ The parameterization of the
velocity oneform if written with respect to reference metric is given by 
\[
\mathbf{u}^{(1)}=-d\eta +\frac{1}{J}\mathbf{g\circ g}_{0}^{-1}\circ *_{%
\mathbf{g}_{0}}\hat{d}\mathbf{A}_{r}^{(1)}. 
\]
This means that $J\mathbf{u}^{(1)}$ will only contribute to the above
divergence term in the continuity equation for the Jacobian through the
acoustic potential. If we want a parameterization where the rotational part
does not contribute in the divergence term of the mass continuity equation,
we have to use $J_{id},\mathbf{g}_{id}$ . Anyhow, we pull back the velocity
oneform and relate it to the generator as a oneform at the reference level
and find 
\begin{eqnarray}
\mathbf{\hat{\psi}}_{t}^{(1)} &=&-\hat{d}\hat{\eta}+*_{\mathbf{g}_{0}}\hat{d}%
\mathbf{\hat{A}}_{r}^{(1)}-\mathbf{u}_{0}^{(1)}=-\hat{d}\tilde{\eta}+*_{%
\mathbf{g}_{0}}\hat{d}\mathbf{\tilde{A}}_{r}^{(1)},  \label{eq32g} \\
\mathbf{\hat{\psi}}_{t}^{d} &\equiv &-\hat{d}\tilde{\eta},\,\,\mathbf{\hat{%
\psi}}_{t}^{r}\equiv *_{\mathbf{g}_{0}}\hat{d}\mathbf{\tilde{A}}_{r}^{(1)}, 
\nonumber \\
\mathbf{\hat{\psi}}_{t} &=&\mathbf{\hat{\psi}}_{t}^{d}+\mathbf{\hat{\psi}}%
_{t}^{r}\,.  \nonumber
\end{eqnarray}
Here we have to choose which generator is first and last corresponding to
the composition of e.g. $\mathbf{\psi }=\mathbf{\psi }^{r}\circ \mathbf{\psi 
}^{d}$ or vica versa.

The above decomposition into divergent and rotational generators shows us
the need to close our system of equations by representing the generators
according to the type chosen. Obviously, the divergent diffeomorphisms
should only have one parameter related to the density structure while the
rotational diffeomorphisms should have only two degrees of freedom related
to the vorticity structure. We notice that if e.g. $\mathbf{\psi =}\exp
(\epsilon \mathbf{\psi }_{1}),$ the generator $\mathbf{\psi }_{\epsilon }=%
\mathbf{\psi }_{1}$ and similar relations can be worked out for other
perturbation parameter relations. Therefore the key is to study the
decomposition $\mathbf{\hat{\psi}}_{\epsilon }=\mathbf{\hat{\psi}}_{\epsilon
}^{d}+\mathbf{\hat{\psi}}_{\epsilon }^{r}$ represented in terms the density
and vorticity structure respectively. We define the density structures $%
\omega =\rho dV_{0},$ $\omega ,_{\epsilon }=\rho ,_{\epsilon }dV_{0},$ $\hat{%
\omega}_{\epsilon }=\hat{\rho}_{\epsilon }dV_{0}\equiv \mathbf{\psi }%
^{*}\omega ,_{\epsilon }$. The pullbacked perturbed density structure is
therefore explicitly related to the density perturbation as $\hat{\rho}%
_{\epsilon }=\mathbf{\psi }^{*}(\frac{\rho ,_{\epsilon }}{J}).$ Above we
have decomposed the velocity oneform into a divergent and rotational part
with respect to an invariant volume element. Here we find it more natural to
consider the dual decomposition of $\bar{\rho}\mathbf{\psi }_{\epsilon },$ $%
\bar{\rho}\mathbf{\psi }_{t}$ and $\bar{\rho}\mathbf{\hat{u}}$
(corresponding to $\omega \otimes \mathbf{\hat{u}}$ e.t.c). In terms of the
perturbed, pullbacked density structure we find that we can represent $\rho
_{0}\mathbf{\hat{\psi}}_{\epsilon }=\mathbf{\psi }_{*}(\bar{\rho}\mathbf{%
\psi }_{\epsilon })$ as $div_{\mathbf{g}_{0}}(\rho _{0}\mathbf{\hat{\psi}}%
_{\epsilon })=-\hat{\rho}_{\epsilon }\,.$ Therefore we have that modulo a
rotational part we can define $\rho _{0}\mathbf{\hat{\psi}}_{\epsilon
}^{d(1)}\equiv -\hat{d}\nabla _{\mathbf{g}_{0}}^{-2}\hat{\rho}_{\epsilon }\,$%
\thinspace \thinspace and the rotational part can be represented as $\rho
_{0}\mathbf{\hat{\psi}}_{\epsilon }^{r(1)}=*_{\mathbf{g}_{0}}\hat{d}\mathbf{%
\hat{A}}_{\epsilon }^{(1)}.$ In terms of this parameterization the divergent
part of the compatibility equation which we found to be equivalent with the
perturbed continuity equation can be formulated at the pullbacked level as 
\begin{eqnarray}
(\frac{\partial }{\partial t}+\mathcal{L}^{(3)}(\mathbf{v}_{0}))\hat{\rho}%
_{\epsilon }+div_{\mathbf{g}_{0}}(\mathbf{\hat{v}}_{\epsilon }\rho _{0})
&=&0,  \label{eq32h} \\
\mathbf{\hat{v}}_{\epsilon } &\equiv &\mathbf{\psi }_{*}^{-1}\mathbf{v}%
,_{\epsilon }\text{ or}\,\,\mathbf{\hat{v}}_{\epsilon }^{(1)}\equiv \mathbf{%
\psi }^{*}\mathbf{v}^{(1)}\mathbf{,}_{\epsilon }\,\,.  \nonumber
\end{eqnarray}
Equation's (\ref{eq32h}) and (\ref{eq32e}) (can be given in perturbed form,
but that deserves a separate study) contains the description of the acoustic
mode and the interaction with the rotational mode and the kinetic
fluctuations. In the linearized case this correspond to a longitudinal wave
equation with kinetic and rotational effects in any background fluid and
kinetic state. What has to be done is a detailed study of the deformation
properties of the stresstensor discussed below with respect to the divergent
and rotational diffeomorphisms and the incoherent kinetic transformations.

Let us now discuss the rotational mode in more detail. We define $\rho
\equiv \mathbf{\psi }^{d}\circ \rho _{r}\,\,,\,\,\rho _{r}\equiv \mathbf{%
\psi }^{r}\circ \rho _{0}\,\,$and $\frac{\partial \rho _{r}}{\partial t}%
+\nabla \cdot (\mathbf{u}_{r}\rho _{r})=0\,,\,\mathbf{\hat{u}=\psi }_{t}^{d}+%
\mathbf{\psi }_{*}^{d}\mathbf{u}_{r}\,,\,\mathbf{u}_{r}\equiv \mathbf{\psi }%
_{t}^{r}+\mathbf{\psi }_{*}^{r}\mathbf{u}_{0}=\mathbf{\bar{u}}_{c}+\mathbf{%
\psi }_{*}^{r}\mathbf{u}_{0}^{d}\,,\,\mathbf{u}_{c}=\mathbf{\psi }_{*}^{d}%
\mathbf{\bar{u}}_{c}.$ We now find the following equivalent compatibility
equations for the rotational generators with respect to the pseudogroup
connected to the density $\rho _{r}$%
\begin{eqnarray}
\mathbf{\psi }_{\in }^{r},_{t}-\mathbf{u}_{r},_{\in }+[\mathbf{u}_{r},%
\mathbf{\psi }_{\in }^{r}] &=&\mathbf{0\,,}  \label{eq32i} \\
\text{equivalent to }\mathbf{\psi }_{\in }^{r},_{t}-\mathbf{\bar{u}}%
_{c},_{\in }+[\mathbf{\bar{u}}_{c},\mathbf{\psi }_{\in }^{r}] &=&\mathbf{0\,,%
}  \nonumber \\
\text{or\thinspace \thinspace \thinspace }(\rho _{r}\mathbf{\psi }_{\in
}^{r(1)}),_{t}-(\rho _{r}\mathbf{u}_{r}^{(1)}),_{\in }+*_{\mathbf{g}_{0}}%
\hat{d}(*_{\mathbf{g}_{0}}(\rho _{r}\mathbf{\psi }_{\in }^{r(1)}\wedge 
\mathbf{u}_{r}^{(1)}))\, &=&\mathbf{0\,,}  \nonumber \\
\text{or }\mathbf{\psi }_{\in }^{r(1)},_{t}-\mathbf{\bar{u}}_{c}^{(1)},_{\in
}+*_{\mathbf{g}_{c}}\hat{d}(*_{\mathbf{g}_{c}}(\mathbf{\psi }_{\in
}^{r(1)}\wedge \mathbf{u}_{c}^{(1)})) &=&0\mathbf{.}  \nonumber
\end{eqnarray}
Here $\mathbf{g}_{c}\equiv \mathbf{\psi }^{r*-1}\mathbf{g}_{0}$ is the
fluctuation metric with respect to the rotational part of the diffeomorphism
in contrast to the total fluctuation metric used earlier.\footnote{%
All the quantities and equations introduced here related to the rotational
part can be transformed by acting with $\mathbf{\psi }^{d}$ to obtain the
actual measured quantities$.$The reason why we do this pullback by $\mathbf{%
\psi }^{d}$ is that many of the relations we present will be very
complicated without it. But of course all the relations and quantities can
be transported to the real measured quantites by the inverse action.} The
above equivalence is due to that $(\mathbf{\psi }_{*}^{r}\mathbf{u}%
_{0}^{d}),_{\in }=[\mathbf{\psi }_{*}^{r}\mathbf{u}_{0}^{d},\mathbf{\psi }%
_{\in }^{r}]$ which is valid in general for any background vectorfield. The
last equality is valid assuming by definition that $div_{\mathbf{g}_{c}}(%
\mathbf{\bar{u}}_{c}^{(1)})=div_{\mathbf{g}_{c}}(\mathbf{\psi }_{\in
}^{r(1)})=0.$ In App. C we introduce a description of the rotational
vectorfield on a family of level surfaces in threespace given by $\beta $
defined as

\begin{eqnarray}
\mathbf{\bar{u}}_{c}^{(1)} &=&\mathbf{X}_{\alpha }^{\beta (1)}\equiv *_{%
\mathbf{g}_{c}}\hat{d}(\alpha \hat{d}\beta ),  \label{eq32j} \\
\mathbf{\psi }_{\in }^{r(1)} &=&\mathbf{X}_{\alpha _{\in }}^{\beta
(1)}\equiv *_{\mathbf{g}_{c}}\hat{d}(\alpha _{\in }\hat{d}\beta ),  \nonumber
\\
\mathbf{\psi }_{t}^{r(1)} &=&\mathbf{X}_{\alpha _{t}}^{\beta (1)}\equiv *_{%
\mathbf{g}_{c}}\hat{d}(\alpha _{t}\hat{d}\beta ), \\
\mathbf{X}_{\alpha }^{\beta } &\equiv &\mathbf{g}_{c}^{-1}(*_{\mathbf{g}_{c}}%
\hat{d}(\alpha \hat{d}\beta ))\text{\thinspace e.t.c.}.
\end{eqnarray}

We define a new roational bracket structure with respect to vorticity
situated at the foliations in threespace of $\beta $ for given metric $%
\mathbf{g}$ and invariant volumeelement $dV$ by 
\begin{eqnarray}
\{\alpha ,f\}_{\beta }\,\,dV &\equiv &*_{\mathbf{g}_{c}}\hat{d}(\alpha \hat{d%
}\beta )\wedge *_{\mathbf{g}_{c}}\hat{d}f=\mathbf{<}df,\mathbf{X}_{\alpha
}^{\beta }>dV,  \label{eq32k} \\
\{\alpha ,f\}_{\beta } &=&\mathbf{X}_{\alpha }^{\beta }(f)=*_{\mathbf{g}%
_{c}}(*_{\mathbf{g}_{c}}(\hat{d}\alpha \wedge \hat{d}\beta )\wedge *_{%
\mathbf{g}_{c}}\hat{d}f).
\end{eqnarray}
In App. C we prove that for purely rotational vectorfields one have that 
\begin{eqnarray}
\lbrack \mathbf{X}_{\alpha _{1}}^{\beta },\mathbf{X}_{\alpha _{2}}^{\beta }]
&=&\mathbf{X}_{-\{\alpha _{1},\alpha _{2}\}_{\beta }}^{\beta }\,, \\
\mathbf{X}_{\alpha }^{\beta }(f(\beta )) &=&\{\alpha ,f(\beta )\}=0.
\end{eqnarray}
This show that one can think about the rotational bracket as a noncanonical
Poisson bracket with functions of type $f(\beta )$ as Casimirs.

From the above splitting of the constraint equation for the velocity field,
we find for the rotational part $\mathbf{\bar{u}}_{c}=\mathbf{\psi }_{t}^{r}+%
\mathbf{\psi }_{*}^{r}\mathbf{u}_{0c}=(\mathbf{\psi }_{*}^{r}\circ \mathbf{%
\phi }_{0*}-Id)\frac{\partial }{\partial t},$ $\mathbf{u}_{0c}=(\mathbf{\phi 
}_{0*}-Id)\frac{\partial }{\partial t}$ . Here we will have one
representation for each component in the constraint equation. A different
representation which concentrate on deformations of the rotational bracket
structure of the reference state is $\mathbf{\bar{u}}_{c}=\mathbf{X}_{\alpha
}^{\beta }=\mathbf{X}_{\alpha }^{\beta ^{0}}+\mathbf{X}_{\alpha }^{\tilde{%
\beta}}\equiv \mathbf{u}_{c}^{0}+\mathbf{\bar{\psi}}^{*}\circ (\mathbf{%
\tilde{\phi}}_{*}-Id)\mathbf{u}_{0c}\,+\,\mathbf{\tilde{u}}_{c}\,,\,\beta
^{0}\equiv \mathbf{\bar{\psi}}^{-1*}\beta _{0}$

$(\mathbf{\bar{\psi}}_{*}\circ \mathbf{\tilde{\phi}}_{*}-Id)\frac{\partial }{%
\partial t}\equiv (\mathbf{\bar{\psi}}_{*}-Id)\frac{d}{d\tau _{c}}\equiv 
\mathbf{\bar{\psi}}_{\tau _{c}}=\mathbf{X}_{\bar{\psi}_{\tau _{c}}}^{\beta
^{0}}\,$such that $\mathbf{u}_{c}^{0}\mathbf{=X}_{\alpha }^{\beta ^{0}}=%
\mathbf{\bar{\psi}}_{\tau _{c}}+\mathbf{\bar{\psi}}_{*}\mathbf{u}_{0c}.\,$%
Here $\frac{d}{d\tau _{c}}\equiv \frac{\partial }{\partial t}+\mathbf{\tilde{%
u}}_{c},\,\,\mathbf{\tilde{u}}_{c}\equiv (\mathbf{\tilde{\phi}}_{*}-Id)\frac{%
\partial }{\partial t}\equiv \mathbf{\tilde{\phi}}_{t}=\mathbf{X}_{\alpha }^{%
\tilde{\beta}}-\mathbf{\bar{\psi}}_{*}\circ (\mathbf{\tilde{\phi}}_{*}-Id)%
\mathbf{u}_{0c}\,.\,$We parameterize the reference rotational velocity as $%
\mathbf{u}_{0c}=\mathbf{X}_{\alpha _{0}}^{\beta _{0}}$ and find that $%
\mathbf{X}_{\alpha -\bar{\psi}_{\tau _{c}}-\mathbf{\bar{\psi}}^{-1*}\alpha
_{0}}^{\beta ^{0}}=0\,\,,$ i.e. $\alpha =\bar{\psi}_{\tau _{c}}+\mathbf{\bar{%
\psi}}^{-1*}\alpha _{0}\,\,$mod$(f(\beta ^{0}))\,.\,\,$The perturbational
aspects of this constraint equation with respect to a parameter $\in $ can
be explored by defining $(\mathbf{\bar{\psi}}_{*}\circ \mathbf{\tilde{\phi}}%
_{*}-Id)\frac{\partial }{\partial \in }=(\mathbf{\bar{\psi}}_{*}-Id)\frac{d}{%
d\bar{\in}}\,\equiv \mathbf{X}_{\bar{\psi}_{\bar{\in}}}^{\beta ^{0}},\,\,%
\frac{d}{d\bar{\in}}\equiv \frac{\partial }{\partial \in }+\mathbf{\tilde{%
\phi}}_{\in }\,,\,\mathbf{\tilde{\phi}}_{\in }\equiv (\mathbf{\tilde{\phi}}%
_{*}-Id)\frac{\partial }{\partial \in }\,.$ To avoid perturbations in the
bracket structure we should pull back the constraint equation for the
rotational potential and obtain $\hat{\alpha}=\hat{\psi}_{\tau _{c}}+\alpha
_{0}\,,\,\alpha =\mathbf{\bar{\psi}}^{-1*}\hat{\alpha},\,\bar{\psi}_{\tau
_{c}}=\mathbf{\bar{\psi}}^{-1*}\hat{\psi}_{\tau _{c}}\,.\,$Here we refer the
potentials to the bracketstructure derived from the background foliation $%
\beta _{0}$ and corresponding vectorfields $\mathbf{X}_{\hat{\alpha}}^{\beta
_{0}}.\,\,$We therefore find that analogous to the theory we have developped
before for density equations 
\begin{eqnarray}
\mathbf{\tilde{\phi}}_{\in },_{t} &=&[\mathbf{\tilde{\phi}}_{\in },\mathbf{%
\tilde{u}}_{c}]+\mathbf{\tilde{u}}_{c},_{\in },  \label{eq32l} \\
\hat{\psi}_{\bar{\in}},_{\tau _{0}} &=&-\{\hat{\alpha},\hat{\psi}_{\bar{\in}%
}\}_{\beta _{0}}+\hat{\alpha},_{\bar{\in}\,\,},  \nonumber \\
\frac{d}{d\tau _{0}} &=&\frac{d}{d\tau _{c}}+\{\alpha _{0},\cdot \}_{\beta
_{0}},  \nonumber \\
\bar{\psi}_{\bar{\in}} &\equiv &\mathbf{\bar{\psi}}^{-1*}\hat{\psi}_{\bar{\in%
}}  \nonumber
\end{eqnarray}
Equation's (\ref{eq32l}) and (\ref{eq32d}) (can also be perturbed with
respect to $\epsilon $ )\thinspace give now a description of both the
vorticity structure and the related diffeomorphism in interaction with
kinetic fluctuations and the longitudinal fluctuations. The above equation
for the vorticity (\ref{eq32d}) with zero right hand side for the purely
rotational case is the direct generalization of the potential description of
the vorticity equation on fixed two dimensional surfaces to convected level
surfaces foliating threespace defined by $\beta ^{0}$ and the rotational
potential $\alpha ^{0}.$ In this case $\mathbf{\bar{u}}_{c}=\mathbf{u}%
_{c}^{0}=\mathbf{X}_{\alpha }^{\beta ^{0}}$ and $\mathbf{\tilde{\phi}=Id}$
since the rotational equation becomes the defining equation for a Lie
pseudogroup related to the rotational structure 
\begin{eqnarray}
(\frac{\partial }{\partial t}+\mathcal{L(}\mathbf{\hat{u}))u}_{c}^{(1)} &=&%
\mathbf{0,}  \label{eq32m} \\
(\frac{\partial }{\partial t}+\mathcal{L(}\mathbf{\hat{u}))\pi }^{(2)} &=&%
\mathbf{0.}  \nonumber
\end{eqnarray}
The rotational diffeomorhisms are still described by eq. \ref{eq32l}, but
with trivial $\mathbf{\tilde{\phi}=Id.\,\,}$In this case the vorticity
equation is analogous to the Vlasov equation with $\alpha \,$playing the
role of the hamiltonian given by $*_{\mathbf{g}_{0}}\mathbf{\hat{\pi}}%
^{(2)}=*_{\mathbf{g}_{0}}\hat{d}\mathbf{X}_{\hat{\alpha}}^{\beta _{0}(1)}=%
\hat{d}*_{\mathbf{g}_{0}}\hat{d}(\hat{\alpha}\hat{d}\beta _{0}).$ If we
invert this operator we find an expression for the rotational potential $%
\hat{\alpha}$ as a functional of the vorticity analogous to noncanonical
hamiltonians on constrained surfaces in classical mechanics. This also
explains why this type of models pops up in many applications in fluid and
plasma physics. Notice that our composition principle for the velocity
induces in principle a finite or infinite series of diffeomorphisms and
divergent and rotational potentials and a corresponding series of
longitudinal and rotational compatibility equations and bracket structures.
The same comments apply to the kinetic generator structure and the
deformations of the electromagnetic fields and vectorpotentials and indeed
the fluctuation metric itself. This completes our discussion of the
parameterization of the hybrid fluid kinetic theory.

\subsubsection{Transformation properties of electromagnetism}

To be complete we also have to formulate Maxwell's equations in such a way
that they are invariant with respect to diffeomorphisms in four space which
fixes time and transform in a natural way. The standard formulation of
Maxwell's equations in four space is with respect to a fixed metric 
\begin{eqnarray}
dF^{(2)} &=&0\,,  \label{eq33} \\
\ast _{g_{0}}d*_{g_{0}}F^{(2)} &=&4\pi j^{(1)},  \nonumber \\
j^{(1)} &=&-\frac{1}{c}j_{k}dx^{k}+\rho dt\equiv J\bar{j}^{(1)}, \\
\ast _{g_{0}}d*_{g_{0}}j^{(1)} &=&0.
\end{eqnarray}
Notice that the homogenous equation transforms in a natural way with respect
to diffeomorphisms in four space while the inhomogenous part of the equation
does not. We have that $dF^{(2)}=d\psi ^{-1*}\hat{F}^{(2)}=\psi ^{-1*}d\hat{F%
}^{(2)}=0.\,$We rectify this in a similar way as for the momentum equation
by introducing $\,\frac{1}{J}*_{g_{0}}d*_{g_{0}}J=*_{g}d*_{g}=\frac{1}{J_{id}%
}*_{g_{id}}d*_{g_{id}}J_{id}.$ Here $g_{id}$ is the metric presented in a
frame of reference where -$\det (g_{id})=1$ and $g_{0}=\psi _{0}^{-1*}g_{id}$%
, $J_{id}=-\det (g).$We now find the following form of inhomogenous
Maxwell's equations 
\begin{eqnarray}
\ast _{g}d*_{g}\bar{F}^{(2)} &=&4\pi \bar{j}^{(1)},  \label{eq34} \\
\bar{F}^{(2)} &=&F^{(2)}/J\,,\,\bar{j}^{(1)}=j^{(1)}/J  \nonumber \\
\ast _{g}d*_{g}\bar{j}^{(1)} &=&0.
\end{eqnarray}

When the Maxwell equations are presented in this way (or even better if one
tensor it with the four volumeform $dV_{g}$ to take into account that it is
relation between densities) they transform in the natural way under
diffeomorphisms as 
\begin{eqnarray}
F^{(2)} &=&\psi ^{-1*}\hat{F}^{(2)},\,\bar{F}^{(2)}=\psi ^{-1*}(\check{F}%
^{(2)})\,,\,\check{F}^{(2)}\equiv \hat{F}^{(2)}\,J(\psi )),  \label{eq35} \\
\mathcal{F} &=&\bar{F}^{(2)}\otimes dV_{g}=F^{(2)}\otimes
dV_{0}=F_{id}^{(2)}\otimes dV_{id},  \nonumber \\
\mathcal{J} &=&\bar{j}^{(1)}\otimes dV_{g}=j^{(1)}\otimes dV_{0}  \nonumber
\\
\mathcal{F} &=&\psi ^{-1*}\mathcal{\hat{F}\,}\,\,,\,\,\mathcal{\hat{F}}=\hat{%
F}^{(2)}\otimes \psi ^{*}dV_{0\,},\,\psi *dV_{0}=J(\psi )dV_{0}, \\
\bar{j}^{(1)} &=&\psi ^{-1*}\hat{j}^{(1)}.
\end{eqnarray}
The upshot is that the new current and field quantities correspond to
current density form and the field density form, $\mathcal{J}$ and $\mathcal{%
F}$ and they transform according to these.\footnote{%
From a measurement point of view it also make sense to study the
transformation properties of density forms since we always have to measure
with respect to some volume and timeinterval.} In case we do not have a
euclidean reference metric, it can be an advantage to refer to an orthogonal
frame instead with $F^{(2)}=F_{id}^{(2)}/(-\det (g_{0}))^{\frac{1}{2}}$ and $%
\bar{F}^{(2)}=\bar{F}_{id}^{(2)}/(-\det (g))^{\frac{1}{2}}.$ It is possible
to introduce electromagnetic equations with respect to $\mathcal{J}$ and $%
\mathcal{F}$, but it requires covariant derivatives which will be beyond the
scope of our presentation. We are now ready to pullback our electromagnetic
equations to the reference level and we find 
\begin{eqnarray}
d\hat{F}^{(2)} &=&0,  \label{eq36} \\
\ast _{g_{0}}d*_{g_{0}}\check{F}^{(2)} &=&4\pi \hat{j}^{(1)},  \nonumber \\
\ast _{g_{0}}d*_{g_{0}}\hat{j}^{(1)} &=&0.  \nonumber
\end{eqnarray}
\thinspace The same type of equation will be fullfilled for the reference
fields and currents, $F_{0}^{(2)}$ and $j_{0}^{(1)}.$ If we define $%
F^{(2)}=\psi ^{-1*}(F_{0}^{(2)}+\tilde{F}^{(2)}),\,\tilde{F}^{(2)}\equiv 
\hat{F}^{(2)}-F_{0}^{(2)},\,\grave{F}^{(2)}\equiv \check{F}%
^{(2)}-F_{0}^{(2)}\,,\,\tilde{j}^{(1)}\equiv \hat{j}^{(1)}-j_{0}^{(1)}\,\,,$
we find electromagnetic equations suitable for perturbation with respect
to\thinspace one diffeomorphism 
\begin{eqnarray}
d\tilde{F}^{(2)} &=&0,  \label{eq37} \\
\ast _{g_{0}}d*_{g_{0}}\grave{F}^{(2)} &=&4\pi \tilde{j}^{(1)},  \nonumber \\
\ast _{g_{0}}d*_{g_{0}}\tilde{j}^{(1)} &=&0.  \nonumber
\end{eqnarray}

In case we have multiple fluids, we would could perform the above procedure
with respect to several diffeomorphisms and fluctuation metrics\footnote{%
The same remark applies to the stresstensor which transformation properties
is discussed below. Although we will not explicitly discuss multiple fluids
it is not a major complication from a formal point of view only for the
complexity of the presentation. Therefore we have decided to leave this
point for explicit, future applications.}.\thinspace Notice, that relations
like $A^{(1)}=\Psi ^{-1*}\hat{A}^{(1)},$ e.t.c. are not trivial when written
out componentwise, especially since we propose to use a fluctuating metric
to obtain contravariant tensors. We will come back to a more elaborate study
of the transformation and perturbation properties of electromagnetism in a
separate paper elsewhere.

\subsubsection{Transformation properties of the stress tensor and current
density}

Let us write our expression for the stresstensor and current density as a
covariant twotensor and oneform respectively and study their transformation
properties with respect to kinetic theory. 
\begin{eqnarray*}
\mathbf{\bar{P}}^{(2)} &=&\int \frac{1}{Jm}(\mathbf{p}^{(1)}-m\mathbf{w}%
^{(1)})\otimes (\mathbf{p}^{(1)}-m\mathbf{w}^{(1)})f(z,t)d^{3}\mathbf{p,} \\
\mathbf{\bar{j}}^{(1)} &=&\int \frac{-e}{J}(\mathbf{p}^{(1)}-m\mathbf{w}%
^{(1)})f(z,t)d^{3}\mathbf{p} \\
\mathbf{w}^{(1)} &=&\mathbf{\hat{u}}^{(1)}+\frac{e}{c}\mathbf{A}^{(1)}%
\mathbf{.}
\end{eqnarray*}
We now find the pullbacked expressions 
\begin{eqnarray}
\mathbf{\hat{P}}^{(2)} &=&\mathbf{\psi }^{*}\mathbf{\bar{P}}^{(2)}=\int 
\frac{1}{m}(\mathbf{p}^{(1)}-m\mathbf{\hat{w}}^{(1)})\otimes (\mathbf{p}%
^{(1)}-m\mathbf{\hat{w}}^{(1)})\tilde{f}(z,t)d^{3}\mathbf{p\,\,,}
\label{eq38} \\
\mathbf{\hat{j}}^{(1)} &=&\mathbf{\psi }^{*}\mathbf{\bar{j}}^{(1)}=\int 
\frac{1}{m}(\mathbf{p}^{(1)}-m\mathbf{\hat{w}}^{(1)})\tilde{f}(z,t)d^{3}%
\mathbf{p\,.}  \nonumber
\end{eqnarray}

\subsection{Physical coordinates}

Our theory will not be complete before we formulate the kinetic theory in
physical variables also. It is possible to work in canonical coordinates at
least when there is no background, magnetic field as we have indicated in
our work on OC theory$^{\text{7}}$ . But it requires doing the
transformation from canonical distribution function to physical distribution
function as an active symplectic transformation on distributions (see
App.A). However, with a magnetic field this transformation is not so
valuable since it is not a perturbation. Moreover, there is the question for
which coordinates one should specify the reference distribution. The reason
why I have been reluctant to give up the canonical formal approach is that
in physical coordinates the Poisson bracket itself will be perturbed.
However, in section A6.3, we have described new physical coordinates which
gives the Poisson bracket in terms of background magnetic fields only . We
call this the \textit{interaction picture} for the Vlasov equation. A direct
extension of our work on OC$^{7}$ theory would be to specify the OC
coordinates in interaction variables. The obvious advantage is that now it
is possible to do perturbation expansions without perturbing the invariant
bracket and the transformation from interaction distribution function to
physical distribution function is now really a perturbation. The relation
between the interaction distribution and the physical and canonical
distributions is given by $\hat{f}=\phi _{c1}^{-1*}f^{i},\,f^{i}=\phi
_{c0}^{-1*}f$ ,where the the transformation $\phi _{c1}^{-1*}$ and $\phi
_{c0}^{-1*}$are described in eq.(\ref{eqa12b}). The relation between the
interaction physical coordinates and canonical or physical coordinates is
given by $\mathbf{\hat{p}}=\mathbf{p-}\frac{e}{c}\mathbf{A}_{0},\,\mathbf{p}%
_{p}\mathbf{=\hat{p}}-\frac{e}{c}\mathbf{A}_{1}$ which is nothing else than
the shift transformations described by $\phi _{c0}^{-1*}$ and $\phi
_{c1}^{-1*}.$ Here we use the gauge $\phi _{1}=0$ so $\mathbf{A}%
_{1}=-c\int\limits^{t}\mathbf{E}_{1}(t^{\prime })dt^{\prime }$ can be given
a gauge invariant meaning as proportional to the accumulated, perturbed
electric field vector referred to some fixed time.

The standard Vlasov equation for the distribution function, $\hat{f},$ in
euclidean physical coordinates may be rewritten in a form more suitable for
our purposes,

\begin{eqnarray*}
&&\frac{d\hat{f}}{d\tau ^{^{\prime }}}+\{\hat{f},H^{p}\}=0, \\
\frac{d}{d\tau ^{^{\prime }}} &\equiv &\frac{\partial }{\partial t}+\mathbf{%
J\cdot (-}\frac{e}{c}\frac{\partial \mathbf{A}^{(1)}}{\partial t}), \\
H^{p} &\equiv &\frac{\mathbf{p}_{p}^{2}}{2m}+e\phi , \\
\mathbf{J} &\equiv &\frac{\partial }{\partial \mathbf{x}}\wedge \frac{%
\partial }{\partial \mathbf{p}_{p}}+\frac{e}{c}\mathbf{B}_{\mathbf{p}}^{(2)},
\\
\{\hat{f},H^{p}\} &\equiv &\mathbf{J:(}d_{P}\hat{f},d_{P}H^{p}).
\end{eqnarray*}
Here $\mathbf{J,\{,\},A}^{(1)}$ and $\mathbf{B}^{(2)}=d\mathbf{A}^{(1)}$ are
the contravariant Poissontensor in euclidean physical coordinates, the
Poisson bracket in physical coordinates, the covariant vectorpotential, the
covariant magnetic field twoform. The Poisson bracket is written in standard
form in eq.(\ref{eqqa12}). $\frac{d}{d\tau ^{^{\prime }}}$ $,d_{P}$ and $d$
are a convective derivative with respect to the vectorpotential part of the
electric field, the exterior derivative in six dimensional phase space and
the exterior derivative in three dimensional space. In App. A we show that $%
\mathbf{J=}\phi _{c*}\mathbf{J}_{c}$ is the pushforward map of the standard
canonical Poissontensor. The definition of the pushforward map is given in
App. A acting on a vectorfield, but the action is trivially extended to
higher degree contravariant tensors by applying the given action on each
tensorindices. We have chosen to represent our tensors with respect to the
standard basis in euclidean coordinates, i.e. $dx_{i},\,dp_{i}$ for
covariant forms and $\frac{\partial }{\partial x_{i}}\,,\,\frac{\partial }{%
\partial p_{i}}$ for contravariant vectorfields with obvious extensions for
higher degree tensors. With this notation the antisymmetric covariant
twotensor $\mathbf{B}^{(2)}=d\mathbf{A}^{(1)}=\frac{\partial A_{k}}{\partial
x_{j}}dx_{j}\wedge dx_{k}$ is pulled down to a contravariant,antisymmetric
multivector identified with the corresponding multivector in momentum space $%
\mathbf{B}_{\mathbf{p}}^{(2)}=$ $\frac{\partial A_{k}}{\partial x_{j}}\frac{%
\partial }{\partial p_{j}}\wedge \frac{\partial }{\partial p_{k}}\,$.%
\footnote{%
In a noneuclidan space we would have to be careful with the metric. All this
can be worked out with respect to an underlying Riemannian metric space, but
we have chosen to postpone the description of this more general theory.}
Notice that with a timedependent vectorpotential it is not possible to treat
single particle dynamics as generated by a hamiltonian in six dimensional
euclidean physical phase space except for in a convected sense. However, in
eight dimensional extended phase space or in six dimensional canonical
coordinates the dynamics is described by hamiltonian generators. We show in
App.B6.3 that the parameterization can still be presented in the new
interaction picture by the following analogous generator equation if we take
time independent background electromagnetic fields $\mathbf{B}_{0},\mathbf{E}%
_{0}=-\nabla \phi _{0}$

\begin{eqnarray*}
\frac{\partial f^{i}}{\partial t}+\{f^{i},H^{i}\} &=&0, \\
\{f^{i},H^{i}\} &=&\mathbf{J}_{0}:(d_{P}f^{i},d_{P}H^{i}\} \\
H^{i} &=&\psi _{t}^{i}+(\psi ^{i})^{-1*}H_{0}^{i} \\
H^{i} &=&\frac{(\mathbf{\hat{p}}-\frac{e}{c}\mathbf{A}_{1})^{2}}{2m}-\gamma
_{0\,,}\;H_{0}^{i}=\frac{\mathbf{\hat{p}}^{2}}{2m}-\gamma _{0}\,, \\
\gamma _{0} &=&-e\phi _{0}.
\end{eqnarray*}
If the background electromagnetic fields are timedependent, we can transform
our Vlasov equation and canonical generator equation to interaction
coordinates by

\begin{eqnarray*}
\frac{df^{i}}{d\tau _{0}^{^{\prime }}}+\{f^{i},H^{i}\}_{n0}
&=&0,\,\,\{f^{i},H^{i}\}_{n0}\equiv \mathbf{J}_{0}:(d_{P}f^{i},d_{P}H^{i}\},
\\
H^{i} &=&\psi _{\tau _{0}^{^{\prime }}}^{i}+(\psi ^{i})^{-1*}H_{0}^{i}\,, \\
\psi _{\tau _{0}^{^{\prime }}}^{i} &\equiv &(\psi _{*}^{i}-Id)\frac{d}{d\tau
_{0}^{^{\prime }}}, \\
\frac{d}{d\tau _{0}^{^{\prime }}} &\equiv &\frac{\partial }{\partial t}+%
\mathbf{J}_{0}\cdot (-\frac{e}{c}\frac{\partial A_{0}^{(1)}}{\partial t}), \\
\mathbf{J}_{0} &=&\phi _{c0}^{-1*}\mathbf{J}_{c}=\frac{\partial }{\partial 
\mathbf{x}}\wedge \frac{\partial }{\partial \mathbf{\hat{p}}}+\frac{e}{c}%
\mathbf{B}_{0\mathbf{p}}^{(2)} \\
\mathbf{J}_{c} &=&\frac{\partial }{\partial \mathbf{x}}\wedge \frac{\partial 
}{\partial \mathbf{p}}=( 
\begin{array}{ll}
\mathbf{0} & \mathbf{I} \\ 
-\mathbf{I} & \mathbf{0}
\end{array}
).
\end{eqnarray*}

Here $\mathbf{B}_{0}^{(2)}=d\mathbf{A}_{0}^{(1)},\mathbf{A}_{0}^{(1)},\,%
\mathbf{J}_{0}\,$and $\mathbf{J}_{c}$ are the reference magnetic field
twoform, the vectorpotential oneform, the interaction Poissontensor and the
canonical Poissontensor.

For those who worry about the gaugeinvariance of the above equations, we
formulate equivalent gaugeinvariant equations by taking out the potential in
the hamiltonian generators and add a corresponding term to the convective
derivatives. We thereby obtain a convective derivative correponding to the
acceleration of the electric field with obvious form of the Vlasov equation
which is gaugeinvariant\footnote{%
In a general metric we must study covariant derivatives and parallel
translation to give this an invariant meaning.}

\begin{eqnarray*}
\frac{d}{d\tau } &=&\frac{\partial }{\partial t}+\mathbf{J\cdot (-}e\mathbf{E%
}^{(1)})\,,\,H_{p}\equiv \frac{\mathbf{p}_{p}^{2}}{2m}, \\
\frac{d}{d\tau _{0}} &\equiv &\frac{\partial }{\partial t}+\mathbf{J}%
_{0}\cdot (-e\mathbf{E}_{0}^{(1)}),\text{ }\bar{H}^{i}\equiv \frac{(\mathbf{%
\hat{p}-}\frac{e}{c}\mathbf{A}_{1})^{2}}{2m}, \\
\bar{H}^{i} &=&\psi _{\tau _{0}}^{i}+(\psi ^{i})^{-1*}\bar{H}_{0}^{i} \\
\psi _{\tau _{0}}^{i} &\equiv &(\psi _{*}^{i}-Id)\frac{d}{d\tau _{0}}.
\end{eqnarray*}

To interpret the above introduction of new noncanonical coordinates we now
give a theorem which show in which sense we can generalize our results in
canonical coordinates to any coordinates independent if they are
noninertially, convected in phase space with respect to the canonical
coordinates or not. Since the definition of what is a hamiltonian flow or
not cannot depend on the coordinate system, we suggest to call the flow
hamiltonian if it can be transferred to a a hamiltonian flow by a
noncanonical (not preserving the Poisson tensor) map as indicated in the
theorem below.

\begin{theorem}
The transformation of the hamiltonian generator, $\psi _{t},$ in canonical
coordinates given by exact oneform $d_{P}\psi _{t}\equiv \omega _{c}\cdot
((\psi _{*}-Id)\frac{\partial }{\partial t})$ to other noncanonical
coordinates (in general timedependent) $\mathbf{\phi }(t):z\longrightarrow Z=%
\mathbf{\phi }(t)(z)$ is given by the action of the corresponding inverse
seven dimensional map which fixes time (c.f. discussion in App. A) $\phi
^{-1}(Z,t)\equiv (\mathbf{\phi }^{-1}(t)(Z),t)$ as

\begin{eqnarray*}
d_{p}\bar{\psi}_{\tau } &\equiv &\omega \cdot ((\bar{\psi}_{*}-Id)\frac{d}{%
d\tau }), \\
\frac{d}{d\tau } &\equiv &\phi _{*}\frac{\partial }{\partial t}=\frac{%
\partial }{\partial t}+\mathbf{X}_{t\,},\,\,\mathbf{X}_{t}\equiv (\phi
_{*}-Id)\frac{\partial }{\partial t}\,, \\
\omega &\equiv &\mathbf{\phi }^{-1*}\omega _{c}\,, \\
\bar{\psi} &\equiv &Ad(\phi )\psi =\phi \circ \psi \circ \phi ^{-1}.
\end{eqnarray*}
The corresponding noncanonical Poissontensor and Poissonbracket are given by 
\begin{eqnarray*}
\mathbf{J} &=&\mathbf{\phi }_{*}\mathbf{J}_{c}\,, \\
\{f,g\}_{n} &=&\mathbf{J:(}d_{P}f,d_{P}g)\,.
\end{eqnarray*}
Moreover, if $\mathbf{\psi }$ is a canonical, Poisson preserving map with
respect $\mathbf{J}_{c},\,\{,\}$ such that $\mathbf{\psi }_{*}\mathbf{J}_{c}=%
\mathbf{J}_{c}\,$and $\mathbf{\psi }^{-1*}\{f,g\}=\{\mathbf{\psi }^{-1*}f,%
\mathbf{\psi }^{-1*}g\}\,$, then $\mathbf{\bar{\psi}}\equiv Ad(\mathbf{\phi
(t))\psi }$ is Poisson preserving with respect to $\mathbf{J,}$ $\{,\}_{n%
\text{ }}$ such that $\mathbf{\bar{\psi}}_{*}\mathbf{J=J}$ and $\mathbf{\bar{%
\psi}}^{-1*}\{f,g\}_{n}=\{\mathbf{\bar{\psi}}^{-1*}f,\mathbf{\bar{\psi}}%
^{-1*}g\}_{n}\,.$
\end{theorem}

\proof%
Let us asume that the vectorfield $\mathbf{X}_{\psi _{t}}\equiv \mathbf{J}%
_{c}\cdot (d_{P}\psi _{t})=(\psi _{*}-Id)\frac{\partial }{\partial t}$ is
hamilonian generated with generator $\psi _{t}$. Then transform the exterior
derivative of the hamiltonian by pullback of the above map either as a
result of an active transformation of the oneform or a passive coordinate
transformation (c.f. the discussion below) is given by 
\begin{eqnarray*}
\phi ^{-1*}d_{P}\psi _{t} &=&d_{P}(\phi ^{-1*}\psi _{t})\equiv d_{P}\bar{\psi%
}_{\tau } \\
&=&\phi ^{-1*}(\omega _{c}((\psi _{*}-Id)\frac{\partial }{\partial t}%
))=\omega (\phi _{*}(\psi _{*}-Id)\frac{\partial }{\partial t}) \\
&=&\omega ((\phi _{*}\circ \psi _{*}\circ \phi _{*}^{-1}-Id)\phi _{*}\frac{%
\partial }{\partial t})=\omega (((Ad(\phi )\psi )_{*}-Id)\frac{d}{d\tau }) \\
&=&\omega ((\bar{\psi}_{*}-Id)\frac{d}{d\tau }). \\
\frac{d}{d\tau } &\equiv &\phi _{*}\frac{\partial }{\partial t}=\frac{%
\partial }{\partial t}+\mathbf{X}_{t}\,, \\
\mathbf{X}_{t} &\equiv &(\phi _{*}-Id)\frac{\partial }{\partial t}\,, \\
\omega &=&\phi ^{-1*}\omega _{c}=\mathbf{\phi }^{-1*}\omega _{c}, \\
&\Rightarrow &\mathbf{J=}\phi _{*}\mathbf{J}_{c}=\mathbf{\phi }_{*}\mathbf{J}%
_{c}.
\end{eqnarray*}
Here we are treating tensors in phase space as embedded in phase space
extended with time such that we freely interchange the action of seven
dimensional map $\phi $ fixing time and the timedependent six dimensional
map $\mathbf{\phi }$ on such tensors. This will also compress notation and
proofs. The action of tensors with even a fixed time component fixes such
components, but give extra contributions to the phase space components
through terms like $\mathbf{(}\phi _{*}-Id)\frac{\partial }{\partial t}\,.$
Notice that for the case that $\mathbf{\phi }$ is not timedependent, we have
that $\mathbf{X}_{t}=\mathbf{0}$ and $\frac{d}{d\tau }=\frac{\partial }{%
\partial t}\,,\bar{\psi}_{\tau }=\bar{\psi}_{t}$ while the Poissontensor,
Poisson bracket e.t.c. is still as prescribed above.

We now observe that vectorfield with only phase space components 
\[
(\bar{\psi}_{*}-Id)\frac{d}{d\tau }=X_{\bar{\psi}_{t}}=\mathbf{J\cdot (}d_{P}%
\bar{\psi}_{\tau }), 
\]
is indeed hamiltonian with respect to the Poissontensor $\mathbf{J}$ since $%
\mathbf{J\cdot }\omega =Id.$

The Poisson bracket is transformed by a non Poissonpreserving map as 
\begin{eqnarray*}
\mathbf{\phi }^{-1*}\{f,g\} &=&\mathbf{\phi }^{-\acute{1}*}(\mathbf{J}%
_{c}:(d_{P}f,d_{P}g)=\mathbf{\phi }^{-1*}\circ \mathbf{J}_{c}\circ \mathbf{%
\phi }^{*}:(\mathbf{\phi }^{-1*}d_{P}f,\mathbf{\phi }^{-1*}d_{P}g) \\
&=&\mathbf{J:(}d_{P}\hat{f},d_{P}\hat{g})=\{\hat{f},\hat{g}\}_{n},\,\,\hat{f}%
=\mathbf{\phi }^{-1*}f\,,\,\hat{g}=\mathbf{\phi }^{-1*}g\,,
\end{eqnarray*}
since $\mathbf{\phi }_{*}\mathbf{T=\phi }^{-1*}\circ \mathbf{T\circ \phi }%
^{*}$ for a contravariant tensor [\cite{geom}].

If $\mathbf{\psi }$ is a canonical Poisson preserving map such that $\mathbf{%
\psi }_{*}\mathbf{J}_{c}=\mathbf{J}_{c}.$ Then it follows that a non Poisson
preserving map $\mathbf{\phi }$ acts as 
\begin{eqnarray*}
\mathbf{\phi }_{*}\circ \mathbf{\psi }_{*}\mathbf{J}_{c} &=&(\mathbf{\phi }%
_{*}\circ \mathbf{\psi }_{*}\circ \mathbf{\phi }_{*}^{-1})\circ \mathbf{\phi 
}_{*}\mathbf{J}_{c}, \\
&\Rightarrow &\mathbf{\bar{\psi}}_{*}\mathbf{J=J,\,\,\,\,\bar{\psi}=}Ad%
\mathbf{(\phi )\psi .}
\end{eqnarray*}
\endproof%

\begin{remark}
All the formal expansions we introduced earlier is valid simply by replacing 
$\mathbf{\psi =}\exp (\mathbf{X}_{w)}\,,\mathbf{X}_{w}\equiv \mathbf{J}%
_{c}\cdot d_{P}w\rightarrow \mathbf{\bar{\psi}=}\exp (\mathbf{X}_{\bar{w}%
})\,,\,\mathbf{X}_{\bar{w}}\equiv \mathbf{J\cdot }d_{P}\bar{w},$ $\psi
_{t}=i\exp (\mathcal{L}_{w})w_{,t}\,,$ $\mathcal{L}_{w}=\{w,\cdot
\}\rightarrow \bar{\psi}_{\tau }=i\exp (\mathcal{L}_{\bar{w}})\frac{d\bar{w}%
}{d\tau }\,,\,\mathcal{L}_{\bar{w}}=\{\bar{w},\cdot \}_{n}\,,$

$H\rightarrow \bar{H}=\mathbf{\phi }^{-1*}H=H\circ \mathbf{\phi }^{-1},$ $%
f\rightarrow \bar{f}=\mathbf{\phi }^{-1*}f=\mathbf{\bar{\psi}}^{-1*}\bar{f}%
_{0}\,,$ $\bar{f}_{0}\equiv \mathbf{\phi }^{-1*}f_{0}$

$\frac{\partial f}{\partial t}+\{f,H\}=0\rightarrow \frac{d\bar{f}}{d\tau }%
+\{\bar{f},\bar{H}\}_{n}=0;$

$H=\psi _{t}+\psi ^{-1*}H_{0}\,$,$\,\bar{H}=\bar{\psi}_{\tau }+\bar{\psi}%
^{-1*}\bar{H}_{0}$

$\psi _{\epsilon },_{t}+\{\psi _{\epsilon },H\}+H,_{\epsilon }=0,\rightarrow 
\bar{\psi}_{\bar{\epsilon}},_{\tau }+\{\bar{\psi}_{\bar{\epsilon}},\bar{H}%
\}_{n}+\bar{H},_{\bar{\epsilon}}=0$

Here the notation $\bar{\epsilon}.$ is used to take care of situations where
the map $\mathbf{\phi }$ also depend on the perturbation parameter $\epsilon 
$. Therefore, analogous to the above notation for the time generator we have
that in a phase space extended by both time and $\epsilon $ 
\[
\frac{d}{d\bar{\epsilon}}\equiv \phi _{*}\frac{\partial }{\partial \epsilon }%
=\frac{\partial }{\partial \epsilon }+(\phi _{*}-Id)\frac{\partial }{%
\partial \epsilon }=\frac{\partial }{\partial \epsilon }+\mathbf{X}%
_{\epsilon }. 
\]
In the case that the coordinate map does not depend the parameter $\epsilon $%
, we can replace $\bar{\epsilon}$ by $\epsilon $ above. This is in agreement
with the above interaction picture philosophy where we suggest to let the
transformation with respect to canonical variables and consequently the
Poisson tensor be unperturbed. After the transformation to the interaction
picture we suggest to introduce additional transformations to e.g.
gyrokinetic, driftkinetic or oscillation center variables due to adiabatic
or exact symmetries in the problem. These applications and desciption of
nonlinear perturbation theory in general is outside the scope of this
article since it requires the introduction of noneuclidean metric.
\end{remark}

The hybrid fluid kinetic theory can therefore now be presented in
interaction physical coordinates by doing the following changes in the
canonical theory 
\begin{eqnarray*}
f^{i} &=&(\psi ^{i})^{-1*}f_{0}^{i}=(\bar{\psi}^{i})^{-1*}\circ (\tilde{\psi}%
^{i})^{-1*}\tilde{f}_{0}^{i}\,, \\
\tilde{f}_{0}\Sp i  \endSp &=&\bar{\psi}_{0}^{i*}f_{0}^{i}\,,
\end{eqnarray*}

\begin{eqnarray}
\overline{\psi }_{\tau _{0}^{\prime }}^{i}(z,t) &=&\overline{\psi }_{1\tau
_{0}^{\prime }}^{i}(z,t)+\overline{\psi }_{2\tau _{0}^{\prime }}^{i}(\mathbf{%
x},t)=\mathbf{\hat{p}}\cdot \widehat{\mathbf{u}}+\overline{\psi }_{2\tau
_{0}^{\prime }}^{i}\;,  \label{eq27b} \\
\overline{\psi }_{2\tau _{0}^{\prime }}^{i} &=&-m\frac{\widehat{\mathbf{u}}%
^{2}}{2}-\frac{e}{c}\mathbf{A}_{1}\cdot \widehat{\mathbf{u}}\;, \\
\overline{\psi }_{\epsilon }^{i}(z,t) &=&\overline{\psi }_{1\epsilon
}^{i}(z,t)+\psi _{2\epsilon }^{i}(\mathbf{x},t)=\mathbf{\hat{p}}\cdot 
\mathbf{\psi }_{\epsilon }+\overline{\psi }_{2\epsilon }^{i}\;,  \nonumber \\
\overline{\psi }_{1\tau _{0}^{\prime }} &\equiv &\mathbf{\hat{p}}\cdot 
\mathbf{\widehat{u}\;,\;}\overline{\psi }_{1\epsilon }\equiv \mathbf{\hat{p}}%
\cdot \mathbf{\psi }_{\epsilon }\;.
\end{eqnarray}

The hybrid fluid kinetic theory can now be repeated and we find analogously
to our earlier derivation in canonical coordinates 
\begin{eqnarray}
\hat{H}^{i} &=&H^{i}-\bar{\psi}_{\tau _{0}^{\prime }}^{i}=\frac{(\mathbf{%
\hat{p}-}m\mathbf{\hat{w}}_{1})^{2}}{2m}-\gamma _{0}\,,  \label{eq27c} \\
\tilde{\psi}_{t}^{i} &=&\tilde{H}^{i}-(\tilde{\psi}^{i})^{-1*}\tilde{H}%
_{0}^{i}\,,\,\tilde{H}^{i}\equiv \bar{\psi}^{\acute{\imath}*}\hat{H}^{i}\,\,,%
\tilde{H}_{0}^{i}\equiv \bar{\psi}_{0}^{\acute{\imath}*}\hat{H}_{0}^{i} 
\nonumber \\
\tilde{f}^{i} &=&(\tilde{\psi}^{i})^{-1*}\tilde{f}_{0}^{i}\,,  \nonumber \\
\mathbf{\hat{w}}_{1} &=&\mathbf{\hat{u}+}\frac{e}{cm}\mathbf{A}_{1}. 
\nonumber
\end{eqnarray}

The fluid part of the hybrid fluid kinetic theory are of course the same as
before while in interaction physical coordinates we have the resonant
particle kinetic equation with the above parameterization given as

\begin{equation}
\frac{d\tilde{f}^{i}}{d\tau _{0}^{\prime }}+\{\tilde{f}^{i},\tilde{H}%
^{i}\}_{n0}=0.  \label{eq27d}
\end{equation}
The corresponding action principle for the resonant particle part will
follow trivially from App. C by doing the above replacements in interaction
physical coordinates.

We are now ready to replace the above fluid symplectic and resonant particle
generators by generators in laboratory frame which are more suitable for
perturbations and linearization which it is our aim to do. We write our new
parameterization as $f^{i}=(\hat{\psi}^{i})^{-1*}\check{f}^{i}=(\hat{\psi}%
^{i})^{-1*}(\check{\psi}^{i})^{-1*}f_{i}^{0}.$ Here $\check{f}^{i}$ is the
resonant particle distribution in the laboratory frame defined such that $%
\int \check{f}^{i}d^{3}\mathbf{\hat{p}=}\rho _{0}\,\,\,$ $(\frac{\partial
\rho _{0}}{\partial t}+\nabla \cdot (\mathbf{u}_{0}\rho _{0})=0)$ . The
discussion of the properties of this representation of the follows trivially
from what we have proved before. The fluid kinetic generator we now define
as 
\begin{eqnarray}
\hat{\psi}_{\tau _{0}^{\prime }}^{i} &\equiv &\mathbf{\hat{p}\cdot \psi }%
_{t}+\hat{\psi}_{2\tau _{0}^{\prime }}^{i}(\mathbf{x,}t)\,,  \label{eq27e} \\
\hat{\psi}_{2\tau _{0}^{\prime }}^{i}(\mathbf{x,}t) &=&-(\frac{m}{2}\mathbf{%
\psi }_{t}^{2}+\frac{e}{c}\mathbf{A}_{1}\cdot \mathbf{\psi }_{t})\,, \\
\hat{\psi}_{\epsilon }^{i} &\equiv &\mathbf{\hat{p}\cdot \psi }_{\epsilon
}\,+\hat{\psi}_{\epsilon }^{i}(\mathbf{x,}t).  \nonumber
\end{eqnarray}

The resonant particle distribution in the laboratory frame is then described
by the following Liouville equation 
\begin{eqnarray}
\frac{d\check{f}^{i}}{d\tau _{0}^{\prime }}+\{\check{f}^{i},\check{H}%
^{i}\}_{n0} &=&0\,,  \label{eq27f} \\
\frac{df_{0}^{i}}{d\tau _{0}^{\prime }}+\{f_{0}^{i},H_{0}^{i}\}_{n0} &=&0, 
\nonumber \\
\check{\psi}_{\tau _{0}^{\prime }}^{i} &=&\check{H}^{i}+\breve{\psi}_{\tau
_{0}^{\prime }}^{i}-(\check{\psi}^{i})^{-1*}H_{0}^{i}\,,  \nonumber \\
\check{H}^{i} &\equiv &\hat{\psi}^{*}\breve{H}^{i}\,,  \nonumber \\
\breve{H}^{i} &\equiv &\frac{(\mathbf{\hat{p}-}\frac{e}{c}\mathbf{A}_{1}-%
\mathbf{\psi }_{t})^{2}}{2m}-\gamma _{0}.  \nonumber
\end{eqnarray}
The above form of the equations and parameterization are the one which is
suitable for perturbation since everything is now really expressed in
interaction physical coordinates and a near identity, symplectic fluid
generator which is directly related to the fluid generator $\mathbf{\psi }%
_{t}$. Moreover, there is no reason to change the fluid part of the theory
earlier developped since it is already expressed in terms of the fluid
generator.

We can relate our new kinetic fluid generator to the earlier one by putting $%
\hat{\psi}\equiv \bar{\psi}\circ \breve{\psi}^{-1}.$ We observe that for the
reference state $\bar{\psi}_0=\breve{\psi}_0.$ We now find that one can
express

\begin{eqnarray*}
\hat{\psi}^{-1*}\breve{\psi}_{\tau _{0}^{\prime }}^{i} &=&-\hat{\psi}_{\tau
_{0}^{\prime }}^{i}+\bar{\psi}_{\tau _{0}^{\prime }}^{i} \\
&=&(\mathbf{\hat{p}-(}m\mathbf{\psi }_{t}+\frac{e}{c}\mathbf{A}_{1}))\cdot 
\mathbf{\psi }_{*}\mathbf{u}_{0}-\frac{m}{2}(\mathbf{\psi }_{*}\mathbf{u}%
_{0})^{2}, \\
\hat{\psi}^{-1*}\breve{\psi}_{\epsilon }^{i} &=&-\hat{\psi}_{\epsilon }^{i}+%
\bar{\psi}_{\epsilon }^{i}=-\hat{\psi}_{2\epsilon }^{i}(\mathbf{x},t)+\bar{%
\psi}_{2\epsilon }^{i}(\mathbf{x},t)=0.
\end{eqnarray*}
The reason why we can take $\bar{\psi}_{\epsilon }^{i}=\hat{\psi}_{\epsilon
}^{i}$ is that the fluid compatibility relations for $\mathbf{\psi }_{t},%
\mathbf{\psi }_{\epsilon }$ or $\mathbf{\psi }_{t},\mathbf{\hat{u}}$ are
consistent with the kinetic compatibility relations $\overline{\psi }%
_{\epsilon ,\tau _{0}^{\prime }}+\{\overline{\psi }_{\epsilon },\overline{%
\psi }_{\tau _{0}^{\prime }}\}-\overline{\psi }_{\tau _{0}^{\prime
}},_{\epsilon }=0$ and $\widehat{\psi }_{\epsilon ,\tau _{0}^{\prime }}+\{%
\widehat{\psi }_{\epsilon },\widehat{\psi }_{\tau _{0}^{\prime }}\}-\widehat{%
\psi }_{\tau _{0}^{\prime }},_{\epsilon }=0.$ Also the fluid description of $%
\bar{\psi}_{2\epsilon }^{i}$ can not depend on which momentum coordinates we
are using.

Some remarks about the form of the action principle developped in App.B with
canonical coordinates is needed. The fluid part of the action principle, $%
\mathcal{A}_{F}$ , can be kept unchanged. In interaction physical
coordinates the resonant particle action principle in the fluid frame takes
a form analogous to the one in App.C 
\begin{equation}
\mathcal{A}_{r}=\int f^{i}(\bar{\psi}_{\tau _{0}^{\prime }}^{i}+(\bar{\psi}%
^{i})^{-1*}\tilde{\psi}_{\tau _{0}^{\prime }}^{i}-H^{i})d^{6}\hat{z}dt=\int 
\tilde{f}^{i}(\tilde{\psi}_{\tau _{0}^{\prime }}^{i}-\bar{\psi}^{i*}\hat{H}%
^{i})d^{6}\hat{z}dt.  \label{eq27g}
\end{equation}
We can develop an extended action principle for interaction physical
coordinates in the way we did in App.C or equivalently in terms of the new
laboratory frame canonical generators.The variation which gives the stress
tensor for the fluid theory still has to be performed with respect to the
fluid-kinetic action term $\mathcal{A}_{I}\equiv -\int f^{i}\hat{H}^{i}d^{6}%
\hat{z}dt$ which we observe correspond to the negative time integrated
internal energy. However, we can vary the resonant particle distribution
with respect to the parameterization in the laboratory frame if we so wish.
A variational equivalent form is therefore if one explicitly want to use the
laboratory frame representation 
\begin{equation}
\mathcal{A}_{r}=\int \check{f}^{i}(\check{\psi}_{\tau _{0}^{\prime }}^{i}-%
\check{H}^{i}-\breve{\psi}_{\tau _{0}^{\prime }}^{i})d^{6}\hat{z}dt\,.
\label{eq27h}
\end{equation}

\section{ Conclusion}

We have described how to parameterize the solutions of the Vlasov-Maxwell
equations by using canonical transformations with respect to a reference
state. The canonical transformations are described by an equation for the
timelike hamiltonian generator relating it to the local hamiltonian and the
reference hamiltonian transformed by the Poisson preserving transformations.
Our main emphasis is on that the solutions of the Vlasov equation has got a
composition principle for the transformations parameterizing the solutions
with respect to a given reference solution of the Vlasov equation. In
particular there is a specific infinite dimensional group of transformations
on phase space extended by time- i.e. a pseudogroup, leaving the reference
distribution invariant. To see that the Vlasov equation defines a
pseudogroup one has to choose transformations on space/time which fixes time
, i.e. a family of canonical transformations on phase space parametrized by
time. The pseudogroup which parameterize the solution of the Vlasov equation
is then given by the pseudogroup of all canonical transformations defined
modulo the transformations which leave the particle density invariant. We
also show how Maxwell's equations can be formulated such that it transforms
in a natural way with respect to diffeomorphisms.

We suggest that the composition principle coming from the underlying
pseudogroup are of fundamental importance. The principle is such that it is
possible to specify new a priori information in the mathematical structure
of the Vlasov equation. Although, the introduction of such a priori
information constrains the experimental situations and the physical
processes for which the theory is applicable, the theory has a gain in the
possibility of modelling kinetic processes in complicated background states.
The model is such that any perturbation theory based on it will preserve the
number of particles in an invariant way since we are basically using
mathematical entities which are not coordinate dependent. Our philosophy is
therefore that in any modelling effort such introduction of new a priori
information is a necessary first step. In fact, since we have replaced the
continuity equation by a constraint equation for Poisson maps on phase space
parameterized by time, the introduction of such a priori information will
change\emph{\ the structure of our equations}. Here we specifically use the
composition principle to make the Vlasov-Maxwell equations into a hybrid
fluid kinetic theory and in addition to formulate the equations for the
divergent and rotational modes in any background. (This application by no
means exhaust the vast number of applications for this fundamental principle
in averaging, separation and perturbation techniques. We are currently
exploring a few of these both in plasma and fluid theory.) The success of
the technique in this case is due to that we are able to identify a
canonical transformation which after integration over the momentum
coordinates coincides with the parameterization of the fluid density. By
this trick we are able to define new particle density coordinates and
corresponding parameterization which contains no deviation on average from
the density and momentum of the reference state. However, the new density
contains resonant particle effects and higher order fluid moments. We
succeed in presenting the theory in new physical coordinates which we call
interaction physical coordinates and the fluid kinetic generators in a form
which is such that it is suitable for near identity transformations. \emph{\
Thus we have obtained a theory where} \emph{kinetic and fluid effects are
naturally separated. }

Such a decomposition of kinetic theory into hybrid fluid kinetic theory is
bound to influence the way one define and think about physical phenomena in
plasmas like resonant particle distribution, dispersion relation, wave-
resonant particle distribution effects, instability e.t.c. Work are under
way to separate higher order fluid moments like stress. The linearized
equations can be presented by Hermitian operators with respect to an
indefinite inner product, but now with a mixed fluid kinetic inner product.
If the system contain some extra symmetries (exact or approximate) either in
the reference state or in an intermediate state it might be convenient to
introduce other coordinates than the interaction physical, euclidean
coordinates we have introduced. Examples of such symmetries are gyrophase
and wave phase symmetry or simply an ignorable coordinate in the plasma
description. This is outside the scope of the present work since one has to
think carefully about noneuclidean, invariant descriptions for kinetic
theory.

\section{Appendix A The pseudogroup connected to continuity equations}

Pseudogroups are infinite dimensional groups.We will not give the detailed
definition of infinite dimensional Lie groups since we would then have to
introduce much more mathematical machinery than we intend to do here, but
the interested reader may find it in the references (5,6). For our purpose
we will only notice that some equations in physics define two type
pseudogroups: one type which fixes the physical field in question and
another one which deforms and parameterize solutions of the the
corresponding physical equation. Mathematically this means that the
equations in question allows an infinite dimensional symmetry taking
solutions into solutions.

We define pseudogroups with smooth structure in the following way:

\textbf{Def. }The totality $\Gamma $ , of smooth maps on a space $M$ form a
pseudogroup $\Gamma \equiv \{\phi :M\rightarrow M\mid \phi \in C^\infty
(M,M)\}$ if

i) $f\circ g\in \Gamma $ when $f,g\in \Gamma $ ,

ii) $id$ , the identity map is in $\Gamma $ ,

iii) $f\circ g^{-1}\in \Gamma $ when $f,g\in \Gamma .$

\vspace{1.0in}

The continuity equation in fluid mechanics 
\[
\frac{\partial \rho }{\partial t}+\nabla \cdot (\mathbf{v}\rho )=0\text{ } 
\]
, can be replaced by :

\textbf{Theor. }The continuity equation can be parameterized by a family
(with respect to timeparameter $t$) of transformations on space $M$ , $%
\mathbf{\psi (}t\mathbf{)}:M\rightarrow M$ , 
\begin{equation}
\rho (t)=\mathbf{\psi (}t)\mathbf{\bullet }\rho _0(t)\equiv J(t)\rho
_0(t)\circ \mathbf{\psi }^{-1}(t).  \label{eqqa1}
\end{equation}

Here $J(t)\equiv \mid \frac{\partial \mathbf{\psi }(t)^{-1}}{\partial 
\mathbf{x}}\mid $ is the Jacobian of the map $\mathbf{\psi }^{-1}(t)$ . The
velocity field defines an equation for the parameterization with respect to
the reference velocity field by

\begin{eqnarray}
\mathbf{v}(t) &\equiv &\mathbf{\psi (}t\mathbf{)\bullet v}_0(t)+\mathbf{k}%
\equiv \mathbf{\psi }_t+\mathbf{\psi }(t)_{*}\mathbf{v}_0(t)+\mathbf{k}%
\hspace{1.0in},  \label{eqqa2} \\
\nabla \cdot \mathbf{(k}\rho (t)) &=&0\;,\;\mathbf{k=\psi }(t)_{*}\mathbf{k}%
_0\;, \\
\mathbf{\psi }_t &\equiv &\frac{\partial \mathbf{\psi (}t\mathbf{)}}{%
\partial t}\circ \mathbf{\psi }(t)^{-1}\;,  \nonumber \\
\mathbf{\psi }(t)_{*}\mathbf{v}_0(t) &\equiv &(\mathbf{v}_0\cdot \nabla 
\mathbf{\psi (}t))\circ \mathbf{\psi }(t)^{-1}.  \nonumber
\end{eqnarray}

\textbf{Proof:}

Here the map $\mathbf{\psi }(t)_{*}$ defined above is the standard
pushforward map. We parameterized the density at a shifted time by 
\[
\rho (t+s)=J_{\mathbf{\psi }_0(t,s)\circ \mathbf{\psi }(t)}(\mathbf{\psi }%
_0(t,s)^{*-1}\circ \mathbf{\psi }(t)^{*-1})\rho _0(t+s)\;, 
\]

where we have decomposed $\mathbf{\psi }(t+s)=\mathbf{\psi }_0(t,s)\circ 
\mathbf{\psi }(t)$ so that $\mathbf{\psi }_t=\frac{\partial \mathbf{\psi }%
_0(t,s)}{\partial s}\mid _{s=0}.$ The following relations follows 
\begin{eqnarray*}
\frac{\partial J_{\mathbf{\psi }_0(t,s)}}{\partial s} &\mid &_{s=0}=-\nabla
\cdot \mathbf{\psi }_t\;, \\
J_{\mathbf{\psi }_0(t,s)\circ \mathbf{\psi (}t)} &=&(J_{\mathbf{\psi (}%
t)}\circ \mathbf{\psi }_0(t,s)^{-1})J_{\mathbf{\psi }_0(t,s)}\;, \\
\frac \partial {\partial s}\rho (t)\circ \mathbf{\psi }_0(t,s)^{-1} &\mid
&_{s=0}=-\mathbf{\psi }_t\cdot \nabla \rho (t)\;, \\
\frac{\partial \rho _0(t+s)}{\partial s} &\mid &_{s=0}=-\nabla \cdot (%
\mathbf{v}_0\rho _0(t))\;.
\end{eqnarray*}

These relations immediately gives us that 
\[
\frac{\partial \rho (t)}{\partial t}=\frac{\partial \rho (t+s)}{\partial s}%
\mid _{s=0}=-\rho (t)\nabla \cdot \mathbf{\psi }_{t}-\mathbf{\psi }_{t}\cdot
\nabla \rho (t)-\mathbf{\psi (}t\mathbf{)\bullet (}\nabla \cdot (\mathbf{v}%
_{0}\rho _{0}(t))).
\]

Since we have the identity 
\[
\mathbf{\psi }(t)^{-1*}(\nabla \cdot \mathbf{u})=\frac 1{^{J_{\mathbf{\psi }%
(t)}}}\nabla \cdot (J_{\mathbf{\psi }(t)}\,\mathbf{\psi (}t)_{*}\mathbf{u}%
)\;, 
\]
one obtains that 
\begin{eqnarray*}
\frac{\partial \rho (t)}{\partial t} &=&-\nabla \cdot (\mathbf{v}\rho
(t))=-\nabla \cdot ((\mathbf{\psi }_t+\mathbf{\psi }(t)_{*}\mathbf{v}_0)\rho
(t))\;, \\
\mathbf{v} &=&\mathbf{\psi }_t+\mathbf{\psi }(t)_{*}\mathbf{v}_0+\mathbf{k\;.%
}
\end{eqnarray*}
Here $\mathbf{k}$ is any vectorfield such that $\nabla \cdot (\mathbf{k}\rho
(t))=0.$ Such vectorfields can also be parameterized with respect to the
reference density according to the above identity for divergences 
\begin{eqnarray*}
\mathbf{k} &=&\mathbf{\psi }(t)_{*}\mathbf{k}_{0\;,} \\
\nabla \cdot (\mathbf{k}_0\rho _0(t)) &=&0\;.
\end{eqnarray*}
In the following we will often not write this additional freedom explicit.

\textbf{End of proof.}

We see that in the case that the reference velocity is zero the above
parametrization is equivalent to the Lagrangian description of fluids , but
transported back to the Eulerian velocity $\mathbf{v}(t)$ by the inverse
mapping. One can verify that the parameterization of the continuity equation
is compatible with composition of families of smooth maps since

\begin{eqnarray}
J_{\mathbf{\phi }(t)\mathbf{\circ \psi (}t)} &=&(J_{\mathbf{\psi }(t)}\circ 
\mathbf{\phi (}t\mathbf{)}^{-1})J_{\mathbf{\phi }(t)}\;,  \label{eqqa3} \\
(\mathbf{\phi }(t)\circ \mathbf{\psi }(t))\bullet \rho _0(t) &=&\mathbf{\psi
(}t)\bullet (\mathbf{\phi }(t)\bullet \rho _0(t))\;,  \nonumber \\
\mathbf{v}(t) &=&(\mathbf{\phi }(t)\circ \mathbf{\psi }(t))\bullet \mathbf{v}%
_0(t)=\mathbf{\phi }(t)\bullet (\mathbf{\psi }(t)\bullet \mathbf{v}_0(t)) 
\nonumber \\
&=&\mathbf{\phi }_t+\mathbf{\phi }(t)_{*}(\mathbf{\psi }_t+\mathbf{\psi (}%
t)_{*}\mathbf{v}_0(t))\;.  \nonumber
\end{eqnarray}

The family of transformations defined above does not conform with the
definition of pseudogroups because of the timedependence. However, if we
look at our family of transformations as transformations on space-time $%
X=M\times \Bbb{R}$ which fixes time $\phi (\mathbf{x,}t)=(\mathbf{\phi (}t)(%
\mathbf{x),}t)$ they can be viewed as member of a pseudogroup. Moreover, the
velocity in this extended space is naturally defined as $v=(\mathbf{v}(\cdot
),1)$ while the family of density maps in three space is transcribed to the
density $\rho _{0}$ in fourspace which has the same value evaluated at
corresponding points and time. Now it is possible to express the above
parameterizations in a more compressed form as

\begin{eqnarray}
\rho &=&\phi \bullet \rho _0\equiv J\,(\rho _0\circ \phi ^{-1})\;,
\label{eqqa4} \\
v &=&\phi _{*}v_0\equiv (v_0\cdot \nabla \phi )\circ \phi ^{-1}\;,  \nonumber
\\
J &=&\mid \frac{\partial \phi ^{-1}}{\partial (\mathbf{x},t)}\mid \,=\,\mid 
\frac{\partial \mathbf{\phi (}t\mathbf{)}^{-1}}{\partial \mathbf{x}}\mid
\,=J(t).  \nonumber
\end{eqnarray}

Here we have used $\nabla $ as the gradient operator both in three and four
space.When we in addition identify the continuity equation in four space $%
\nabla \cdot (v\rho )=0$ as an infinitesimal Lie equation, it is clear how
the density structure define a pseudogroup $\Gamma _{d}$ which leaves the
density in four space invariant$^{5,6}$ 
\begin{equation}
\Gamma _{d}\equiv \{\phi \in C^{\infty }(X,X)\mid \phi \bullet \rho =\rho
\;,\phi (\mathbf{x},t)=(\mathbf{\phi }(t)(\mathbf{x)},t)\}\;.  \label{eqqa5}
\end{equation}

The above pseudogroup is a finite Lie equation and the infinitesimal version
of it corresponds to the continuity equation obtained by taking the
infinitesimal map $\phi =Id+\epsilon v\cdot \nabla ....$ in the finite Lie
equation. The parametrization of the solution space of the continuity
equation as given by eq.( \ref{eqqa4} ) with respect to a reference
distribution is generated by the pseudogroup of all smooth transformations
on space-time which fix time $\Gamma _t\equiv \{\phi \in C^\infty (X,X)\mid
\phi (\mathbf{x},t)=(\mathbf{\phi }(t)(\mathbf{x),}t)\}$ .

\textbf{Def.} We define the density leaf fixed by a density $\rho _0$ as $%
P_0=\{\rho \mid \rho =\phi \bullet \rho _0\,,\forall \phi \in \Gamma _t\}$ .

We notice that it is also possible to define a pseudogroup of smooth
transformations which leaves $\rho _0$ invariant $\Gamma _{0d}$ which in
fact can be transported to every element in the density leaf by $\Gamma
_{\rho d}\equiv \{\tilde{\psi}\equiv \phi \circ \psi \circ \phi ^{-1}\mid
\psi \in \Gamma _{0d}\;,\phi \in \Gamma _t\}$. In this sense it is possible
to look at $\Gamma _{0d}$ as a type of generalized gaugetransformations with
respect to a given leaf and moreover it will be sufficient to generate the
leaf by $\bar{\Gamma}\equiv \Gamma _t$ modulo$\Gamma _{0d}$ with the above
transportation.(In fact it corresponds to that one has to introduce a
semidirect product as group product in $\bar{\Gamma}.$) We will not pursue
the generalization of gauge transformations on the density leaf any further
in this paper as it also will need more mathematical background than we
intend to show here.

\subsection{Geometric interpretation of the fluid generator}

In our derivation of the parameterization the new vectorfield $\mathbf{\psi }%
_t$ appeared as the derivative with respect to the near identity map $%
\mathbf{\psi }_0(t,s)$ . This map have the property that $\mathbf{\psi }%
_0(t,0)=Id$ . Now, we shall think of our transformations with respect to a
specific density leaf $P_0$ given by a reference density $\rho _0.$ Then it
is realized that $\rho (t+s)=\mathbf{\psi }_0(t,s)\bullet \rho (t)$.
Therefore for $s=0$,\ $\mathbf{\psi }_0$ is the identity map at the point $%
\rho \in P_0$ . Consequently, one can also think about the vectorfield $%
\mathbf{\psi }_t$ as a point in the space of vectorfields $\mathcal{X}_\rho $
at the point $\rho .$ At the reference density we have the reference space
of vectorfields $\mathcal{X}_{\rho _0}$ . It turns out that all vectorfields
can be pulled back to the space of reference vectorfields. This is done by
defining the related near identity map $\mathbf{\hat{\psi}}_0$ by 
\[
\mathbf{\psi }(t+s)=\mathbf{\psi }(t)\circ \mathbf{\hat{\psi}}_0(t,s)\;. 
\]
Again we see that $\hat{\psi}_0(t,0)=Id$ . This time we have that 
\[
\mathbf{\psi }(t)^{-1}\bullet \rho (t+s)\equiv \hat{\rho}_0(t,s)=\mathbf{%
\hat{\psi}}_0(t,s)\bullet \rho _0(t)\text{ .} 
\]
Therefore one realizes that the corresponding vectorfield $\mathbf{\hat{\psi}%
}_t\equiv \frac{\partial \mathbf{\hat{\psi}}_0(t,s)}{\partial s}\mid
_{s=0}\in \mathcal{X}_{\rho _0}.$ Moreover, it is verified that 
\begin{eqnarray}
\mathbf{\psi }_t &=&\mathbf{\psi (}t)_{*}\mathbf{\hat{\psi}}_t=(\mathbf{\hat{%
\psi}}_t\cdot \nabla \mathbf{\psi }(t))\circ \mathbf{\psi }(t)^{-1}\;,
\label{eqqa5b} \\
\mathbf{v} &=&\mathbf{\psi }(t)_{*}(\mathbf{\hat{\psi}}_t+\mathbf{v}_0)\;. 
\nonumber
\end{eqnarray}
Notice that a sum of vectorfields at the reference space of vectorfields
ordered according to the connected composition, corresponds to a total
vectorfield at $\rho $ by

\begin{eqnarray}
\mathbf{\hat{\psi}}_t &=&\sum\limits_{i=1}^n\mathbf{\hat{\psi}}_{t,i}\;,
\label{eqqa5c} \\
\mathbf{\psi (}t) &=&\prod\limits_{i=1}^n\mathbf{\psi }_i(t)\circ \;, 
\nonumber \\
\mathbf{\psi }_t &=&\mathbf{\psi }_{t,1}+\mathbf{\psi }_1(t)_{*}\mathbf{\psi 
}_{t,2}+...+(\prod\limits_{i=1}^{n-1}\mathbf{\psi }_i(t)_{*}\circ )\mathbf{%
\psi }_{t,n}\;  \nonumber \\
\mathbf{\psi }_{t,i} &=&(\prod\limits_{j=1}^{n-1}\mathbf{\psi }%
_j(t)_{*}\circ )\mathbf{\hat{\psi}}_{t,i}\,.
\end{eqnarray}
This is in fact our fundamental relation which tells us how to extend the
above parameterization of the solution of the continuity equation to a
composition of several and in principle also infinitely many transformations.

\subsection{Variational relations and perturbations}

We need to define perturbations and variations of our quantities with
respect to generators in the space of vectorfields $\mathcal{X}_\rho $ which
as we have demonstrated can be pulled back to $\mathcal{X}_{\rho _0}.$ We do
that by first defining the perturbed family of maps with respect to one
parameter $\epsilon ,$ $\mathbf{\psi (}t,\epsilon )$ such that $\mathbf{\psi 
}(t,0)=\mathbf{\psi }(t).$ From this map we define the near identity maps $%
\mathbf{\psi }_0(t,\epsilon ,\delta )\equiv \mathbf{\psi }(t,\epsilon
+\delta )\circ \mathbf{\psi }(t,\epsilon )^{-1}$ and $\mathbf{\hat{\psi}}%
_0(t,\epsilon ,\delta )\equiv \mathbf{\psi }(t,\epsilon )^{-1}\circ \mathbf{%
\psi }(t,\epsilon +\delta ).$ One realizes as before that on the density
leaf $P_0$ these maps are deformations from densities $\rho (t,\epsilon )$
and $\rho _0(t)$ respectively. This follows by defining the perturbed
density $\rho (t,\epsilon +\delta )\equiv \mathbf{\psi }(t,\epsilon +\delta
)\bullet \rho _0(t)=\mathbf{\psi }_0(t,\epsilon ,\delta )\bullet \rho
(t,\epsilon )$ and $\hat{\rho}_0(t,\epsilon ,\delta )\equiv \mathbf{\psi (}%
t,\epsilon )^{-1}\bullet \rho (t,\epsilon +\delta )=\mathbf{\hat{\psi}}%
_0(t,\epsilon ,\delta )\bullet \rho _0(t).$ We can now define the generating
vectorfields for the defined $\epsilon $ perturbation as 
\begin{eqnarray}
\mathbf{\psi }_\epsilon &\equiv &\frac{\partial \mathbf{\psi }_0(t,\epsilon
,\delta )}{\partial \delta }\mid _{\delta =0}\,\in \mathcal{X}_{\rho \;}\;,
\label{eqab6} \\
\mathbf{\hat{\psi}}_\epsilon &\equiv &\frac{\partial \mathbf{\hat{\psi}}%
_0(t,\epsilon ,\delta )}{\partial \delta }\mid _{\delta =0}\,\in \mathcal{X}%
_{\rho _0}\;.  \nonumber
\end{eqnarray}

Instead of one parameterfamilies of deformations, one could define many
parameter families of deformations or simply deformed maps $\mathbf{\tilde{%
\psi}}(t)$ of $\mathbf{\psi }(t)$ without any reference to any parameters.
We define the corresponding near identity maps $\mathbf{\psi }_0(t)\equiv 
\mathbf{\tilde{\psi}}(t)\circ \mathbf{\psi }(t)^{-1}$ and $\mathbf{\hat{\psi}%
}_0(t)\equiv \mathbf{\psi (}t)^{-1}\circ \mathbf{\tilde{\psi}}(t)$ and the
deformed densities $\tilde{\rho}(t)\equiv \mathbf{\psi }_0(t)\bullet \rho
(t),\,\hat{\rho}_0(t)\equiv \mathbf{\hat{\psi}}_0(t)\bullet \rho _0(t)$. The
corresponding generating variations of these maps are then defined as the
infinitesimal vectorfields 
\begin{eqnarray}
\delta \mathbf{\psi } &\equiv &\mathbf{(}\delta \mathbf{\tilde{\psi}}%
(t))\circ \mathbf{\psi }(t)^{-1}\in \mathcal{X}_\rho \;,  \label{eqab7} \\
\delta \mathbf{\hat{\psi}} &\equiv &\mathbf{\psi (}t)^{-1}\circ \delta 
\mathbf{\tilde{\psi}}(t)\in \mathcal{X}_{\rho _0}\;.  \nonumber
\end{eqnarray}

We understand from the above that to formulate variational principles on a 
\emph{density leaf has certain nonconventional aspects due to that the
fields involved are generated by the action of maps on densities}. We give
the following identities needed to do variations on density and the
pushforward velocity. These can be verified by doing infinitesimal
variations of the corresponding parameterized fields.

\begin{eqnarray}
\delta \rho (t) &=&-\nabla \cdot (\delta \mathbf{\psi }\rho (t))\;,
\label{eqab8} \\
\delta (\mathbf{\psi (}t)_{*}\mathbf{v}_{0}) &=&[\delta \mathbf{\psi },%
\mathbf{\psi }(t)_{*}\mathbf{v}_{0}]\;,  \nonumber \\
\frac{\partial \rho (t,\epsilon )}{\partial \epsilon } &=&-\nabla \cdot (%
\mathbf{\psi }_{\epsilon }\rho (t,\epsilon ))\;,  \nonumber \\
\frac{\partial (\mathbf{\psi (}t,\epsilon )_{*}\mathbf{v}_{0})}{\partial
\epsilon } &=&[\mathbf{\psi }_{\epsilon },\mathbf{\psi (}t,\epsilon )_{*}%
\mathbf{v}_{0}]\;.  \nonumber
\end{eqnarray}
Here the symbol $[\cdot ,\cdot ]$ is the usual vectorfield bracket which is
defined as $[\mathbf{X,Y]}\equiv \mathbf{X\cdot }\nabla \mathbf{Y-Y\cdot }%
\nabla \mathbf{X}$ \ . The deformation generating vectorfields and the
generating vectorfields in the timelike direction satisfy a certain
compatibility condition in the space $\mathcal{X}_{\rho }$ .

\textbf{Theor. }When $\delta \mathbf{\psi ,\psi }_{\epsilon }$ , $\mathbf{%
\psi }_{t},\mathbf{k}\in \mathcal{X}_{\rho },$ we have the compatibility
conditions 
\begin{eqnarray}
&&\delta \mathbf{\psi },_{t}-\delta \mathbf{\psi }_{t}-[\delta \mathbf{\psi
,\psi }_{t}]+\mathbf{k}=\mathbf{0}\;,  \label{eqab9} \\
&&\mathbf{\psi }_{\epsilon },_{t}-\mathbf{\psi }_{t},_{\epsilon }-[\mathbf{%
\psi }_{\epsilon }\mathbf{,\psi }_{t}]+\mathbf{k}=\mathbf{0}\;  \nonumber \\
\nabla \cdot (\mathbf{k}\rho ) &=&0.
\end{eqnarray}

For $\delta \mathbf{\hat{\psi},\;\hat{\psi}}_\epsilon ,\;\mathbf{\hat{\psi}}%
_t\in \mathcal{X}_{\rho _0}$ , we have the compatibility conditions

\begin{eqnarray}
&&\delta \mathbf{\hat{\psi}},_{t}-\delta \mathbf{\hat{\psi}}_{t}+[\delta 
\mathbf{\hat{\psi},\hat{\psi}}_{t}]+\mathbf{k}_{0}=\mathbf{0}\;,
\label{eqab10} \\
&&\mathbf{\hat{\psi}}_{\epsilon },_{t}-\mathbf{\hat{\psi}}_{t},_{\epsilon }+[%
\mathbf{\hat{\psi}}_{\epsilon }\mathbf{,\hat{\psi}}_{t}]+\mathbf{k}_{0}=%
\mathbf{0\,,}  \nonumber \\
\nabla \cdot (\mathbf{k}_{0}\rho ) &=&0,\;\,\mathbf{k}=\mathbf{\psi (}t)_{*}%
\mathbf{k}_{0}.
\end{eqnarray}

\textbf{Proof:} The proof of the above result follows simply by writing out
the compatibility conditions for the two equal variations $\delta \frac{%
\partial }{\partial t}\tilde{\rho}(t)\mid _{\tilde{\rho}=\rho }=\frac{%
\partial }{\partial t}\delta \tilde{\rho}(t)\mid _{\tilde{\rho}=\rho }$ . If
we use the above formulas, it is obtained that 
\[
\nabla \cdot ((\delta \mathbf{\psi },_{t}-\delta \mathbf{\psi }_{t}-[\delta 
\mathbf{\psi ,\psi }_{t}])\rho (t))=0\;, 
\]
from which the first identity follows.The second compatibility condition
follows similarly from that $\frac{\partial ^{2}}{\partial t\partial
\epsilon }\rho (t,\epsilon )=\frac{\partial ^{2}}{\partial \epsilon \partial
t}\rho (t,\epsilon )\;.$ The last compatibility conditions in $\mathcal{X}%
_{\rho _{0}}$ either follows from pulling back the above compatibility
conditions in $\mathcal{X}_{\rho }$ to $\mathcal{X}_{\rho _{0}}$ or by
studying the compatility condtions for the deformed density at $\rho _{0}$,
i.e.$\delta \frac{\partial }{\partial t}\hat{\rho}_{0}(t)\mid _{\hat{\rho}%
_{0}=\rho _{0}}=\frac{\partial }{\partial t}\delta \hat{\rho}_{0}(t)\mid _{%
\hat{\rho}_{0}=\rho _{0}}$. In either case one finds that 
\[
\nabla \cdot ((\delta \mathbf{\hat{\psi}},_{t}-\delta \mathbf{\hat{\psi}}%
_{t}+[\delta \mathbf{\hat{\psi},\hat{\psi}}_{t}])\rho _{0}(t))=0. 
\]
The derivation for the $\epsilon -$parameterized case is similar.

\subsection{Parameterization of the Vlasov equation}

A special case of continuity equations are the Liouville equation and the
Vlasov equation on the phase space of space and momentum. In this case we
will have to deal with Hamiltonian vectorfields $X_{H}$ and reference
Hamiltonian vectorfields $X_{H_{0}}$ . The conservation law for the particle
density $f$ on phase space in this case is given by the Vlasov equation
(must be supplied by the definition of the Hamiltonian in question and the
Maxwell's equations$^{8}$)

\begin{eqnarray}
\frac{\partial f}{\partial t}+X_H\cdot \nabla f &=&0\;,\text{ or}
\label{eqqa6} \\
\frac{\partial f}{\partial t}+\{f,H\} &=&0\;.  \nonumber
\end{eqnarray}
where $\{,\}$ is the Poisson bracket.

The volumeform in 6 dimensional phase space $P$ with coordinates $Z$ is
given by the expression $dV=J_{V}d^{6}Z$ where $J_{V}\equiv \mid \frac{%
\partial \phi _{V}^{-1}}{\partial Z}\mid $ is the Jacobian of the map $\phi
_{V}^{-1}:Z\rightarrow z=\phi _{V}^{-1}(Z)$ to a standard system where $%
dV=d^{6}z$ . For our purposes we will only use standard Euclidean space with
physical or canonical momentum coordinates which both will have
volumeelement in the standard form. For physical and canonical coordinates
with respect to Euclidean space we have since the vectorfields $X_{H}$
preserve phase space volume that $J_{V}=1$ and $\nabla \cdot (X_{H})=0$ . In
more general coordinate systems, which is needed in gyrokinetic and
oscillation center kinetic theory the conservation of phase space volume can
be expressed by the conservation law 
\begin{equation}
\frac{\partial J_{V}}{\partial t}+\nabla \cdot (X_{H}J_{V})=0\;.
\label{eqqa7}
\end{equation}

If we combine these two equations we obtain the continuity equation in phase
space for the quantity $\rho =J_Vf\;$ 
\begin{equation}
\frac{\partial (J_Vf)}{\partial t}+\nabla \cdot (X_HJ_Vf)=0\;.  \label{eqqa8}
\end{equation}

In an analogous way as above we can now parameterize

\begin{eqnarray}
\rho (t) &=&\psi (t)\bullet \rho _0(t)\;,\;\psi (t)\in C^\infty (P,P)\;,
\label{eqqa9} \\
\rho _0(t) &=&J_{V0}(t)f_0(t)\;, \\
J_V(t) &=&\psi (t)\bullet J_{V0}(t)\;,  \nonumber \\
f(t) &=&f_0(t)\circ \psi (t)^{-1}\equiv (\psi (t)^{-1})^{*}f_0(t)\;, 
\nonumber \\
X_H(t) &=&\frac{\partial \psi (t)}{\partial t}\circ \psi (t)^{-1}+\psi
(t)_{*}X_{H_0}(t)\;.  \nonumber
\end{eqnarray}

The Vlasov equation has additional structure for situations when the flow is
described by Hamiltonian vectorfields. The maps which is generated by
Hamiltonian flows are Poisson preserving maps and we must therefore take
this into account.

\begin{enumerate}
\item  Define the space of Poisson preserving maps as $\mathcal{F}=\{\psi
\in C^{\infty }(P,P)\mid \psi ^{*}\{f,g\}=\{\psi ^{*}f,\psi ^{*}g\}$ for any 
$f,g\in C^{\infty }(P,P)\}$ .
\end{enumerate}

We then have the following theorem:

\textbf{Theorem }For $\psi (t)\in \mathcal{F}$ we have

\begin{eqnarray}
\psi (t)_{*}X_{H_0} &=&X_{\psi (t)^{*-1}H_0}\;,  \label{eqqa10} \\
\frac{\partial \psi (t)}{\partial t}\circ \psi (t)^{-1} &=&\psi (t)_{*}X_{%
\hat{\psi}_t}=X_{\psi (t)^{*-1}\hat{\psi}_t}=X_{\psi _t}\;,  \nonumber \\
\psi _t &\equiv &\psi ^{*-1}\hat{\psi}_t\;,  \nonumber \\
X_{\hat{\psi}_t} &\equiv &\frac{\partial \hat{\psi}_0(t,s)}{\partial s}\mid
_{s=0}\;,  \nonumber \\
X_{\psi _t} &\equiv &\frac{\partial \psi _0(t,s)}{\partial s}\mid _{s=0\;,} 
\nonumber \\
\psi (t+s) &=&\psi (t)\circ \hat{\psi}_0(t,s)=\psi _0(t,s)\circ \psi (t), 
\nonumber \\
H &=&\psi _t+\psi (t)^{-1*}H_0\;.
\end{eqnarray}

\textbf{Remark:}

Notice that both the map $\psi _0$ and $\hat{\psi}_0$ are identity maps for $%
s=0.$ Therefore we can regard the Hamiltonian vectorfield $X_{\hat{\psi}}$
as an element of the Lie algebra at the reference density structure
corresponding to the pseudogroup $\Gamma _t$ restricted to Poisson
preserving maps. It is not our purpose here to study this Lie algebra and
its correspondence to our generalized gaugegroup (i.e.the Lie pseudogroup
keping the density fixed) since it is best formulated with some more exact
mathematical machinery available than we have presently assumed.

\textbf{Proof:}

Let the vectorfield $\psi (t)_{*}X_{H0}$ act on a function on phase space $%
f\in C^\infty (P,\Bbb{R)}$ . One can convince oneself that in this case one
has the alternative expression (see \cite{geom})for this vectorfield when
one think of it as an operator acting as directional derivative, i.e. $%
X(f)\equiv (X\cdot \nabla )f$ as is commonly done 
\[
(\psi (t)_{*}X_{H_0})f\equiv \psi (t)^{*-1}\circ X_{H_0}\circ \psi
(t)^{*}\circ f=\psi (t)^{*-1}\{\psi (t)^{*}f,H_0\}.\; 
\]

Since the map is Poisson preserving we immediately get the result 
\[
(\psi (t)_{*}X_{H_0})f=\{f,\psi (t)^{*-1}H_0\}=(X_{\psi (t)^{*-1}H_0})f\;. 
\]

Consequently we have proven the first of the above results up to a possible
Casimir generated vectorfield $X_C$ such that $\{C,g\}=0,$ $\forall g\in
C^\infty (P,\Bbb{R)}.$ For symplectic maps and canonical coordinates there
are no Casimir for the Poisson bracket while in general noncanonical
coordinates there will be Casimirs. However, in our case this present no
problem since we are only interested in functions restricted to the Poisson
leaf generated by a reference $f^0.$ On such a leaf the Casimir is fixed and
there will be no loss in generality to assume the above identity up to any
function commuting with the Poisson leaf density $f$ related to a reference
density $f^0$.

One can easily prove that $\psi _0(t,s)$ and $\hat{\psi}_0(t,s)$ are Poisson
maps since $\psi (t)$ and $\psi (t+s)$ are Poisson maps. The phase space
functions $\hat{\psi}_t$ and $\psi _t$ are the Hamiltonians corresponding to
the Hamilonian vectorfields defined by 
\begin{eqnarray*}
\frac{\partial \hat{\psi}_0(t,s)}{\partial s} &\mid &_{s=0}=\frac{\partial
\psi (t)^{-1}\circ \psi (t+s)}{\partial s}\mid _{s=0}\equiv X_{\hat{\psi}%
_t}\;, \\
\frac{\partial \psi _0(t,s)}{\partial s} &\mid &_{s=0}=\frac{\partial \psi
(t+s)\circ \psi (t)^{-1}}{\partial s}\mid _{s=0}\equiv X_{\psi _t}\;.
\end{eqnarray*}

From these definitions we derive that

\[
\frac{\partial \psi (t)}{\partial t}=\frac{\partial \psi (t+s)}{\partial s}%
\mid _{s=0}=(X_{\hat{\psi}_{t}}\cdot \nabla )\psi (t). 
\]

and therefore one deduce that $X_{\psi _{t}}=\psi (t)_{*}X_{\hat{\psi}_{t}}$
. Together with the first equality we then obtain that $X_{\psi
_{t}}=X_{\psi (t)^{*-1}(\hat{\psi}_{t})}$ . It then follows that we can put $%
\psi _{t}=\psi (t)^{*-1}\hat{\psi}_{t}$ up to any function poisson commuting
with the leaf density $f$ corresponding to a reference density $f_{0}.$ 
\textbf{End of proof.}

In canonical coordinates $z=(\mathbf{x},\mathbf{p})$ , where $\mathbf{p}=%
\mathbf{p}_{p}+\frac{e}{c}\mathbf{A}$ and $\mathbf{p}_{p},\mathbf{A}$ are
the physical momentum and vectorpotential, the Hamiltonian vectorfield is
given by $X_{H}=\mathbf{J}_{c}\cdot d_{P}H=(\frac{\partial H}{\partial 
\mathbf{p}},-\frac{\partial H}{\partial \mathbf{x}})$ in canonical,
euclidean coordinates. The transformed Hamiltonian is given in the same form
since the symplectic and the Poisson tensor does not change by canonical
transformations, $X_{\psi (t)^{*-1}(H_{0})}=\mathbf{J}_{c}\cdot d_{P}(\psi
(t)^{-1*}H_{0})=(\frac{\partial \psi (t)^{*-1}H_{0}}{\partial \mathbf{p}}$ $%
,-\frac{\partial \psi (t)^{*-1}H_{0}}{\partial \mathbf{x}})$ . In physical
coordinates based on Euclidean space the symplectic tensor depends on the
magnetic field which change also has to be specified, i.e. $\mathbf{B}%
_{0}\rightarrow \mathbf{B}$ .Therefore we have that

\begin{eqnarray*}
X_{H_{0}} &=&\mathbf{J}_{0}\cdot d_{P}H_{0}=(\frac{\partial H_{0}}{\partial 
\mathbf{p}},-\frac{\partial H_{0}}{\partial \mathbf{x}}-\frac{\mathbf{B}_{0}%
}{c}\times \frac{\partial H_{0}}{\partial \mathbf{p}})\;, \\
X_{\psi (t)^{-1*}H_{0}} &=&\mathbf{J\cdot d}_{P}(\psi (t)^{-1*}H_{0}) \\
&=&(\frac{\partial \psi (t)^{-1*}H_{0}}{\partial \mathbf{p}},-\frac{\partial
\psi (t)^{-1*}H_{0}}{\partial \mathbf{x}}-\frac{\mathbf{B}}{c}\times \frac{%
\partial \psi (t)^{-1*}H_{0}}{\partial \mathbf{p}})\;.
\end{eqnarray*}

We have not given explicitly how the magnetic field changes under the action
of the pseudogroup of smooth transformations on space time here. The answer
to this question follows from the same infinite dimensional symmetry for
electromagnetic fields which are responsible for the usual gauge
parameterizations. We will explore this more general parameterization of the
electromagnetic fields in a forthcoming paper.

The connection between the canonical distribution function and the physical
distribution function in physical coordinates ($\mathbf{x,\hat{p}}=m\mathbf{%
v)}$ is given by a canonical tranformation in the radiation gauge as 
\begin{eqnarray}
\hat{f} &=&\phi _{c}^{-1*}f\;,  \label{eqqa11} \\
\phi _{c}^{-1*} &=&\exp (X_{c})\;,  \nonumber \\
X_{c} &=&\frac{e}{c}\mathbf{A\cdot }\frac{\partial }{\partial \mathbf{p}}=-%
\mathbf{J\cdot }\Bbb{(}\frac{e}{c}\mathbf{A}^{(1)})  \nonumber
\end{eqnarray}
Here $\mathbf{A}^{(1)}$ is the vectorfield $\mathbf{A}$ lifted to a oneform.
Notice that the vectorfield $X_{c}$ is phase space volume and Poisson
preserving, but it is not generated by a hamiltonian.The explicit form of
this vectorfield in other cordinates will follow from the tranformation
properties of the Poisson tensor, $\mathbf{J}$. The above transformation is
nothing else than the shift transformation from physical $\mathbf{p}_{p}$ to
canonical $\mathbf{p.}$\textbf{\ }

\begin{itemize}
\item  \textbf{Lemma} The canonical transformation $\phi _{c}$ transforms
the bracket between two functions $f,g\;$on canonical phase space to the
physcical bracket between the corresponding functions $\hat{f},\hat{g}$ on
physical phase space. 
\[
\phi _{c}^{-1*}\{f,g\}=\{\hat{f},\hat{g}\}_{n}\;=\mathbf{J:(}d_{P}\hat{f}%
,d_{P}\hat{g}). 
\]
\end{itemize}

Here the physical bracket is given in it's standard euclidean form

\begin{equation}
\{\hat{f},\hat{g}\}_{n}\equiv \frac{\partial \hat{f}}{\partial \mathbf{x}_{p}%
}\cdot \frac{\partial \hat{g}}{\partial \mathbf{p}_{p}}-\frac{\partial \hat{f%
}}{\partial \mathbf{p}_{p}}\cdot \frac{\partial \hat{g}}{\partial \mathbf{x}%
_{p}}+\frac{e\mathbf{B}}{c}\cdot (\frac{\partial \hat{f}}{\partial \mathbf{p}%
_{p}}\times \frac{\partial \hat{g}}{\partial \mathbf{p}_{p}})\;.
\label{eqqa12}
\end{equation}

\proof%
\textbf{\ }The Poisson tensor in canonical coordinates can be expressed by
the multivector $\mathbf{J}_{c}=\frac{\partial }{\partial \mathbf{x}}\wedge 
\frac{\partial }{\partial \mathbf{p}}\,.$ The action of the pullback map
gives $\phi _{c}^{-1*}\{f,g\}=\phi _{c}^{-1*}(\mathbf{J}%
_{c}:(d_{P}f,d_{P}g))=(\phi _{c}^{-1*}\circ \mathbf{J}_{c}\circ \phi
_{c}^{*}):(\phi _{c}^{-1*}d_{P}f,\phi _{c}^{-1*}d_{P}g).$ The exterior
derivative operator commutes with the pullback operator $\phi
_{c}^{-1*}d_{P}f=d_{P}\hat{f}$ and for contravariant tensors $\phi _{*}%
\mathbf{T=}\phi ^{*-1}\circ \mathbf{T\circ }\phi ^{*}$ . Therefore one have
that $\phi _{c}^{-1*}\{f,g\}=\mathbf{J:(}d_{P}\hat{f},d_{P}\hat{g})\,,\,%
\mathbf{J}\,=\,\phi _{c*}\mathbf{J}_{c}\,=\frac{\partial }{\partial \mathbf{x%
}_{p}}\wedge \frac{\partial }{\partial \mathbf{p}_{p}}+\frac{e}{c}(\frac{%
\partial A_{i}}{\partial \mathbf{x}_{p}^{j}}-\frac{\partial A_{j}}{\partial 
\mathbf{x}_{p}^{i}})\frac{\partial }{\partial \mathbf{p}_{p}^{i}}\wedge 
\frac{\partial }{\partial \mathbf{p}_{P}^{j}}.$ This expression is identical
to the standard particle Poissontensor in euclidean physical phase space
variables given above.

$%
\endproof%
.$

We immidiately notice two major problems with this bracket. It is not
compatible with the reference distribution, $f^{0}$ since the Vlasov
equation for that has to be expressed with respect to background
electromagnetic fields. Secondly, it is not compatible with perturbation
theory either since then one would have to do a perturbation expansion of
the bracket itself. This completely destroys the ideas we advocated for
above using canonical fixed brackets as a tool for invariant expansions. To
resolve this in our opinion fundamental problem in plasma physics, we
suggest to define a new physical distribution function given by the
background fields $f^{i}\equiv \phi _{c0}^{-1*}f$ . The bracket for these
kind of distribution functions are now transformed to the same form as in
eq.(\ref{eqqa12}), but with $\mathbf{B}\rightarrow \mathbf{B}_{0}.$ This
distribution function is still a gaugeinvariant distribution function since
we will use the gauge, $\phi _{1}=0$ where $\mathbf{A}_{1}=-c\int\limits^{t}%
\mathbf{E}_{1}(t^{\prime })dt^{\prime }$ has a physical meaning in terms of
the timeintegrated perturbed electric field,$\;\mathbf{E}_{1}\equiv \mathbf{%
E-E}_{0}$ . The relation between the physical distribution function and the
interaction distribution function is given by

\begin{eqnarray}
\hat{f} &=&\phi _{c1}^{-1*}f^{i}\,,  \label{eqa12b} \\
\phi _{c1}^{-1*}\cdot &=&\exp (X_{c_{1}})\,,  \nonumber \\
\mathbf{\ }X_{c1} &=&\frac{e}{c}\mathbf{A}_{1}\cdot \frac{\partial }{%
\partial \mathbf{p}}=-\mathbf{J}_{0}\cdot \Bbb{(}\frac{e}{c}\mathbf{A}%
_{1}^{(1)}))
\end{eqnarray}
$\mathbf{J}_{0}$ is the euclidean physical coordinates Poisson tensor with $%
\mathbf{B\rightarrow B}_{0}.\,$ It seems fitting to call this description of
the Vlasov fields the \textit{interaction picture since it is now possible
to separate background and fluctuating quantities in an invariant way
suitable for perturbation theory}.

\section{Appendix B Hybrid fluid-kinetic action principle}

We will in this appendix study the action principle for the hybrid
fluid-kinetic theory. In two other works$^{4,8}$ we have elaborated on the
action principles for the Vlasov equation and the ideal fluid equations
respectively. Our approach is based on varying the generators of the
underlying infinite dimensional group acting on the respective densities.
The basic method is quite different from the approach of Larsson$^{1,2}$
which is using canonical conjugate variables on the accessible leaf.
However, our method can be revised to introduce canonical conjugate
variables with certain differences since our action also explicitly takes
into account the group composition law and the compatibility conditions. The
action principle for the Vlasov equation is (the Maxwell equation has it's
own action principle which in fact also can be parameterized by an infinite
dimensional group) 
\begin{equation}
\mathcal{A}_{p}=\int f(\psi _{t}-H)d^{6}zdt.  \label{eqb1}
\end{equation}
The compatibility condition for the parameterized phase space density can
then be used to formulate a revised action principle for densities which
depend on an additional formal perturbation parameter $\epsilon $, i.e. $%
f(t,\epsilon )$%
\begin{equation}
\mathcal{A}_{p}^{(1)}=\int_{0}^{1}\int f(\psi _{\epsilon ,t}+\{\psi
_{\epsilon },H\}-H_{,\epsilon })d^{6}zdtd\epsilon  \label{eqb2}
\end{equation}
In eq.\ref{eqb1} we treat $f$ as parameterized by symplectic transformations
with respect to a reference state $f^{0}.$ The variation is nonstandard in
the sence that the action is varied and sought stationary with respect to
the the infinitesimal Hamiltonian generator $\delta \psi $ such that $\delta
f=\{\delta \psi ,f\}.$ (However, this action imply standard variational
principles by the introduction of the revised variational principle through
one parameter variations.) Moreover, the variation of the generator $\psi
_{t}$ is determined through the compatibility relation for the variation $%
\delta \frac{\partial f}{\partial t}=\frac{\partial }{\partial t}\delta f$
which leads to the compatibility condition for variations

\begin{eqnarray}
\delta \psi ,_{t}-\delta \psi _{t}+\{\delta \psi ,\psi _{t}\} &=&0\func{mod}%
k\;,  \label{eqb3} \\
k &=&\psi ^{-1*}k^{0},\;\{k,f\}=0\;.  \nonumber
\end{eqnarray}
The function $k$ has no influence on the variations. Note that if the
variations are restricted to a one parameter group, the above compatibility
condition is equivalent to the one we have derived before since then $\delta
\psi =\psi _{\epsilon }\delta \epsilon $ . This means that for both action
principles one obtain the Vlasov equation by the variations 
\[
\frac{\delta \mathcal{A}_{p}}{\delta \psi }=\frac{\delta \mathcal{A}%
_{p}^{(1)}}{\delta \psi _{\epsilon }}=-f_{,_{t}}-\{f,H\}=0\;. 
\]
In the revised action principle it is possible to introduce a variation with
respect to $\delta f$ keeping $\psi _{\epsilon }$ fixed since variation with
respect to the one parameter generator $\psi _{\epsilon }$ is only a
subvariation. The variation with respect to $f$ then gives the compatibility
condition as one of the variational equations. 
\[
\frac{\delta \mathcal{A}_{p}^{(1)}}{\delta f}=\psi _{\epsilon
,t}-H,_{\epsilon }+\{\psi _{\epsilon },H\}=0\;. 
\]
In fact, there is no reason why one could not introduce many(even infinite)
parameter groups if this is suitable for the problem at hand.\ If we want,
we could also give up the explicit parameterization of $f$ through
symplectic transformations in the revised action principle and formulate a
canonical field theory for canonical conjugate variables $\psi _{\epsilon
},\;f$ as Larsson$^{1,2}$ do. In this case one would have to introduce $%
f,_{\epsilon }=\{\psi _{\epsilon },f\}$ as an additional constraint. In our
paper$^{4}$ we do this by introducing a Lagrange multiplier, but also by
embedding the problem in a larger double symplectic space which includes
also the $\epsilon $ -dynamics. For some purposes this might be a somewhat
restrictive point of view, e.g. if one want to derive model equations based
on several layers of transformations as we want to do.

We are now in a position to formulate a new hybrid fluid kinetic action
principle where we restrict one part of the symplectomorphism, $\overline{%
\psi }$, to correspond to what we have found in section 3 to be equivalent
to volume density preserving transformations in space (both parameterized by
time even if we do not explicitly indicate it). The second part of the
composition corresponds to an incoherent kinetic transformation $\widetilde{%
\psi }$ due to higher order Hamiltonian generators than linear in the
momentum coordinate. The total hybrid fluid-kinetic action restricted to
such a composition takes the form

\begin{eqnarray*}
\mathcal{A}_H &=&\mathcal{A}_r+\mathcal{A}_F\;, \\
\mathcal{A}_r &=&\int f(\overline{\psi }_t+\overline{\psi }^{-1*}\widetilde{%
\psi }_t-H)d^6zdt=\int f(\overline{\psi }^{-1*}\widetilde{\psi }_t-%
\widetilde{H})d^6zdt\;.
\end{eqnarray*}

Here the Hamiltonian $\tilde{H}$ and the related $\hat{H}$ is defined in
eq.( \ref{16b}). Up to variational equivalence (which after all is what is
important in a variational principle), we can freely move the action of a
symplectic transformation between a phase space density $f=\phi ^{-1*}%
\widehat{f}$ with suitable decaying properties in infinity and a multiplying
phase space function $\int fgd^{6}zdt\Longleftrightarrow \int \widehat{f}%
(\phi ^{*}g)d^{6}zdt.$ Therefore equivalently the incoherent part of the
action principle restricted to fluid orbits can be written 
\begin{equation}
\mathcal{A}_{r}=\int \widetilde{f}(\widetilde{\psi }_{t}-\overline{\psi }^{*}%
\widehat{H})d^{6}zdt  \label{eqb4}
\end{equation}
Such changes between variational equivalent forms of variational principles
will later on be freely done without further mentioning. The fluid part of
the action principle has the form$^{8}$(here the density parameterization is
given in the action principle) 
\begin{eqnarray}
\mathcal{A}_{F} &=&\int \rho (\mathbf{w\cdot \widehat{u}-}\frac{\mathbf{u}%
^{2}}{2})d^{4}x,  \label{eqb5} \\
\mathbf{w} &=&\mathbf{u}+\frac{e}{mc}\mathbf{A\;,}  \nonumber \\
\rho &=&\mathbf{\psi \bullet }\rho ^{0}, \\
\mathbf{\widehat{u}} &=&\mathbf{\psi }_{t}+\mathbf{\psi }_{*}\mathbf{u}_{0}.
\end{eqnarray}

\begin{remark}
The fluid action,$\mathcal{\;A}_{F}=\mathcal{\overline{A}}_{F}+\mathcal{A}%
_{I},$ can further devided into one part which is simply the momentum space
integrated $\overline{\mathcal{A}_{F}}=\int (\int f\overline{\psi }_{t}d^{3}%
\mathbf{p)}d^{4}x$ if we identify $\frac{\widehat{\mathbf{u}}^{2}}{2}$ with $%
\frac{\mathbf{u}^{2}}{2}$ and one part which could be identified as the
electromagnetic/fluid interaction part, $\mathcal{A}_{I}=\int \rho \frac{e}{%
mc}\mathbf{A\cdot \widehat{u}}d^{4}x$.
\end{remark}

\begin{lemma}
The above fluid action is again nonstandard in the sense that the variations
has to be done with respect to a infinitesimal variation of the fluid
generator $\delta \mathbf{\psi }$ for the quantities which are parameterized
with respect to the reference fluid state $\rho ^{0},\;\mathbf{u}^{0}$ 
\begin{eqnarray}
\delta \rho &=&-\nabla \cdot (\delta \mathbf{\psi }\rho )\;,  \label{eq6b} \\
\delta \mathbf{\psi ,}_{t} &=&\delta \mathbf{\psi }_{t}+[\delta \mathbf{\psi
,\psi }_{t}]\;,  \nonumber \\
\delta \widehat{\mathbf{u}} &=&\delta \mathbf{\psi },_{t}-[\delta \mathbf{%
\psi },\mathbf{u]\;.}  \nonumber
\end{eqnarray}
The variations with respect to $\mathbf{u}$ and the electromagnetic
potential are standard. We have in our earlier work$^{8}$, found by using
the above relations that 
\begin{eqnarray*}
\frac{\delta \mathcal{A}_{F}}{\delta \mathbf{\psi }} &=&-\rho (\frac{%
\partial \mathbf{u}}{\partial t}+\widehat{\mathbf{u}}\cdot \nabla \mathbf{u-}%
\frac{1}{m}\mathbf{f}_{L})\;, \\
\mathbf{f}_{L} &\equiv &\frac{e}{m}(\mathbf{E+}\frac{1}{c}\widehat{\mathbf{u}%
}\times \mathbf{B)\;,} \\
\mathbf{E} &\equiv &-\frac{1}{c}\frac{\partial \mathbf{A}}{\partial t},\;%
\mathbf{B=}\nabla \times \mathbf{A\;,} \\
\frac{\delta \mathcal{A}_{F}}{\delta \mathbf{u}} &=&\rho (\widehat{\mathbf{u}%
}-\mathbf{u)}=\mathbf{0}\Rightarrow \mathbf{u=\widehat{u}\;.}
\end{eqnarray*}
It also seems natural to call the term $\mathcal{A}_{I}=-\int f\widehat{H}%
d^{6}zdt$ which correspond to the internal energy for the fluid-kinetic
interaction part of the action. The variation of this part of the action
with respect to the fluid generator gives the divergence of the stresstensor
needed to complete the fluid momentum equation. In this formulation of the
action principle the mass density continuity equation is implicitly given by
parameterization of density. 
\[
\frac{\delta \mathcal{A}_{I}}{\delta \mathbf{\psi }}\mathbf{=}\int \frac{%
\delta \overline{\psi }}{\delta \mathbf{\psi }}\{f,\hat{H}\}d^{3}\mathbf{p=-}%
\int \mathbf{p}\{f,\widehat{H}\}d^{3}\mathbf{p=-}\nabla \cdot (\Bbb{P})\;. 
\]
Here we have used the obvious lemma valid for phase space densities with a
suitable decay in infinity and an appropriate class of phase space
observables which $g(z,t)$ belongs to
\end{lemma}

\begin{lemma}
$\int \mathbf{G(x},t)\{g,f\}d^{3}\mathbf{p=}-\nabla \cdot \mathbf{(}\int 
\frac{\partial g}{\partial \mathbf{p}}fd^{3}\mathbf{p)G(x},t).$
\end{lemma}

The variation with respect to the infinitesimal generator $\delta \widetilde{%
\psi }$ gives us the reduced Liouville equation on fluid orbits with respect
to similar variational rules as we discussed in eq.(\ref{eqb3}) 
\begin{equation}
\frac{\delta \mathcal{A}_H}{\delta \widetilde{\psi }}=-\frac{\partial 
\widetilde{f}}{\partial t}-\{\widetilde{f},\overline{\psi }^{*}\widehat{H}%
\}=0\;.  \label{eqb7}
\end{equation}

\subsection{Revised hybrid fluid-kinetic action principle}

From the parameterized version of the hybrid fluid-kinetic action principle
it is possible to derive a revised action principle in the same way as we
did above for the Vlasov action principle. This is done simply by assuming
that $f,\;\widetilde{f},\;\mathbf{\psi }\;$and $\widetilde{\psi }$ depend on
an additional formal parameter $\epsilon $. We then find the revised hybrid
fluid-kinetic action principle

\begin{eqnarray}
\mathcal{A}_{H}^{(1)} &=&\int\limits_{0}^{1}\int (\mathbf{\psi }_{\epsilon
}\cdot (-\rho (\frac{\partial \mathbf{u}}{\partial t}+\widehat{\mathbf{u}}%
\cdot \nabla \mathbf{u-}\frac{1}{m}\mathbf{f}_{L}\mathbf{-}\underline{\nabla
\cdot \Bbb{P}})  \label{eqb8} \\
&&-(\mathbf{\psi }_{\epsilon }\cdot \mathbf{w)(}\frac{\partial \rho }{%
\partial t}+\nabla \cdot (\mathbf{\hat{u}}\rho )) \\
&&+\rho (\mathbf{u,}_{\epsilon }+\mathbf{\psi }_{\epsilon }\cdot \nabla 
\mathbf{u)}\cdot (\widehat{\mathbf{u}}-\mathbf{u)+}\rho \frac{e}{cm}\mathbf{%
A,}_{\epsilon }\cdot \widehat{\mathbf{u}})d^{4}xd\epsilon \\
&&+\int\limits_{0}^{1}\int \widetilde{f}(\widetilde{\psi }_{\epsilon ,t}+\{%
\widetilde{\psi }_{\epsilon }+\underline{\overline{\psi }^{*}\overline{\psi }%
_{\epsilon }},\overline{\psi }^{*}\widehat{H}\}-\overline{\psi }^{*}\frac{%
\partial \widehat{H}}{\partial \epsilon })d^{6}zdtd\epsilon \;.  \nonumber
\end{eqnarray}
The underlined terms in the fluid and kinetic part of the action are
interchangeable forms of the same term. With this revised action principle
we obtain the same equations as above by varying with respect to $\mathbf{%
\psi }_{\epsilon },\;\mathbf{u,}_{\epsilon }\;$and $\widetilde{\psi }%
_{\epsilon }..$

\section{App. C Rotation and divergence defined in an invariant way}

One way to define the rotation and divergence of a fluid element in an
invariant way is through the Hodge star operation relative to a metric [\cite
{geom}]. Another more intrinsic way is through the Lie derivative of a
volume element. Our interest in this is motivated by the the need to
formulate physical equations and here fluid dynamics in such a way that they
transform naturally with respect to diffeomorphisms. From a practical point
of view this is needed to formulate perturbation theory with low complexity
and new models. However, from a more fundamental point of view there is a
need for an intrinsic description of observable quantities like rotation and
divergence of the flow of a fluid element. The definition of the Hodge star
operator with respect to an invariant volume element $dV=JdV_{0}=\mathbf{%
\psi }^{-1*}dV_{0},$ give that one can easily check that it must behave
naturally with respect to diffeomorphisms 
\begin{eqnarray}
\alpha \wedge *_{\mathbf{g}}\beta &\equiv &\mathbf{g}^{-1}\mathbf{(}\alpha
,\beta )dV=<\alpha ,\mathbf{g}^{-1}(\beta )>dV  \label{eqc1} \\
\ast _{\mathbf{g}} &=&\mathbf{\psi }^{-1*}\circ *_{\mathbf{g}_{0}}\circ 
\mathbf{\psi }^{*}, \\
\mathbf{g} &=&\mathbf{\psi }^{-1*}\mathbf{g}_{0}.  \nonumber
\end{eqnarray}
Here $*_{\mathbf{g}}$and $*_{\mathbf{g}_{0}}$ is the Hodge star operation
with respect to $\mathbf{g}$ and $\mathbf{g}_{0}$ respectively and $\alpha $
and $\beta $ are forms in $\Lambda ^{k}T^{*}M$ for some $k$ $(k=1,2,3$ for
us in three dimensional space). By abuse of notation we will use the same
notation for the metric $\mathbf{g}$ and its inverse $\mathbf{g}^{-1}$ as
for the maps induced by them, e.g. here $\mathbf{g}^{-1}\beta \equiv \mathbf{%
g}^{-1}(\cdot ,\beta )$ is a contravariant multivectorfield in $\Lambda
^{k}T $ $M$. Moreover, $<\cdot ,\cdot >$ is the standard contraction between 
$\Lambda ^{k}T^{*}M$ and it's dual space $\Lambda ^{k}TM.$ A more direct
description of the action of the Hodge star operation is formulated by

\begin{description}
\item  \textbf{Lemma}

\item  $*_{\mathbf{g}}v=*_{\mathbf{g}_{0}}(J\mathbf{g}_{0}\circ \mathbf{g}%
^{-1}(v)),$

\item  $*_{\mathbf{g}}v=\frac{1}{J}\mathbf{g\circ g}_{0}^{-1}\circ *_{%
\mathbf{g}_{0}}v,$

\item  where $v$ is a form in $\Lambda ^{k}T^{*}M$ ,$k=1,2,3.$

\proof%
With respect to the metric $\mathbf{g}_{0}$ we have that $\alpha \wedge *_{%
\mathbf{g}_{0}}\beta \equiv \mathbf{<\alpha ,g}_{0}^{-1}(\beta )>dV_{0}.$
Therefore one find that $\alpha \wedge *_{\mathbf{g}}v=<\alpha ,\mathbf{g}%
_{0}^{-1}((\mathbf{g}_{0}\circ \mathbf{g}^{-1})v)>dV=<\alpha ,\mathbf{g}%
_{0}^{-1}(J(\mathbf{g}_{0}\circ \mathbf{g}^{-1})v)>dV_{0}$
\end{description}

$=\alpha \wedge *_{\mathbf{g}_{0}}(J(\mathbf{g}_{0}\circ \mathbf{g}^{-1})v%
\dot{)}.$ Since this equality is valid for all k-forms $\alpha ,$ the first
part of the lemma is proved. For the second part of the lemma we use that we
can write $v=*_{\mathbf{g}}v^{\prime }=*_{\mathbf{g}_{0}}v_{0}^{\prime }$
where $v^{\prime }=*_{\mathbf{g}}v$ and $\ v_{0}^{\prime }=*_{\mathbf{g}%
_{0}}v.$ Here we have used that in three space (same equality up to sign in
some other dimension) $*_{\mathbf{g}}\circ *_{\mathbf{g}}=*_{\mathbf{g}%
_{0}}\circ *_{\mathbf{g}_{0}}=1.$ The first part of the lemma then implies
that $v_{0}^{\prime }=*_{\mathbf{g}_{0}}v=J(\mathbf{g}_{0}\circ \mathbf{g}%
^{-1})v^{\prime }$ and consequently $v^{\prime }=*_{\mathbf{g}}v=\frac{1}{J}(%
\mathbf{g\circ g}_{0}^{-1})*_{\mathbf{g}_{0}}v $ . This proves the second
part of the lemma.

We are now in a position to state Hodge decomposition (we do not consider
singular contributions) with respect to a general metric as

\begin{theorem}
For $v,\,A,\,\eta $ as forms in $\Lambda ^{k}T^{*}M,$ $\Lambda ^{k-1}T^{*}M,$
$\Lambda ^{n-k-1}T^{*}M$ respectively ($n=3$ for $M$ three dimensional), we
have that 
\begin{eqnarray}
v &=&d\eta +*_{\mathbf{g}}dA=d\eta +\frac{1}{J}\mathbf{g\circ g}%
_{0}^{-1}\circ *_{\mathbf{g}_{0}}A,  \label{eqc2} \\
J\mathbf{g}^{-1}(v) &=&J\mathbf{g}^{-1}(d\eta )+\mathbf{g}_{0}^{-1}(*_{%
\mathbf{g}_{0}}dA),  \nonumber \\
v &=&\mathbf{\psi }^{-1*}(d\hat{\eta}+*_{\mathbf{g}_{0}}d\hat{A}),\,\eta =%
\mathbf{\psi }^{-1*}\eta ,\,\,A=\mathbf{\psi }^{-1*}\hat{A}.  \nonumber
\end{eqnarray}
\end{theorem}

The proof follows from direct use of the above lemma.

We will now specialize to oneforms and define divergence and rotation with
respect to a transformed metric $\mathbf{g.}$

\begin{definition}
$div_{\mathbf{g}}(\mathbf{v)\equiv }*_{\mathbf{g}}d*_{\mathbf{g}}\mathbf{v}%
^{(1)},$\thinspace \thinspace $curl_{\mathbf{g}}(\mathbf{v)\equiv g}^{-1}(*_{%
\mathbf{g}}d\mathbf{v}^{(1)}),\,\,\mathbf{v}^{(1)}=\mathbf{g(v).}$ Here $%
\mathbf{v}$ is a vectorfield in $TM,$ $\mathbf{v:}M\mathbf{\rightarrow }TM.$
\end{definition}

This definition leads to the following theorem

\begin{theorem}
\begin{eqnarray}
div_{\mathbf{g}}(\mathbf{v)} &=&\frac{1}{J}div_{\mathbf{g}_{0}}(J\mathbf{%
v)=\psi }^{-1*}(div_{\mathbf{g}_{0}}(\mathbf{\hat{v})),\,\,\,v=\psi }_{*}%
\mathbf{\hat{v},}  \label{eqc3} \\
curl_{\mathbf{g}}\mathbf{v} &=&\frac{1}{J}curl_{\mathbf{g}_{0}}(\mathbf{g}%
_{0}^{-1}\circ \mathbf{g(v))=\psi }_{*}(curl_{\mathbf{g}_{0}}\mathbf{\hat{v}%
).}  \nonumber
\end{eqnarray}
\end{theorem}

\proof%
We prove this by applying the above definition for divergence and curl. $%
div_{\mathbf{g}}(\mathbf{v)=*}_{\mathbf{g}}d*_{\mathbf{g}}\mathbf{v}%
^{(1)}=*_{\mathbf{g}}d*_{\mathbf{g}_{0}}(J\mathbf{v}_{0}^{(1)})=\frac{1}{J}%
*_{\mathbf{g}_{0}}d*_{\mathbf{g}_{0}}(J\mathbf{v}_{0}^{(1)})=\frac{1}{J}div_{%
\mathbf{g}_{0}}(J\mathbf{v),\,\,v}_{0}^{(1)}=\mathbf{g}_{0}(\mathbf{v).}$ On
the other hand, we have that $\mathbf{*}_{\mathbf{g}}d*_{\mathbf{g}}\mathbf{v%
}^{(1)}=\mathbf{\psi }^{-1*}(*_{\mathbf{g}_{0}}d*_{\mathbf{g}_{0}}\mathbf{%
\hat{v}}_{0}^{(1)})=\mathbf{\psi }^{-1*}(div_{\mathbf{g}_{0}}(\mathbf{\hat{v}%
)),\,\,\,v}_{0}^{(1)}=\mathbf{g}_{0}(\mathbf{\hat{v}).}$ Similarly for curl
we have that $curl_{\mathbf{g}}(\mathbf{v)=g}^{-1}(*_{\mathbf{g}}d\mathbf{v}%
^{(1)})=\frac{1}{J}\mathbf{g}^{-1}\circ \mathbf{g\circ g}_{0}^{-1}(*_{%
\mathbf{g}_{0}}d(\mathbf{g}_{0}(\mathbf{g}_{0}^{-1}\circ \mathbf{g}(\mathbf{v%
}))))$

$=\frac{1}{J}\mathbf{g}_{0}^{-1}(*_{\mathbf{g}_{0}}d(\mathbf{g}_{0}(\mathbf{g%
}_{0}^{-1}\circ \mathbf{g}(\mathbf{v}))))=\frac{1}{J}curl_{\mathbf{g}_{0}}(%
\mathbf{g}_{0}^{-1}\circ \mathbf{g}(\mathbf{v})).$ On the other hand we have
that $\mathbf{g}^{-1}(*_{\mathbf{g}}d\mathbf{v}^{(1)})=\mathbf{\psi }_{*}(%
\mathbf{g}_{0}^{-1}(*_{\mathbf{g}_{0}}d\mathbf{\hat{v}}^{(1)}))=\mathbf{\psi 
}_{*}(curl_{\mathbf{g}_{0}}(\mathbf{\hat{v})),}$ and the theorem is proved.

A more geometric way to study divergence independent of metric is by Lie
derivative of the invariant volumeform with respect to the velocity field $%
\mathcal{L}(\mathbf{v)}dV=di_{\mathbf{v}}dV\equiv div_{\mathbf{g}}(\mathbf{v)%
}dV=Jdiv_{\mathbf{g}}(\mathbf{v)}dV_{0}=di_{\mathbf{v}}(JdV_{0})=di_{(J%
\mathbf{v)}}dV_{0}=\mathcal{L}(J\mathbf{v)}dV_{0}=div_{\mathbf{g}_{0}}(J%
\mathbf{v)}dV_{0}.$ On the other hand we have that $\mathcal{L}(\mathbf{v)}%
dV=\mathbf{\psi }^{-1*}(\mathcal{L}(\mathbf{\hat{v})}dV_{0})=\mathbf{\psi }%
^{-1*}(div_{\mathbf{g}_{0}}(\mathbf{\hat{v}))}JdV_{0}.$ We therefore observe
that a divergence free vectorfield or purely rotational is simply a
vectorfield $\mathbf{v}_{c}$ in the kernel of the Lie operator, i.e. (by the
way an infinitesimal Lie equation by App. A) $\mathcal{L}(\mathbf{v}%
_{c})dV=div_{\mathbf{g}}(\mathbf{v}_{c})dV=div_{\mathbf{g}_{0}}(J\mathbf{v}%
_{c})dV_{0}=0.$ This is consistent with a parameterization of a rotational
vectorfield for an invariant volume element as $\mathbf{v}_{c}=curl_{\mathbf{%
g}}(\mathbf{A)=}\frac{1}{J}curl_{\mathbf{g}_{0}}\mathbf{A}_{0}=\mathbf{\psi }%
_{*}(curl_{\mathbf{g}_{0}}(\mathbf{\hat{A}))}$ where $\mathbf{A}_{0}\equiv 
\mathbf{g}_{0}^{-1}\circ \mathbf{g(A),\,\,A=\psi }_{*}\mathbf{\hat{A}\,.}$
The Hodge decomposition then give us that the velocity field can be
decomposed with respect to an invariant volume element as

\begin{equation}
\mathbf{v}^{(1)}=-d\eta +*_{\mathbf{g}}d\mathbf{A}^{(1)}=-d\eta +\frac{1}{J}%
\mathbf{g\circ g}_{0}^{-1}(*_{\mathbf{g}_{0}}d\mathbf{A}^{(1)}).
\label{eqc4}
\end{equation}

However, with respect to the reference metric $\mathbf{g}_{0}$ we could
consider the decomposition of $J\mathbf{v}^{(1)}$ or $J\mathbf{v.}$ In fact
with respect to a reference state where the fluid is fixed and homogenous $%
(\rho _{0}^{f}$ is constant and $J^{f}$ is the Jacobian with respect to the
corresponding diffeomorphism$)$, we could just as well multiply by the
constant mass density and obtain since $\rho =J^{f}\rho _{0}^{f}$ a
decomposition of $\rho \mathbf{v}^{(1)}$ or $\rho \mathbf{v}$ as $\rho 
\mathbf{v=-}\rho \nabla _{\mathbf{g}}\eta +curl_{\mathbf{g}_{0}}(\rho
_{0}^{f}\mathbf{g}_{0}^{-1}(\mathbf{A}^{(1)}))\,,\,\,\,\nabla _{\mathbf{g}%
}\eta \equiv \mathbf{g}_{0}\circ \mathbf{g}^{-1}(d\eta \dot{)}.\,$This
decomposition is interesting since it is exactly the one we need in
connection with the discussion of the pseudogroup defined by the continuity
equation defined in App.A.

\subsection{Rotational bracket structure}

We will represent a rotational vectorfield $\mathbf{v}_{c}$(or one
form\thinspace $\mathbf{v}_{c}^{(1)}$)by a Pfaff decomposition of the one
form $\mathbf{A}^{(1)}$ which in the nonsingular case is $\mathbf{A}%
^{(1)}=\alpha d\beta +d\gamma .$ The rotational vectorfield is then $\mathbf{%
v}_{c}^{(1)}=*_{\mathbf{g}}d(\alpha d\beta ),$ i.e. we can mod out $\gamma .$
We want to think about $\beta $ as a family of level surfaces (foliations)
which the rotational vectorfield is situated on. With this interpretation in
mind we will use the notation $\mathbf{v}_{c}=\mathbf{X}_{\alpha }^{\beta
}\,=\mathbf{g}^{-1}(\mathbf{v}_{c}^{(1)})\,,\,\mathbf{v}_{c}^{(1)}=\mathbf{X}%
_{\alpha }^{\beta (1)}$ for a given metric $\mathbf{g.}$ Analogous with the
Poisson bracket in phase space we define the new rotational bracket on the
foliations defined by $\beta $ as $\{\alpha ,f\}_{\beta }\equiv -\mathbf{X}%
_{\alpha }^{\beta }(f)=*_{\mathbf{g}}(\mathbf{X}_{\alpha }^{\beta (1)}\wedge
*_{\mathbf{g}}df)=<df,\mathbf{g}^{-1}(\mathbf{X}_{\alpha }^{\beta (1)})>\,=%
\mathbf{g}^{-1}(df,\mathbf{X}_{\alpha }^{\beta (1)}).$ The equivalence of
the first two definitions comes through that both of them are by trivial use
of the definitions equivalent to the third expression. We can now find the
following lemma valid for rotational one forms represented with respect to a
given foliation.

\begin{lemma}
\begin{eqnarray}
\ast _{\mathbf{g}}(\mathbf{X}_{\alpha _{1}}^{\beta ^{(1)}}\wedge \mathbf{X}%
_{\alpha _{2}}^{\beta (1)}) &=&\{\alpha _{1},\alpha _{2}\}_{\beta }\,d\beta ,
\label{eqc5} \\
\lbrack \mathbf{X}_{\alpha _{1}}^{\beta },\mathbf{X}_{\alpha _{2}}^{\beta }]
&=&\mathbf{X}_{-\{\alpha _{1},\alpha _{2}\}_{\beta }}^{\beta }\,\,, 
\nonumber \\
\mathbf{X}_{\alpha }^{\beta }(f(\beta )) &=&0\,\,\text{where }f\text{ is
differentiable.}  \nonumber
\end{eqnarray}
\end{lemma}

\proof%
For the proof of the first identity we find that $*_{\mathbf{g}}(\mathbf{X}%
_{\alpha _{1}}^{\beta ^{(1)}}\wedge \mathbf{X}_{\alpha _{2}}^{\beta (1)})=*_{%
\mathbf{g}}(*_{\mathbf{g}}(d\alpha _{1}\wedge d\beta )\wedge \mathbf{X}%
_{\alpha _{2}}^{\beta (1)})=*_{\mathbf{g}}(d\beta \wedge *_{\mathbf{g}}(%
\mathbf{X}_{\alpha _{2}}^{\beta (1)}\wedge d\alpha _{1})).$ Now, we use the
identity $*_{\mathbf{g}}(\mathbf{a}^{(1)}\wedge *_{\mathbf{g}}(\mathbf{b}%
^{(1)}\wedge \mathbf{c}^{(1)}))=\mathbf{g}^{-1}(\mathbf{a}^{(1)},\mathbf{c}%
^{(1)})\,\mathbf{b}^{(1)}-\mathbf{g}^{-1}(\mathbf{b}^{(1)},\mathbf{c}%
^{(1)})\,\mathbf{a}^{(1)}$(analogue to the classical triple crossproduct) to
derive $*_{\mathbf{g}}(\mathbf{X}_{\alpha _{1}}^{\beta ^{(1)}}\wedge \mathbf{%
X}_{\alpha _{2}}^{\beta (1)})=-\mathbf{g}^{-1}(\mathbf{X}_{\alpha
_{2}}^{\beta (1)},d\alpha _{1})d\beta +\mathbf{g}^{-1}(\mathbf{X}_{\alpha
_{2}}^{\beta (1)},d\beta )d\alpha _{1}.$ But we have that $\mathbf{X}%
_{\alpha _{2}}^{\beta }(\beta )=-*_{\mathbf{g}}(*_{\mathbf{g}}(d\alpha
_{2}\wedge d\beta )\wedge d\beta )=0$ which implies $*_{\mathbf{g}}(\mathbf{X%
}_{\alpha _{1}}^{\beta (1)}\wedge \mathbf{X}_{\alpha _{2}}^{\beta
(1)})=\{\alpha _{1},\alpha _{2}\}_{\beta }\,d\beta .$ This statement also
proves the third part of the lemma since $\mathbf{X}_{\alpha }^{\beta
}(f(\beta ))=f^{\prime }(\beta )\mathbf{X}_{\alpha }^{\beta }(\beta )=0.$
The second identity is established by that for purely rotational
vectorfields (c.f. the section about parametrization of hybrid fluid kinetic
theory in the main text) 
\[
\lbrack \mathbf{X}_{\alpha _{1}}^{\beta },\mathbf{X}_{\alpha _{2}}^{\beta
}]^{(1)}=-*_{\mathbf{g}}d(*_{\mathbf{g}}(\mathbf{X}_{\alpha _{1}}^{\beta
(1)}\wedge \mathbf{X}_{\alpha _{2}}^{\beta (1)}))=-*_{\mathbf{g}}d(\{\alpha
_{1},\alpha _{2}\}d\beta )=\mathbf{X}_{-\{\alpha _{1},\alpha _{2}\}}^{\beta
(1)}.
\]
Therefore there is a Lie antihomomorphism between the Lie algebra of
rotational vectorfields and the corresponding Lie algebra with respect to
the rotational bracket on the space of rotational potentials for a given
family of foliations of space. All the usual relations for Lie algebra's
like Jacobi identity e.t.c. follows for the rotational bracket structure
through this antihomorphism.%
\endproof%

\section{Acknowledgement}

The initial stages of this project was started while the author was at the
Center for Advanced Studies, Norwegian Academy of Sciences and Letters in
1994. I have continually discussed with P. Jakobsen and V. Lychagin about
the mathematically aspects of the project. I had preliminary ideas for
parameterization and separation into hybrid fluid-kinetic theory of the
linearized Vlasov-Maxwell equations already in 1993. This was presented in
the proceedings of the Oslo Astrophysics Minisymposium, 1993. However, the
real initialization of this project has to be dated back to my sabbatical at
LBL, Berkeley with A.N. Kaufmann and his influential theoretical plasma
physics group.

\end{document}